\documentclass[11pt]{article}
\usepackage{fullpage}
\usepackage{cite}
\usepackage{amsmath}
\usepackage{amssymb}
\usepackage{graphicx}
\title{}
\date{}
\def\para{\\ [-2mm]}

\def\eqn#1{eq.~(\ref{#1})} \def\Eqn#1{Equation~(\ref{#1})}
\def\eqns#1#2{eqs.~(\ref{#1}) and~(\ref{#2})}

\def \be  {\begin{equation}}
\def \ee  {\end{equation}}
\def \ba  {\begin{eqnarray}}
\def \ea  {\end{eqnarray}}
\newcommand{\nn}{\nonumber}

\def\ie{{i.e.,~}}
\def\viz{{viz.,~}}
\def\eg{{e.g.,~}}
\def \Tr {\mathop{\rm Tr}\nolimits}

\def\id{  {\mathsf{1}\kern -3pt \mathsf{l} } }
\def\IZ{\relax\ifmmode\mathchoice
{\hbox{\cmss Z\kern-.4em Z}}{\hbox{\cmss Z\kern-.4em Z}}
{\lower.4pt\hbox{\cmsss Z\kern-.4em Z}}
{\lower1.2pt\hbox{\cmsss Z\kern-.4em Z}}\else{\cmss Z\kern-.4em Z}\fi}
\newcommand{\Z}{\mathsf{Z}\kern -5pt \mathsf{Z}}
\newcommand{\unit}{\mathsf{1}\kern -3pt \mathsf{l}}

\def\ve {\varepsilon}

\def\barpsi{{\bar\psi}}
\def\kslash{ \rlap{/}k  }
\def\kaslash{ \rlap{/}{k_a}  }
\def\Kslash{ \rlap{\,/}K  }
\def\epsslash{ \rlap{/}{\ve_3}  }
\def\epsaslash{ \rlap{/}{\ve_a}  }
\def\V {\gamma}

\def\cA {  {\cal A} }
\def\cAn {  {\cal A}_n }

\def\cN {  {\cal N} }
\def\cO {  {\cal O} }

\def\bc{  {\bf c} }
\def\bn{  {\bf n} }

\def\ta {\textsf{a}}
\def\tb {\textsf{b}}
\def\tc {\textsf{c}}
\def\td {\textsf{d}}
\def\te {\textsf{e}}
\def\ti {\textsf{i}}
\def\tj {\textsf{j}}
\def\tk {\textsf{k}}

\def\I { {I'} }
\def\alp{\alpha}
\def\bet{\alpha'}
\def\Sai{ S_{a,i} }
\def\SaB{ S_{a,B} }
\def\SaC{ S_{a,C} }
\def\aIvr{  { (a,\I,v,r) } }
\def\aIvrr{  { (r) } }
\def\aIvrt{  { (3) } }
\def\aIvrf{  { (4) } }
\def\ccr { c_{(r)} } 
\def\nr { n_{(r)} } 
\def\crp { c'_{(r)} } 
\def\nrp { n'_{(r)} } 
\def\sumr{  \sum_{r=1}^3 }
\def\sumrr{  \sum_{r=1}^4 }
\def\deltatwo { \delta_2 ~ }
\def\cAbi{ {\cal A}^{{\rm scalar}} }

\begin{document}

\titlepage
\begin{flushright}
BOW-PH-162\\
MCTP-16-17\\
\end{flushright}

\vspace{3mm}

\begin{center}
{\Large\bf\sf
BCJ relations from a new symmetry of gauge-theory amplitudes
}

\vskip 1.5cm

{\sc
Robert W. Brown$^a$
and Stephen G. Naculich$^{b,c}$
}

\vskip 0.5cm
$^a${\it
Department of Physics\\
Case Western Reserve University\\
Cleveland, OH 44106 USA
}

\vskip 0.5cm
$^b${\it
Department of Physics\\
Bowdoin College\\
Brunswick, ME 04011 USA
}

\vskip 0.5cm
$^c${\it
Michigan Center for Theoretical Physics (MCTP)\\
Department of Physics\\
University of Michigan\\
Ann Arbor, MI 48109 USA
}

\vspace{5mm}
{\tt
rwb@case.edu,  naculich@bowdoin.edu
}
\end{center}

\vskip 1.5cm

\begin{abstract}

We introduce a new set of symmetries 
obeyed by tree-level gauge-theory amplitudes 
involving at least one gluon.
The symmetry acts as a momentum-dependent shift on the color factors
of the amplitude. 
Using the radiation vertex expansion,
we prove the invariance under this color-factor shift of 
the $n$-gluon amplitude,
as well as amplitudes involving 
massless or massive particles 
in an arbitrary representation of the gauge group
with spin zero, one-half, or one.
The Bern-Carrasco-Johansson relations are a 
direct consequence of this symmetry. 

We also introduce the cubic vertex expansion of an amplitude, 
and use it to derive a generalized-gauge-invariant constraint 
on the kinematic numerators of the amplitude.
We show that the amplitudes of the bi-adjoint scalar theory 
are invariant under the color-factor symmetry, and 
use this to derive the null eigenvectors of the propagator matrix.

We generalize the color-factor shift to loop level,
and prove the invariance under this shift 
of one-loop $n$-gluon amplitudes in any theory 
that admits a color-kinematic-dual representation
of numerators.
We show that the one-loop color-factor symmetry 
implies known relations among the integrands of one-loop color-ordered amplitudes.

\end{abstract}

\vspace*{0.5cm}

\vfil\break

\section{Introduction}
\setcounter{equation}{0}

In 2008, Bern, Carrasco, and Johansson discovered
a novel set of linear relations satisfied by 
tree-level color-ordered amplitudes in gauge theories\cite{Bern:2008qj}.
They arrived at these relations 
by writing the tree-level $n$-gluon amplitude 
as a sum over $(2n-5)!!$ diagrams assembled from cubic vertices
\be
\cAn ~=~ \sum_i {c_i ~ n_i \over d_i } 
\label{cubicdecompintro}
\ee
where the color factor $c_i$ associated with the diagram
is composed of group theory structure constants $f_{\ta\tb\tc}$,
the denominator $d_i$ consists of the product of the inverse propagators
associated with the diagram,
and the kinematic numerator $n_i$ 
depends on the momenta and polarizations of the gluons.
All contributions from diagrams with quartic vertices are redistributed
among the cubic diagrams.
By virtue of the Jacobi identity
$ 
 f_{\ta \tb \te} f_{\tc \td \te} 
+f_{\ta \tc \te} f_{\td \tb \te} 
+f_{\ta \td \te} f_{\tb \tc \te}=0 
$
satisfied by the structure constants, 
the color factors $c_i$ obey 
a set of Jacobi relations of the form
\be
c_i + c_j + c_k ~=~ 0  \,.
\ee
Because of these linear dependences, 
the kinematic numerators $n_i$ are not uniquely defined, but can undergo
{\it generalized gauge transformations}
$n_i \to n_i + \delta n_i$
which leave \eqn{cubicdecompintro} unchanged
\cite{Bern:2010ue,Bern:2010yg}.
The authors of ref.~\cite{Bern:2008qj} conjectured that there exists a generalized gauge 
in which the kinematic numerators satisfy the same algebraic relations
as the color factors; in particular,
they can be made to satisfy kinematic Jacobi relations
\be
n_i + n_j + n_k ~=~ 0  \,.
\label{kinematicjacobi}
\ee
{}From this assumption of {\it color-kinematic duality}, 
they demonstrated the existence of 
new relations (subsequently known as BCJ relations) 
satisfied by the color-ordered amplitudes $A(1, \cdots, n)$. 
These relations can be derived from the 
{\it fundamental BCJ relation} (and permutations thereof)
\cite{BjerrumBohr:2009rd,Feng:2010my,Sondergaard:2011iv}
\be
\sum_{b=3}^n \left( \sum_{c=1}^{b-1} k_2 \cdot k_c \right) 
A(1, 3, \cdots, b-1, 2, b, \cdots, n) 
~=~ 0
\label{fundbcjintro}
\ee
where $k_a$ are the (outgoing) momenta of the gluons.
Besides color-kinematic duality, these relations rely on the properties of the propagator 
matrix \cite{Vaman:2010ez}, constructed from the  inverse denominators $1/d_i$ 
(see sec.~\ref{sec:prop} for a precise definition).
Specifically, as a consequence of momentum conservation, 
this $(n-2)! \times (n-2)!$ matrix has rank $(n-3)!$, 
and consequently possesses a set of $(n-3) (n-3)!$ eigenvectors
with eigenvalue zero.
\para

The BCJ relations (\ref{fundbcjintro}) 
were subsequently proven using string-theory techniques
\cite{BjerrumBohr:2009rd,Stieberger:2009hq}
and BCFW on-shell recursion \cite{Feng:2010my,Chen:2011jxa},
providing evidence for the conjecture 
of tree-level color-kinematic duality.
Bern et al. conjectured that color-kinematic duality also applies to the 
integrands of loop-level amplitudes \cite{Bern:2008qj,Bern:2010ue}; 
while not proven, this conjecture has been tested for 
$\cN=4$ supersymmetric Yang-Mills theory through four 
loops \cite{Carrasco:2011mn,Bern:2011rj,Bern:2012uf,Yuan:2012rg,Boels:2012ew,Carrasco:2012ca,Bjerrum-Bohr:2013iza,Bern:2013uka,Bern:2014sna}, 
and for pure Yang-Mills theory through two loops \cite{Bern:2013yya,Nohle:2013bfa}.
Another exciting aspect of the story is that gauge-theory kinematic numerators 
obeying color-kinematic duality can be used to construct gravitational 
amplitudes via the double copy procedure \cite{Bern:2008qj,Bern:2010ue,Bern:2010yg}.
A recent review of all of these developments may be found in ref.~\cite{Carrasco:2015iwa}.
\para

Despite the fact that the BCJ relations for $n$-gluon amplitudes
have been definitively established,
interest in tree-level kinematic numerators continues, 
not least because the numerators that are naturally 
generated by Feynman rules\footnote{
String theory can generate numerators that respect color-kinematic 
duality \cite{Mafra:2011kj}.}
generally do not obey the relations
(\ref{kinematicjacobi}) except in the case of four-point 
amplitudes \cite{Zhu:1980sz,Goebel:1980es}.
Many approaches have been developed to obtain 
kinematic numerators that obey color-kinematic duality
directly from a Lagrangian 
approach \cite{Bern:2010yg,Monteiro:2011pc,BjerrumBohr:2012mg,Fu:2012uy,Boels:2013bi,Tolotti:2013caa,Vaman:2014iwa,Mastrolia:2015maa,Lee:2015upy,Mafra:2015vca,Fu:2016plh}.  
\para

In this paper, we introduce a new set of symmetries 
obeyed by tree-level gauge-theory amplitudes,
associated with each external gluon in the amplitude.\footnote{For bi-adjoint scalar theories,
there is a symmetry for each external massless adjoint scalar in the amplitude.}
These symmetries act on the color factors $c_i$ of the amplitude,
shifting them by momentum-dependent quantities. 
Since color factors do not carry any momentum dependence, 
this is a purely formal operation;
we prove, however, that the tree-level $n$-gluon amplitude is invariant
under these shifts by writing it in an alternative form 
known as the {\it radiation vertex expansion} \cite{Brown:1982xx}.
\para

We then show that the BCJ relations (\ref{fundbcjintro})
follow as an immediate consequence of the color-factor symmetry 
of the $n$-gluon amplitude.
Although BCJ relations have been previously established, 
our results reveal a more direct connection to the 
symmetries of the Lagrangian formulation of gauge theory 
and its Feynman rules (\ie gauge and Poincar\'e invariance)
and provide a basis for generalizations.
\para


Let us describe this symmetry in a bit more detail,
reserving a full description for section \ref{sec:cfs}.
Given a tree-level $n$-gluon color factor $c_i$,  
the choice of one of the external gluon legs $a$ divides 
the diagram in two at its point of attachment.  
Let $\Sai$ denote the subset of the remaining legs on one side
of this point;  
it does not matter which side we choose.
The shift of the color factor $c_i$ associated with gluon $a$ 
must satisfy
\be
\delta_a c_i  ~\propto~  \sum_{c\in \Sai } k_a \cdot k_c  \,.
\label{colorfactorshiftintro} 
\ee
Choosing to sum over the complement of $\Sai$ gives the same result 
(up to sign)
due to momentum conservation. 
The constants of proportionality in \eqn{colorfactorshiftintro} 
are then constrained by requiring that the shifted color factors 
respect all the Jacobi relations satisfied by $c_i$
for any values of the momenta.
\para

\begin{figure} 
\begin{center}
\includegraphics[width=8.0cm]{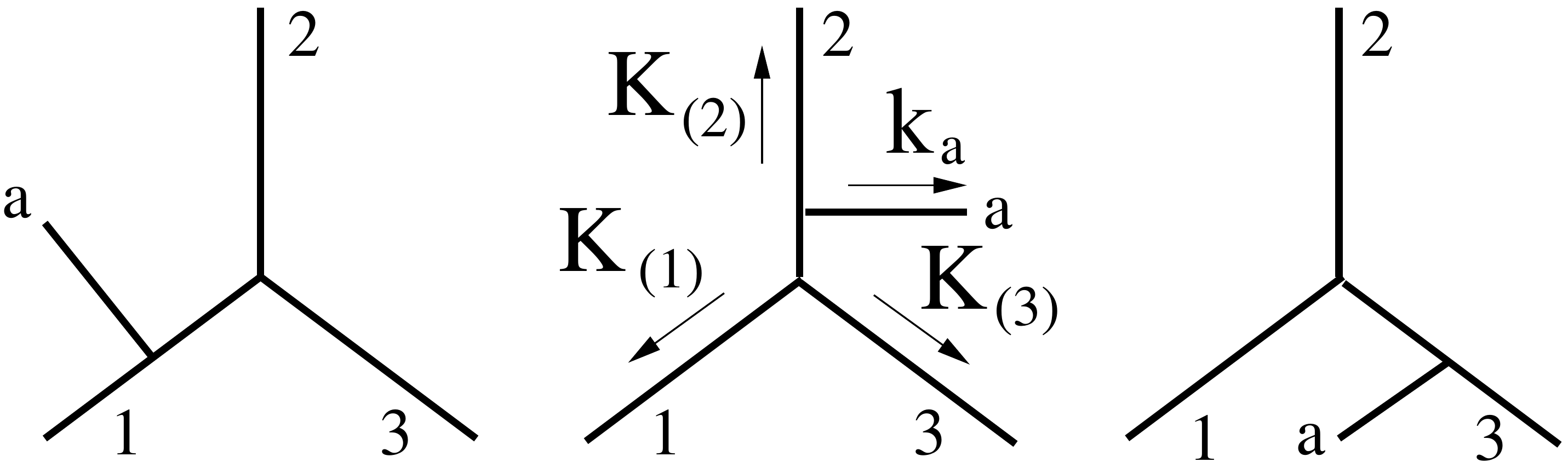}
\caption{Attaching a gluon to the legs of a cubic vertex.
These form parts of the color factors 
$c_{(1)}$,  $c_{(2)}$,   and $c_{(3)}$,  respectively.}
\label{fig:one}
\end{center}
\end{figure}

Consider the case where $a$ 
is one of the legs involved in the Jacobi identity (see fig.~\ref{fig:one}).
Imagine that each of the three graphs in fig.~\ref{fig:one} 
is embedded in a larger tree diagram, the same for each.
Denote the color factors associated with each diagram 
by $\ccr$, where $r=1, 2, 3$.
For example, the color factor $c_{(1)}$ for the figure on the left 
contains $ \cdots f_{\ta_{1} \ta_{a} \tb} f_{\tb \ta_{2} \tb_3} \cdots$,
where the labels on $f_{\ta \tb \tc}$ follow the diagram in clockwise order.  
Flipping $a$ to the other side of a leg changes the sign
of the color factor due to the antisymmetry of $f_{\ta \tb \tc}$. 
As a result of the Jacobi identity, the color factors obey $\sumr \ccr = 0$.
By \eqn{colorfactorshiftintro},  the shifts of these 
color factors are 
\be
\delta_a \ccr  =  \alpha_{(r)}  \,  k_a \cdot K_{(r)} 
\ee
where $K_{(r)}$ is the momentum flowing out of each leg.
Requiring $\sumr \delta_a \ccr = 0$
implies that $\alpha_{(r)}$ is independent of $r$,
as a result  of momentum conservation and masslessness of the gluon.
A more detailed description of the color-factor shifts is 
given in sec.~\ref{sec:cfs}.
\para

The symmetry we have introduced has roots in the 
radiation symmetry \cite{Brown:1983pn}
that underlies the general radiation zero 
theorem \cite{Brodsky:1982sh,Brown:1982xx,Brown:1983vk}.
In theories with local gauge couplings 
and spins $\le 1$, 
all single-photon tree amplitudes vanish if the ratios 
$Q_c /k_a \cdot k_c $ are all equal,\footnote{A universal ratio is restrictive
and few photon amplitudes have zeros in the physical phase space.}
where $k_a$ is the photon momentum, 
and $c$ labels external particles with momentum $k_c$ and charge $Q_c$.
These spin-independent zeros have spin-dependent counterparts 
where $Q_c$ are replaced by numerators $J_c$,
closely related to the kinematic numerators $n_i$ in \eqn{cubicdecompintro}. 
The underlying radiation symmetry refers to invariance under 
$Q_c \to Q_c + \alpha k_a \cdot k_c$
as well as 
$J_c \to J_c + \beta k_a \cdot k_c$
for arbitrary $\alpha$ and  $\beta$.
The extension to nonabelian ``charges'' has also been considered 
and the details behind a nonabelian radiation vertex expansion 
discussed \cite{Brown:1982xx,Brown:1983pn}.
The general color-factor symmetry introduced here,
however, incorporates crucial nonabelian constraints 
(Jacobi relations) on $\alpha$,
which lead to a complete set of BCJ relations,
and have not heretofore been developed.
Nevertheless, since the color-factor symmetry relies on the
presence of massless gauge bosons,
we may regard it as a generalized radiation symmetry.
\para

We also introduce in this paper the  
{\it cubic vertex expansion} of an $n$-point amplitude $\cAn$ 
with respect to one of the gluons $a$.
Consider the set of cubic diagrams $I$ that contribute to the $(n-1)$-point
amplitude of all the particles in $\cAn$ except for gluon $a$.
We show that, for any $a$, 
the amplitude $\cAn$ can be written as a triple sum
over the legs $r$ of the vertices $v$ of the cubic diagrams $I$:
\be
\cAn ~=~ 
\sum_I  
\sum_v  {1 \over \prod_{s=1}^3 d_{(a,I,v,s)}  }
\sumr 
{ c_{(a,I,v,r)} n_{(a,I,v,r)} \over 2 k_a \cdot K_{(a,I,v,r)}   } \,.
\ee
Here  
$d_{(a,I,v,r)}$ is the product of inverse propagators that branch off 
from leg $r$ of vertex $v$ of diagram $I$,
$c_{(a,I,v,r)}$ is the color factor of the $n$-point diagram 
obtained by attaching 
gluon $a$ to leg $r$ of vertex $v$ of diagram $I$
(exactly as in fig.~\ref{fig:one}),
and $n_{(a,I,v,r)}$ is the associated  $n$-point kinematic numerator.
The shift of $c_{(a,I,v,r)}$ associated with gluon $a$ is
$\delta_a \, c_{(a,I,v,r)} ~=~   \alpha_{(a,I,v)} \,k_a \cdot K_{(a,I,v,r)}$,
where, as explained above,  $\alpha_{(a,I,v)} $ is independent of $r$.
Since the alternative radiation vertex expansion shows that 
the amplitude $\cAn$ is invariant under the color-factor shift,
we may conclude from the cubic vertex expansion of $\cAn$ that
\be
\sum_I  \sum_v
{\alpha_{(a,I,v)}  \over \prod_{s=1}^3 d_{(a,I,v,s)}  }
\sumr 
n_{(a,I,v,r)}  ~ = ~ 0 \,.
\label{sumoverdeltaintro}
\ee
Note that this constraint on the kinematic numerators,
less stringent than the kinematic Jacobi relations
(which state that $\sumr n_{(a,I,v,r)}   = 0$ for each vertex),
is nonetheless sufficient to imply the BCJ relations (\ref{fundbcjintro}).
Moreover, unlike the kinematic Jacobi relations,
the condition (\ref{sumoverdeltaintro}) 
is invariant under generalized gauge transformations.
A constraint of precisely the form (\ref{sumoverdeltaintro})
was derived in refs.~\cite{BjerrumBohr:2010zs,Tye:2010dd}
for the five-gluon amplitude 
using the monodromy properties of string theory amplitudes.
\para

We show in this paper that more general gauge-theory amplitudes,
with both gluons and massless or massive particles 
in an arbitrary representation of the gauge group 
and with arbitrary spin $\le 1$,
are also invariant under the color-factor symmetry.
Consequently, the kinematic numerators of these amplitudes 
obey a constraint analogous to \eqn{sumoverdeltaintro}.
We further show that color-factor symmetry 
implies BCJ relations for the color-ordered amplitudes 
of a class of $n$-point amplitudes involving $n-2$ gluons 
and a pair of particles in an arbitrary representation of the gauge group 
and arbitrary spin,
as previously conjectured in refs.~\cite{Naculich:2014naa,Johansson:2015oia}.
\para

BCJ relations for the primitive amplitudes of a more 
general class of amplitudes containing gluons and 
an arbitrary number of pairs of differently flavored fundamentals
(based on a proper decomposition developed 
by Melia \cite{Melia:2013bta,Melia:2013epa,Melia:2015ika}
and Johansson and Ochirov \cite{Johansson:2015oia})
were conjectured by Johansson and Ochirov \cite{Johansson:2015oia},
and subsequently proven using BCFW on-shell recursion 
by de la Cruz, Kniss, and Weinzierl \cite{delaCruz:2015dpa}.
In a sequel to this paper \cite{Brown:2016hck},
we prove that these BCJ relations also follow as a direct consequence of 
the color-factor symmetry.
\para

The amplitudes of the theory of massless bi-adjoint scalars 
with cubic interactions\cite{Cachazo:2013iea}
also exhibit invariance under color-factor symmetry,
as we show using the cubic vertex expansion.   
In this case, the color-factor shifts are associated with 
each massless adjoint scalar in the amplitude.
As a consequence, we demonstrate the reduced rank of the propagator matrix
for the $n$-gluon gauge-theory amplitude
by deriving the set of its null eigenvectors.
\para

Finally, we generalize the cubic vertex expansion and 
color-factor symmetry to loop-level amplitudes 
containing at least one external gluon.
We exhibit an independent set of shifts that 
act on the color factors of one-loop $n$-gluon amplitudes
and which depend on the loop momentum as well as external momenta.
These one-loop amplitudes are invariant under color-factor shifts 
in theories that admit a color-kinematic-dual representation of numerators.
The color-factor symmetry also implies 
certain relations among the integrands of one-loop 
color-ordered amplitudes that were previously 
uncovered in refs.~\cite{Boels:2011tp,Boels:2011mn,Du:2012mt}.
\para

The contents of this paper are as follows.
In sec.~\ref{sec:cfs} we define the color-factor shift for the $n$-gluon amplitude
and derive the BCJ relations as a consequence of the invariance of the amplitude
under this shift.
We also introduce the cubic vertex expansion, and use it to 
derive a generalized-gauge-invariant constraint on the kinematic numerators
of the $n$-gluon amplitude.
We introduce an analogous set of shifts of the kinematic numerators, 
and show that they correspond to a generalized gauge transformation.
In sec.~\ref{sec:fourgluon}, we prove the invariance of the four-gluon amplitude
under the color-factor symmetry, and in 
sec.~\ref{sec:rvegluon}, we extend this to the $n$-gluon amplitude
by using the radiation vertex expansion.
In sec.~\ref{sec:fund}, we define 
the color-factor shift for more general amplitudes, 
and derive the BCJ relations
for the class of amplitudes containing $n-2$ gluons 
and a pair of particles in an arbitrary representation ${\cal R}$. 
In sec.~\ref{sec:fourfund}, we prove the invariance of the 
four-point amplitude with two gluons 
and a pair of massive particles of 
arbitrary spin  $\le 1$ and representation ${\cal R}$
under the color-factor symmetry, 
and in sec.~\ref{sec:rvefund}, 
we extend this to a general $n$-point amplitude containing gluons 
and other particles. 
In sec.~\ref{sec:prop}, we prove the invariance of the amplitudes of the bi-adjoint scalar theory 
under the color factor symmetry, and derive the null eigenvectors of the propagator matrix.
In sec.~\ref{sec:loop}, we generalize the cubic vertex expansion
and color-factor symmetry to loop-level amplitudes,
and derive a constraint on the integrands of one-loop color-ordered amplitudes.
Section \ref{sec:concl} contains a discussion and conclusions.
In appendix \ref{sec:app}, we write the shifts 
for all the color factors of the five-gluon amplitude, 
and derive the explicit constraint on the kinematic numerators 
that follow from the color-factor symmetry.

\section{Color-factor symmetry and its consequences}
\setcounter{equation}{0}
\label{sec:cfs}

We begin this section by introducing the color-factor symmetry 
in the simplest context, the tree-level four-gluon amplitude
\be
\cA_4 ~=~   {c_s n_s \over s} + {c_t n_t \over t} + {c_u n_u \over u} 
\label{fourpointamplitude}
\ee
where 
\be 
c_s ~=~ f_{\ta_{1} \ta_{2} \tb} f_{\tb \ta_{3} \tb_4}\,, \qquad
c_t ~=~ f_{\ta_{1} \ta_{4} \tb} f_{\tb \ta_{2} \tb_3}\,, \qquad
c_u ~=~ f_{\ta_{1} \ta_{3} \tb} f_{\tb \ta_{4} \tb_1}
\ee
and $s$, $t$, and $u$ are Mandelstam variables.
We define the four-point color-factor shift to act 
as\footnote{In the case of the four-gluon amplitude,
the shifts associated with various legs are all the same.} 
\be
c_s ~\to~ c_s + \alpha  ~s,\qquad 
c_t ~\to~ c_t + \alpha  ~t,\qquad 
c_u ~\to~ c_u + \alpha  ~u
\label{fourpointcfs}
\ee
where $\alpha$ is arbitrary.
\Eqn{fourpointcfs} preserves the Jacobi relation
$c_s+ c_t+c_u=0$
by virtue of momentum conservation 
$s+t+u=0$.
\para

The statement that \eqn{fourpointamplitude} is invariant under \eqn{fourpointcfs} implies
the kinematic Jacobi relation
\be
n_s+ n_t+n_u~=~0 \,.
\label{fourpointkinematicjacobi} 
\ee
It is well-known \cite{Zhu:1980sz,Goebel:1980es}
that \eqn{fourpointkinematicjacobi} is satisfied 
in the case of the four-gluon amplitude,
and we will show this explicitly in sec.~\ref{sec:fourgluon}.
This serves as proof of the invariance of the four-gluon amplitude 
under the color-factor shift.
\para

Recall that the kinematic numerators $n_i$ 
are not uniquely defined by \eqn{fourpointamplitude} 
because a generalized gauge transformation
\be
n_s \to n_s + \beta ~s,\qquad 
n_t \to n_t + \beta ~t,\qquad 
n_u \to n_u + \beta ~u
\label{fourpointggt}
\ee
(with $\beta$ arbitrary)
leaves \eqn{fourpointamplitude} unchanged
by virtue of the Jacobi identity $c_s+ c_t+c_u=0$.
In the case of the four-gluon amplitude,
however,
the sum $n_s+n_t+n_u$ is well-defined:
it is invariant under the generalized gauge transformation (\ref{fourpointggt})
due to momentum conservation.
\para

The four-gluon amplitude can be written in terms of color-ordered amplitudes as 
\be
\cA_4 ~=~   c_s A(1,2,3,4) - c_u A(1,3,2,4) \,.
\label{fourgluonbcj}
\ee
Invariance of \eqn{fourgluonbcj} under the shift (\ref{fourpointcfs})
immediately implies
\be
\delta~ \cA_4 ~=~ s A(1,2,3,4)  - u A(1,3,2,4) ~=~  0 
 \ee
which is the four-gluon BCJ relation \cite{Bern:2008qj}. 
\para

\subsection{Color-factor shift for $n$-gluon amplitudes}

\begin{figure}
\begin{center}
\includegraphics[width=6.0cm]{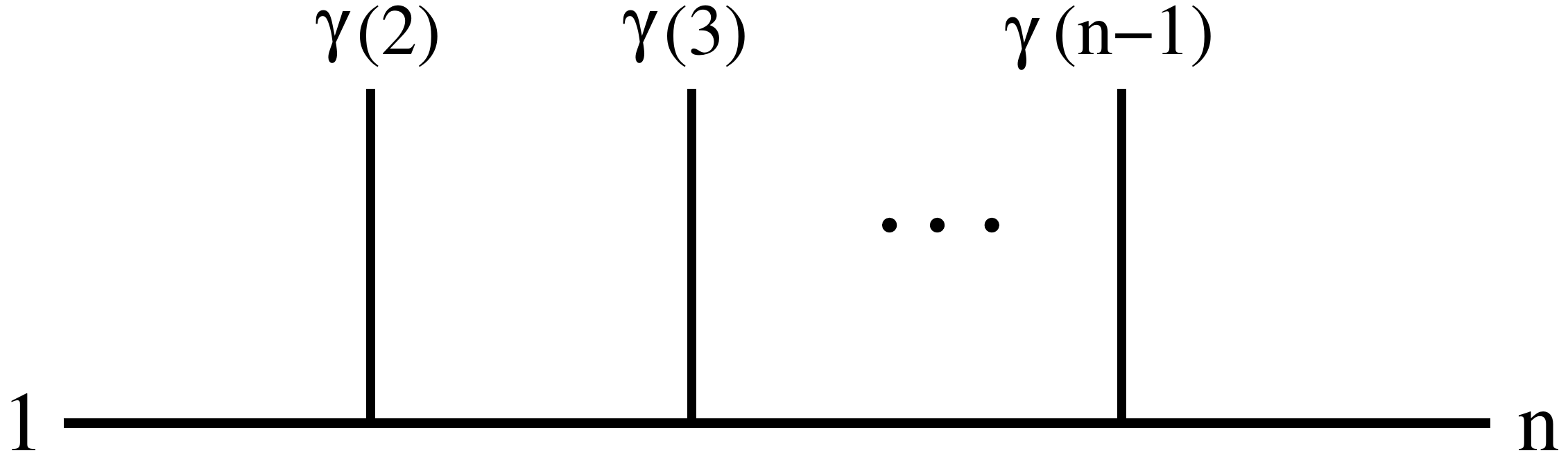}
\caption{Diagram for the half-ladder color factor
$\bc_{1 \gamma(2) \cdots \gamma(n-1)  n }$.}
\label{fig:half}
\end{center}
\end{figure}

Next we turn to tree-level $n$-gluon amplitudes with $n>4$,
which may be written as a sum over diagrams composed of cubic vertices 
(referred to as cubic diagrams) \cite{Bern:2008qj}
\be
\cAn ~=~ \sum_i {c_i ~ n_i \over d_i } \,.
\label{cubicdecomp}
\ee
Associated with each cubic diagram $i$ is a color factor $c_i$
obtained by sewing together structure constants $f_{\ta \tb \tc}$. 
Among these color factors $c_i$ we may identify the subset
of half-ladder color factors $\bc_{\alpha}$  
defined by (see fig. ~\ref{fig:half})
\be
\bc_{\alpha} ~\equiv~  \sum_{\tb_1,\ldots,\tb_{n{-}3}} 
f_{\ta_{\alpha(1)} \ta_{\alpha(2)} \tb_1}
f_{\tb_1 \ta_{\alpha(3)} \tb_2}
\cdots f_{\tb_{n{-}3} \ta_{\alpha(n{-}1)} \ta_{\alpha(n)}} \,, \qquad
\alpha \in S_n \,.
\label{halfladder}
\ee
The color factors $c_i$ are not independent but obey 
a set of Jacobi relations.
Using the procedure outlined in ref.~\cite{DelDuca:1999rs},
the Jacobi identity
$ 
f_{\ta \tb \te} f_{\tc \td \te} +f_{\ta \tc \te} f_{\td \tb \te} +f_{\ta \td \te} f_{\tb \tc \te}=0 
$
may be repeatedly applied to reduce each $c_i$ 
to a linear combination of half-ladder color factors 
\be
c_i ~=~ \sum_{\gamma \in S_{n-2}}  M_{i, 1\gamma n} {\bc}_{1 \gamma n },
\qquad
\bc_{1 \gamma n } ~\equiv~
\bc_{1 \gamma(2) \cdots \gamma(n-1) n } 
\label{dependent}
\ee
where $\gamma$ denotes a permutation of $ \{2, \cdots, n-1\}$.
The $(n-2)!$  half-ladders $\bc_{1 \gamma n}$ form an independent set.
Alternatively, 
$M_{i, 1\gamma n}$ may be computed  
by rewriting $c_i$ using $f_{\ta \tb \tc } = \Tr( [T^\ta, T^\tb] T^\tc )$,
reducing the resulting expression to a linear combination of single traces,
and then identifying the coefficient of
$\Tr(T^{\ta_1} T^{\ta_{\gamma(2)}} \cdots T^{\ta_{\gamma(n-1)}} T^{\ta_n})$
(see \eg ref.~\cite{Naculich:2014rta}).
\para

We now define a set of momentum-dependent shifts,
associated with each external gluon $a$ in the amplitude,
that act on the color factors $c_i$.
The action of the shift $\delta_a c_i$ associated with gluon $a$
is constrained by two requirements:
(I) that it preserve all the Jacobi relations satisfied by $c_i$,
and 
(II) that it satisfy 
\be
\delta_a c_i  ~\propto~  \sum_{c\in \Sai  } k_a \cdot k_c  
\label{colorfactorshift}
\ee
where $\Sai$ denotes the subset of the external particles 
on one side\footnote{Either side gives the same result up to 
sign due to momentum conservation.}
of the point at which $a$ is attached to $c_i$.
In particular, if $c_i$ is a color factor in which gluon $a$ 
is attached to an external leg $b$, the shift is proportional to $k_a \cdot k_b$,
which is an inverse propagator in the associated Feynman diagram.
More generally, \eqn{colorfactorshift} is related (see \eqn{splitprop})  
to the propagators in the Feynman diagram associated with $c_i$. 
\para

Consider the subset of $n$-point color factors  
obtained from a given  $(n-1)$-point cubic diagram $I$
by attaching gluon $a$ to it in all possible ways.
One of these color factors has gluon $a$ attached to external leg $1$ 
of the $(n-1)$-point diagram\footnote{Unless $a=1$}; 
define its shift to be $\alpha_I \, k_a \cdot k_1$.
One may easily verify (using the argument in the introduction)
that the conditions (I) and (II) above 
uniquely fix the coefficients of the shifts of all the other 
color factors in this subset.
The coefficients $\alpha_I$ for different $(n-1)$-point diagrams
are then constrained by Jacobi relations among their color factors. 
\para

\begin{figure} 
\begin{center}
\includegraphics[width=12.0cm]{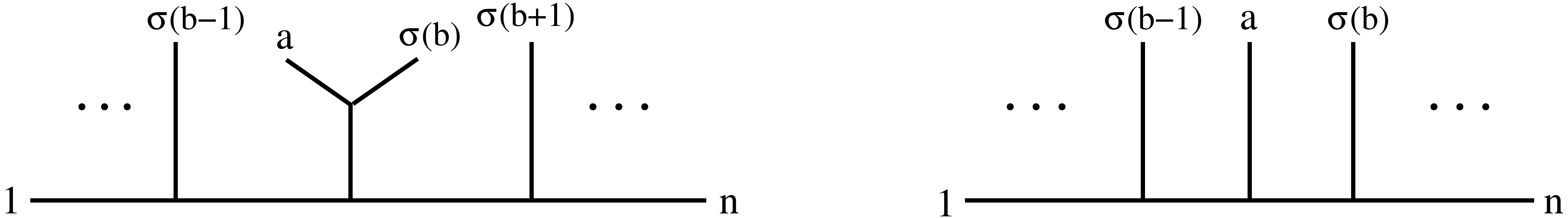}
\caption{Diagrams with color factors 
$ c_{1 \cdots \sigma(b-1) [a\sigma(b)] \sigma(b+1) \cdots n} $
and $ \bc_{1\cdots \sigma(b-1)a\sigma(b) \cdots n}$ } 
\label{fig:two}
\end{center}
\end{figure}

We now demonstrate that there is an $(n-3)!$-parameter family 
of color-factor shifts associated with each gluon $a$ 
in the $n$-gluon amplitude.
First choose $a \in \{2, \cdots, n-1\}$,
and consider the subset of half-ladder color factors
$\bc_{1 a \sigma n}$, 
where $\sigma \in S_{n-3}$ denotes a permutation of $ \{2, \cdots, n-1\} \setminus  \{ a\}$.
We define the color-factor shift associated with gluon $a$ to act on these half ladders as
\be
\delta_a~ \bc_{1 a \sigma(2) \cdots \sigma(n-1)  n } 
~=~ 
 \alpha_{a,\sigma}\,  k_a \cdot k_1  
\label{colorshiftdefinition}
\ee
where $\alpha_{a,\sigma}$ are a set of $(n-3)!$ arbitrary, independent
constants (or functions) for each $a$.
Let $c_{1 \cdots \sigma(b-1) [a\sigma(b)] \sigma(b+1) \cdots n}$ denote
the color factor shown in fig.~\ref{fig:two};
its shift is proportional to $k_a \cdot k_{\sigma(b)}$. 
This together with \eqn{colorshiftdefinition}
and the Jacobi relation
\be
    c_{1 \sigma(2) \cdots \sigma(b-1) [a\sigma(b)] \sigma(b+1) \cdots \sigma(n-1) n}
~=~ 
  \bc_{1 \sigma(2) \cdots \sigma(b-1) a \sigma(b) \cdots  \sigma(n-1) n } 
- \bc_{1 \sigma(2) \cdots \sigma(b) a \sigma(b+1) \cdots  \sigma(n-1) n } 
\ee
implies that $\delta_a$ acts on the 
independent half-ladder color factors  $\bc_{1\gamma n}$ 
as\footnote{In the case $a=2$, replace $\sigma(2)$ with $\sigma(3)$,
and the sum over $c$ should begin with 3.
In the case $a=n-1$, replace $\sigma(n-1)$ with $\sigma(n-2)$.}
\be
\delta_a~ \bc_{1 \sigma(2) \cdots \sigma(b-1) a \sigma(b) \cdots  \sigma(n-1) n } 
~=~ 
 \alpha_{a,\sigma} \left( 
k_a \cdot k_1  +  \sum_{c=2}^{b-1} k_a \cdot k_{\sigma(c)} \right), 
\qquad a, b \in \{2, \cdots, n-1\}, \qquad b \neq a 
\label{halfladdershift}
\ee
consistent with \eqn{colorfactorshift}.
The action on the remaining color factors is given by
\be
\delta_a\, c_i ~=~ \sum_{\gamma \in S_{n-2}}  M_{i, 1\gamma n} 
~\delta_a\, \bc_{1 \gamma  n }  
\label{dependentshift}
\ee
which is also consistent with \eqn{colorfactorshift}, as may be shown 
using the procedure described in ref.~\cite{DelDuca:1999rs}.
\para

The color-factor shifts associated with the gluons 
$\{2, \cdots, n-1\}$  are not all independent.
In particular, the $(n-3)!$-parameter family of
shifts associated with gluon $n-1$
are linear combinations of shifts associated with 
$a \in \{2, \cdots, n-2\}$ 
as a result of momentum conservation.\footnote{
We have verified this numerically through $n=9$, 
but we know the result must be true for all $n$ 
because, as we will see in sec.~\ref{sec:prop},
the color-factor shifts correspond to 
null eigenvectors of the propagator matrix.   
Since the $(n-2)!\times(n-2)!$ propagator matrix has rank $(n-3)!$
\cite{Cachazo:2013iea}
there are at most $(n-3) (n-3)!$ independent null eigenvectors.}
We may also define $(n-3)!$-parameter families of shifts 
associated with gluons 1 and $n$.
These are also not independent of the others.
Thus the dimension of the (abelian) group of color-factor shifts 
is $(n-3) (n-3)!$.

\subsection{Fundamental BCJ relations from the color-factor symmetry}

By using \eqn{dependent}, 
the tree-level $n$-gluon amplitude (\ref{cubicdecomp}) may be rewritten
in the Del Duca-Dixon-Maltoni half-ladder decomposition \cite{DelDuca:1999ha,DelDuca:1999rs}  
\be
\cAn ~=~ \sum_{\gamma \in S_{n-2}}  
\bc_{1 \gamma n } A(1, \gamma(2), \cdots, \gamma(n-1), n)
\label{ddm}
\ee
where the coefficients
\be 
A(1, \gamma(2),  \cdots, \gamma(n-1), n )  
~=~
\sum_i 
{M_{i, 1\gamma n}  ~ n_i 
\over d_i } 
\label{colorordered}
\ee
are color-ordered amplitudes belonging to the Kleiss-Kuijf 
basis \cite{Kleiss:1988ne}.
In the previous subsection, we defined the action of the color-factor symmetry
associated with a given gluon $a$.  
The variation of \eqn{ddm} under the shift associated with $a=2$ gives
\be
\deltatwo  \cAn ~=~ \sum_{\sigma  \in S_{n-3} } \alpha_{2,\sigma}
\sum_{b=3}^n \left( k_1 \cdot k_2 + \sum_{c=3}^{b-1} k_2 \cdot k_{\sigma(c)}  \right) 
A(1, \sigma(3), \cdots, \sigma(b-1), 2, \sigma(b), \cdots, \sigma(n-1), n)  \,.
\ee
In secs.~\ref{sec:fourgluon} and \ref{sec:rvegluon}, we prove that $\cAn$ 
is invariant under this shift. 
Since $\alpha_{2,\sigma}$ are arbitrary and independent, this implies that
\be
\sum_{b=3}^n \left( k_1 \cdot k_2 + \sum_{c=3}^{b-1} k_2 \cdot k_{\sigma(c)}  \right) 
A(1, \sigma(3), \cdots, \sigma(b-1), 2, \sigma(b), \cdots, \sigma(n-1), n) 
~=~ 0
\label{fundbcj} 
\ee
which is the fundamental BCJ relation (\ref{fundbcjintro}).
All other permutations of this relation can be obtained using the invariance
of the amplitude under the color-factor shifts associated with gluons 3 through $n-1$. 
\para

It is known \cite{Bern:2008qj} 
that the BCJ relations reduce the number of independent color-ordered amplitudes
from $(n-2)!$ to $ (n-3)!$.
Not surprisingly, 
the difference between these, $(n-3) (n-3)!$, is the dimension of the 
group of color-factor shifts that leave the amplitude invariant.

\subsection{Cubic vertex expansion}

In order to examine the implications of the 
color-factor symmetry for the kinematic numerators
$n_i$ appearing in the cubic decomposition (\ref{cubicdecomp}),
we introduce in this section the 
{\it cubic vertex expansion} 
of the amplitude with respect to one of the gluons. 
This expansion is similar to, but distinct from, 
the radiation vertex expansion \cite{Brown:1982xx} 
that will be used in secs.~\ref{sec:rvegluon}  and \ref{sec:rvefund}
to prove the invariance of $n$-point gauge-theory amplitudes 
under color-factor shifts.
\para

The cubic decomposition (\ref{cubicdecomp}) 
is a sum over the $(2n-5)!!$ cubic diagrams of an $n$-gluon amplitude,
but for any $a \in \{1, \cdots, n \}$ it can be viewed as a sum over the $(2n-7)!!$ cubic
diagrams of an $(n-1)$-point function with external legs $\{ 1, \cdots, n\} \setminus \{ a\}$,
to each of which gluon $a$ is attached in $2n-5$ different ways.
Let us label these $(n-1)$-point cubic diagrams by $I$
and their denominators by $d_{(a,I)}$.
\para

Each $(n-1)$-point cubic diagram $I$ has $n-3$ vertices,
the set of which we denote by $V_{(a,I)}$.
For each vertex $v \in V_{(a,I)}$, 
we can break $d_{(a,I)}$ into three factors 
$\prod_{r=1}^3 d_{(a,I,v,r)}$,
where $d_{(a,I,v,r)}$ is the product of propagators that branch off from leg $r$ of the vertex.
If leg $r$ is an external leg of the diagram, then $d_{(a,I,v,r)}=1$.
\para

We can attach gluon $a$ 
either to one of the $n-1$ external legs 
or to one of the $n-4$ internal lines of $I$,
yielding altogether $2n-5$ of the terms in the sum (\ref{cubicdecomp}). 
Let $K$ be the momentum running through one of the internal lines of $I$.
Attaching gluon $a$ to this line will replace the factor 
$K^2$ in $d_{(a,I)}$ with $K^2 (K + k_a)^2 $.  
We split the inverse denominator into two terms using the identity
\be
{ 1\over K^2  (K+k_a)^2 }
~=~ 
{1 \over K^2  (2 k_a \cdot K) } 
~+~
{1 \over  (-2 k_a \cdot K) (K+k_a)^2} 
\label{splitprop}
\ee
and we associate each of the terms on the right hand side 
of the equation with one of the two vertices
to which the internal line is connected.
Thus, with this doubling of internal line terms, we now have a total of 
$(n-1) + 2 (n-4) = 3 (n-3)$ terms for each $I$;
namely, one term for each of the legs of each of the $n-3$ vertices of $I$.
We label this term by $(a,I,v,r)$, and write the 
cubic vertex expansion of the $n$-gluon amplitude 
with respect to gluon $a$ as
\be
\cAn ~=~ 
\sum_I  
\sum_{v \in V_{(a,I)}}  {1 \over \prod_{s=1}^3 d_{(a,I,v,s)}  }
\sumr 
{ c_{(a,I,v,r)} n_{(a,I,v,r)} \over 2 k_a \cdot K_{(a,I,v,r)}   }
\label{cve}
\ee
where 
$c_{(a,I,v,r)}$ are the color factors $\ccr$ in fig.~\ref{fig:one}  associated
with each vertex $(a,I,v)$, 
$n_{(a,I,v,r)}$ are the associated kinematic numerators, 
and $K_{(a,I,v,r)}$ denotes the momentum flowing out of leg $r$.
The $c_{(a,I,v,r)}$ and $n_{(a,I,v,r)}$  
are equal to the $c_i$ and $n_i$ in \eqn{cubicdecomp} up to signs
(such that $c_{(a,I,v,r)} n_{(a,I,v,r)}   =  c_i n_i$).
An explicit example of the cubic vertex expansion for the 
five-gluon amplitude is given in appendix A.
\para

The $\pm$ freedom in the definition of 
$c_{(a,I,v,r)}$ is used to make 
the relative signs in the Jacobi relation positive:
\be 
\sumr c_{(a,I,v,r)} ~=~ 0   \,.
 \label{vertexjacobi}
\ee 
The denominators in each triple also sum to zero
\be 
\sumr k_a \cdot K_{(a,I,v,r)}   ~=~ 0
 \label{vertexmomentum}
\ee 
by momentum conservation $k_a + \sumr K_{(a,I,v,r)} = 0$
and the masslessness of the gluon $k_a^2=0$.
{\it A priori}, however, there is no reason for the kinematic numerators 
$n_{(a,I,v,r)} $ associated with each vertex to sum to zero.
We will see in the next subsection, however, that the color-factor symmetry of the amplitude 
leads to a constraint on the sum of kinematic numerators.
\para

\subsection{Constraint on kinematic numerators from the color-factor symmetry}
\label{subsec:kinematicnumeratorconstraint}

Having introduced the cubic vertex expansion 
of the $n$-gluon amplitude with respect to 
gluon $a$, we now consider the effect of a color-factor shift $\delta_a$ on the amplitude. 
The shift associated with gluon $a$ acts on the color factors 
appearing in the cubic vertex expansion (\ref{cve}) 
as
\be
\delta_a ~ c_{(a,I,v,r)} ~=~   \alpha_{(a,I,v)} ~k_a \cdot K_{(a,I,v,r)}
\ee
where
$\alpha_{(a,I,v)}$ is a linear combination of $\alpha_{a,\sigma}$ 
uniquely determined by the Jacobi relations.
These shifts respect \eqn{vertexjacobi} by virtue of \eqn{vertexmomentum}.
The variation of \eqn{cve} under this shift gives
\be
\delta_a ~ \cAn  ~ = ~
{1\over 2} \sum_I  
\sum_{v \in V_{(a,I)}}  
{\alpha_{(a,I,v)}  \over \prod_{s=1}^3 d_{(a,I,v,s)}  }
\sumr 
n_{(a,I,v,r)}    \,.
\label{variation}
\ee
We prove that $\delta_a \cAn =0$ 
in secs.~\ref{sec:fourgluon} and \ref{sec:rvegluon};
hence the color-factor symmetry implies the following 
constraint on the kinematic numerators
\be
\sum_I  
\sum_{v \in V_{(a,I)}}  
{\alpha_{(a,I,v)}  \over \prod_{s=1}^3 d_{(a,I,v,s)}  }
\sumr 
n_{(a,I,v,r)}  ~ = ~ 0 \,.
\label{sumoverdelta}
\ee
Now {\it if} the constants $\alpha_{(a,I,v)}$ were all independent, 
then we could conclude from this argument that $ \sumr n_{(a,I,v,r)} = 0 $,
i.e. that the kinematic numerators necessarily obey Jacobi relations.
This would be in conflict with the well-known fact that the kinematic
numerators obtained from Feynman rules in general do {\it not} satisfy color-kinematic 
duality\footnote{Only in the case $n=4$, 
where there is only one term in the sum (\ref{sumoverdelta}),
may we conclude that $n_s + n_t + n_u=0$.};
indeed this is not a valid inference from \eqn{sumoverdelta} because
the $\alpha_{(a,I,v)}$ are not independent. 
The set of $\alpha_{(a,I,v)}$ for all the vertices of a given diagram $I$
are equal (up to signs) because any two adjacent vertices 
share a common color factor (see the example discussed in the appendix).
In fact, with an appropriate choice of signs for $c_{(a,I,v,r)}$,
the $\alpha_{(a,I,v)}$ may be made independent of $v$.
The $\alpha_{(a,I)}$ for different diagrams $I$
are further constrained by the Jacobi relations among the
color factors of $I$. 
\para

While \eqn{sumoverdelta} does {\it not} imply that the numerators satisfy the kinematic Jacobi relations,
it {\it does} impose a set of (generalized-gauge-invariant) conditions that the color-kinematic violations 
$\Delta_{ijk} = n_i + n_j + n_k$ must satisfy.
We wish to emphasize that,  
while the kinematic Jacobi relations
$n_i + n_j + n_k=0$ are not invariant under generalized gauge 
transformations\footnote{Again except in the case of four-gluon amplitudes.}
(hence the actual claim of color-kinematic duality 
is that {\it there exists} a generalized gauge in which they hold true), 
the conditions (\ref{sumoverdelta}) are
invariant under generalized gauge transformations.
The argument for this is simple.   
A generalized gauge transformation is a transformation 
$n_i \to n'_i$ that leaves the amplitude (\ref{cubicdecomp}) unchanged.  
Hence by starting with $ \cAn = \sum_i (c_i n'_i / d_i)$ and following
the steps above (since the condition $\delta_a \, \cAn=0$ is also gauge invariant), 
we obtain the same result (\ref{sumoverdelta}) except with 
$n_{(a,I,v,r)}$ replaced with $n'_{(a,I,v,r)}$.
\para

To obtain a more explicit form of \eqn{sumoverdelta}, 
we would need to identify all the linear dependences 
among the $\alpha_{(a,I,v)}$
required by the color Jacobi relations.
Previously, we observed that the number of independent color-factor shifts
was $(n-3) (n-3)!$,
parametrized by constants 
$\alpha_{a,\sigma}$, 
where $a=2, \cdots, n-2$ and 
$\sigma \in S_{n-3}$ denotes a permutation of $ \{2, \cdots, n-1\} \setminus  \{ a\}$.
If we were to write $\alpha_{(a,I,v)}$ in terms of these independent constants, 
\eqn{sumoverdelta} would yield $(n-3) (n-3)!$ independent constraints on
the $\Delta_{ijk}$. 
In appendix A, we carry out this procedure for the five-gluon amplitude.
\para

While the BCJ relations (\ref{fundbcj}) were originally derived as a consequence of 
the assumption of color-kinematic duality, it was known 
\cite{BjerrumBohr:2010zs,Tye:2010dd} 
from early on that they are 
equivalent to a set of weaker conditions on the numerators.
The conditions for five-gluon numerators were derived in 
refs.~\cite{BjerrumBohr:2010zs,Tye:2010dd} 
as a consequence of the monodromy properties of string-theory amplitudes. 
These conditions are equivalent to \eqn{sumoverdelta},
as we show in appendix A.
In this section, we have demonstrated 
that both \eqn{sumoverdelta} and the BCJ relations (\ref{fundbcj}) 
are a consequence of the invariance of the amplitude 
under the color-factor symmetry.

\subsection{Kinematic numerator shift symmetry}

We have considered the effect on the amplitude of a shift of the color factors.   
One may ask what effect an analogous shift of the kinematic numerators would 
have.\footnote{These can be considered a generalization of the shifts 
of $J$ considered in refs.~\cite{Brown:1983pn,Brown:1983vk}.}
We show in this section that such a shift is simply a generalized gauge transformation.
\para

Let us define the kinematic shift associated with leg $a$,
where $a \in \{2, \cdots, n-1\}$,
on the half-ladder numerator $\bn_{1\gamma n}$ to be
\be
\delta_a~ \bn_{1 \sigma(2) \cdots \sigma(b-1) a \sigma(b) \cdots  \sigma(n-1) n } 
~=~ 
 \beta_{a,\sigma} \left( 
k_1 \cdot k_a  +  \sum_{c=2}^{b-1} k_a \cdot k_{\sigma(c)} \right), 
\qquad a, b \in \{2, \cdots, n-1\}, \qquad b \neq a
\label{kinematichalfladdershift}
\ee
where $\sigma \in S_{n-3}$ denotes a permutation of $ \{2, \cdots, n-1\} \setminus  \{ a\}$,
and $\beta_{a,\sigma}$ are a set of arbitrary 
constants (or functions).
The action on all other numerators $n_i$ is then defined by 
\be
\delta_a\, n_i ~=~ \sum_{\gamma \in S_{n-2}}  M_{i, 1\gamma n} 
~\delta_a\, \bn_{1 \gamma  n } \,.
\label{kinematicdependentshift}
\ee
Note that we have {\it not} assumed that the $n_i$ obey the Jacobi relations
(\ref{kinematicjacobi}).
However, the numerator shifts defined by \eqns{kinematichalfladdershift}{kinematicdependentshift} 
will satisfy
\be
\delta_a (n_i + n_j + n_k) ~=~ 0
\ee
so if the  $n_i$ do satisfy kinematic Jacobi relations, 
the shifted numerators will continue to do so,
and if they do not, then the neither will the shifted numerators. 
\para

Now consider the cubic vertex expansion of the $n$-gluon amplitude 
\be
\cAn ~=~ 
\sum_I  
\sum_{v \in V_{(a,I)}}  {1 \over \prod_{s=1}^3 d_{(a,I,v,s)}  }
\sumr 
{ c_{(a,I,v,r)} n_{(a,I,v,r)} \over 2 k_a \cdot K_{(a,I,v,r)}   } \,.
\ee
The kinematic shift with respect to gluon $a$ 
acts on the numerators appearing in this expression as 
\be
\delta_a ~ n_{(a,I,v,r)} ~=~   \beta_{(a,I,v)}  \, k_a \cdot K_{(a,I,v,r)}
\ee
and therefore on the amplitude itself as 
\be
\delta_a ~ \cAn  ~ = ~
{1 \over 2} \sum_I  
\sum_{v \in V_{(a,I)}}  
{\beta_{(a,I,v)}  \over \prod_{s=1}^3 d_{(a,I,v,s)}  }
\sumr 
c_{(a,I,v,r)} \,.
\ee
This vanishes courtesy of \eqn{vertexjacobi}; 
hence the $n$-gluon amplitude  is invariant under 
the shift of kinematic numerators. 
This is precisely the definition of a generalized gauge 
transformation \cite{Bern:2010ue,Bern:2010yg}.

\section{Proof of color-factor symmetry for four-gluon amplitudes} 
\setcounter{equation}{0}
\label{sec:fourgluon}

In this section 
we prove that the tree-level four-gluon amplitude 
is invariant under the color-factor symmetry.
In doing so, we develop some results that will be necessary 
for our more general proof of the invariance of the 
$n$-gluon amplitude  in the next section.
\para

The four-gluon amplitude can be constructed from a three-gluon vertex 
by attaching a fourth gluon 
to a propagator emanating from each of the legs of the vertex
or to the vertex itself.
This yields
\be
\cA_4 ~=~ \sumr  {\ccr \nr  \over (k_r + k_4)^2}
\label{fourgluonamplitude}
\ee
where
\be
c_{(1)}  ~=~  f_{\ta_{1} \ta_4\tb} f_{\tb \ta_{2} \ta_{3} } ,\qquad
c_{(2)}  ~=~  f_{\ta_{2} \ta_4\tb} f_{\tb \ta_{3} \ta_{1} } , \qquad 
c_{(3)}  ~=~  f_{\ta_{3} \ta_4\tb} f_{\tb \ta_{1} \ta_{2} }  \,.
\label{fourgluoncolorfactors}
\ee
The kinematic numerator $\nr$ receives contributions 
from the diagram in which gluon 4 is attached to leg $r$ 
as well as from the four-gluon vertex.
The color-factor symmetry (associated with gluon 4) acts on the 
color factors (\ref{fourgluoncolorfactors}) and the four-gluon amplitude 
(\ref{fourgluonamplitude}) as 
\be
\delta_4 \ccr  ~=~ \alpha_4  \, k_4 \cdot k_r 
\qquad \implies \qquad 
\delta_4 \cA_4 ~=~ 
{1\over 2} \alpha_4
\sumr \nr  \,.
\ee
Thus, by showing that $\sumr \nr =0$,
we will establish that $\delta_4 \cA_4 =0$. 
This we now proceed to do.

\subsection{Attaching a gluon to a leg}

The three-gluon vertex is\footnote{Our structure constants 
are normalized by $f_{\ta \tb \tc } = \Tr( [T^\ta, T^\tb] T^\tc )$ 
with $\Tr( T^\ta T^\tb ) = \delta^{\ta\tb}$,
so that $[T^\ta, T^\tb] = f_{\ta\tb\tc} T^\tc$. 
This differs from the standard textbook convention by a factor of $i \sqrt2$.
We use $\eta_{00}=1$ in this paper.}
\be
-   { i g \over \sqrt2} f_{\ta_{1} \ta_{2} \ta_3 } V^{\mu_1 \mu_2 \mu_3} (k_1, k_2, k_3) 
\label{fV}
\ee
where
\be
V^{\mu_1 \mu_2 \mu_3} (k_1, k_2, k_3) ~=~ 
 \eta^{\mu_1\mu_2} (k_2 - k_1)^{\mu_3}
+\eta^{\mu_2\mu_3} (k_3 - k_2)^{\mu_1}
+\eta^{\mu_3\mu_1} (k_1 - k_3)^{\mu_2}
\label{threegluonvertex}
\ee
and $k_a$ are outgoing momenta.
In Feynman gauge, the gluon propagator is $-i \eta_{\mu\nu} \delta_{\ta\tb} /k^2$.
Attaching gluon 4 to leg 1 yields the expression
\begin{align}
 { ig^2 \over 2} 
& 
{f_{\ta_{1} \ta_4\tb} f_{\tb\ta_{2} \ta_{3} } 
\over (k_1 + k_4)^2  }
V^{\mu_1 \mu_4 \nu} (k_1, k_4, -k_1-k_4) 
V_{\nu}^{~\mu_2 \mu_3} (k_1 + k_4, k_2, k_3)
\nn\\
~=~ &
  {i g^2 \over 2} 
{  c_{(1)} \over (k_1 + k_4)^2  }
\Big[   
\eta^{\mu_1 \nu  } k_4^{\mu_4 }
- \eta^{\mu_4 \nu } k_1^{\mu_1 }
- \eta^{\mu_1 \mu_4 } (k_1 + k_4)^{\nu}
\label{fiveterms}
\\
& 
\qquad \quad\qquad\qquad
+ 2 \eta^{\mu_1 \nu} k_1^{\mu_4} 
+ 2 \left( \eta^{\mu_1 \mu_4} k_4^\nu - \eta^{\mu_4 \nu} k_4^{\mu_1}  \right)
 \Big]  
{V_{\nu}^{~\mu_2 \mu_3} (k_1 + k_4, k_2, k_3) } \,.
\nn
\end{align}
The contribution of this diagram to the four-gluon amplitude (\ref{fourgluonamplitude}) 
is obtained by contracting with $\prod_{a=1}^4 \ve_{a\mu_a} $
and dividing by $i$.
The first and second terms in the square brackets vanish  
using $\ve_a \cdot k_a=0$. 
The third term vanishes due to 
\be
(k_1 + k_4)^\nu V_{\nu}^{~\mu_2 \mu_3} (k_1 + k_4, k_2, k_3) \ve_{2\mu_2} \ve_{3\mu_3} ~=~0
\label{Wardone}
\ee
using $k_2^2 = k_3^2 =0$.
The contribution of the remaining terms  of \eqn{fiveterms} 
to the kinematic numerator $n_{(1)}$ can be written
\be
n_{(1)}\Big|_{\rm leg} 
~=~  g^2 \ve_{1\mu_1} 
\left[  
   \ve_4 \cdot k_1 \delta^{\mu_1}_{~\nu }
-i \ve_{4\alpha} k_{4\beta} ( S_1^{\alpha\beta} )^{\mu_1}_{~~ \nu} 
 \right]  
V^{\nu \mu_2 \mu_3} (k_1 + k_4, k_2, k_3)
\ve_{2\mu_2} \ve_{3\mu_3} 
\label{attachleg} 
\ee 
where 
\be
( S_r^{\alpha\beta} )^{\mu_r}_{~~ \nu} ~=~ 
i ( \eta^{\alpha\mu_r} \delta^{\beta}_{~\nu} - \eta^{\beta\mu_r} \delta^\alpha_{~\nu} )
\label{S}
\ee
are the spin-one angular momentum matrices acting on gluon $r$. 
These satisfy the Lorentz algebra commutation relations
\be
[ S_r^{\alpha\beta}, S_r^{\gamma\delta}  ]
~=~ 
-i \left[ 
 \eta^{\alpha\gamma} S_r^{\beta\delta}  
-\eta^{\alpha\delta} S_r^{\beta\gamma}  
-\eta^{\beta\gamma} S_r^{\alpha\delta}  
+ \eta^{\beta\delta} S_r^{\alpha\gamma} 
\right] \,.
\ee

\subsection{Attaching a gluon to a vertex} 

Attaching gluon 4 directly to the three-gluon vertex, we obtain the four-gluon vertex
\begin{align}
V_{\ta_1 \ta_2 \ta_3 \ta_4}^{\mu_1 \mu_2 \mu_3 \mu_4} 
~=~  
{ i g^2 \over 2} \Big[ 
& f_{\ta_{1} \ta_4\tb} f_{\tb \ta_{2} \ta_{3} } 
\left( \eta^{\mu_1 \mu_2} \eta^{\mu_3 \mu_4} -\eta^{\mu_1 \mu_3} \eta^{\mu_2 \mu_4} \right)
\nn\\
+&
f_{\ta_{2} \ta_4\tb} f_{\tb \ta_{3} \ta_{1} } 
\left( \eta^{\mu_2 \mu_3} \eta^{\mu_1 \mu_4} -\eta^{\mu_1 \mu_2} \eta^{\mu_3 \mu_4} \right)
\label{fourgluonvertex}
\\
+&
f_{\ta_{3} \ta_4\tb} f_{\tb \ta_{1} \ta_{2} } 
\left( \eta^{\mu_1 \mu_3} \eta^{\mu_2 \mu_4} -\eta^{\mu_2 \mu_3} \eta^{\mu_1 \mu_4} \right)
\Big]
\,.
\nn
\end{align}
Using \eqn{threegluonvertex}, this can be recast as 
\be
V_{\ta_1 \ta_2 \ta_3 \ta_4}^{\mu_1 \mu_2 \mu_3 \mu_4} 
~=~
-  {i g^2 \over 2} \left( 
 c_{(1)} {\partial \over \partial  k_{1\mu_4}   }
+c_{(2)} {\partial \over \partial  k_{2\mu_4}   }
+c_{(3)} {\partial \over \partial  k_{3\mu_4}   }
\right) 
V^{\mu_1 \mu_2 \mu_3} (k_1, k_2, k_3)  \,.
\label{fourgluonvertexderiv}
\ee
The contribution of this vertex to the four-gluon amplitude is 
obtained by contracting with $\prod_{a=1}^4 \ve_{a\mu_a} $
and dividing by $i$.
This contribution is then parceled out among the three terms in \eqn{fourgluonamplitude};
the contribution to $n_{(1)}$  is 
\be
n_{(1)}\Big|_{\rm vertex}
~=~ - g^2 k_4 \cdot k_1 
\ve_{4\mu_4} 
{\partial \over \partial  k_{1\mu_4}  }  V^{\mu_1 \mu_2 \mu_3} (k_1, k_2, k_3) 
\ve_{1\mu_1} 
\ve_{2\mu_2} 
\ve_{3\mu_3}  \,.
\label{attachvertex}
\ee
The reader may be concerned about the use of \eqn{fourgluonvertexderiv} for the following reason.
The three-gluon vertex (\ref{threegluonvertex}) can be rewritten using 
momentum conservation $\sum_{a=1}^3 k_a = 0$.
For example, we can eliminate $k_3$ from \eqn{threegluonvertex}, writing it as
\be
V^{\mu_1 \mu_2 \mu_3} (k_1, k_2, k_3) ~=~ 
 \eta^{\mu_1\mu_2} (k_2 - k_1)^{\mu_3}
+\eta^{\mu_2\mu_3} (-k_1 - 2 k_2)^{\mu_1}
+\eta^{\mu_3\mu_1} (2k_1 + k_2)^{\mu_2} \,.
\label{altthreegluonvertex}
\ee
The partial derivatives $(\partial/\partial k_{a \mu_4})V(k_1,k_2,k_3)$ 
obtained from \eqn{altthreegluonvertex}
differ from those obtained from \eqn{threegluonvertex}, 
so that \eqn{fourgluonvertexderiv} gives
\begin{align}
V_{\ta_1 \ta_2 \ta_3 \ta_4}^{\mu_1 \mu_2 \mu_3 \mu_4} 
~=~  
{ i g^2 \over 2} \Big[
& f_{\ta_{1} \ta_4\tb} f_{\tb \ta_{2} \ta_{3} } 
\left( 
  \eta^{\mu_1 \mu_2} \eta^{\mu_3 \mu_4} 
+ \eta^{\mu_2 \mu_3} \eta^{\mu_1 \mu_4}
-2\eta^{\mu_1 \mu_3} \eta^{\mu_2 \mu_4} 
\right)
\nn\\
+&
f_{\ta_{2} \ta_4\tb} f_{\tb \ta_{3} \ta_{1} } 
\left( 
2 \eta^{\mu_2 \mu_3} \eta^{\mu_1 \mu_4} 
- \eta^{\mu_1 \mu_2} \eta^{\mu_3 \mu_4} 
- \eta^{\mu_1 \mu_3} \eta^{\mu_2 \mu_4} 
\right)
\Big] \,.
\label{altfourgluonvertex}
\end{align}
Nonetheless \eqn{altfourgluonvertex} is equal to \eqn{fourgluonvertex}
courtesy of the Jacobi relation $\sumr \ccr=0$. 
\Eqn{altfourgluonvertex} 
simply corresponds to a different way of parceling the four-gluon vertex 
among the color factors $\ccr$,
and the expressions $\nr$ will differ by a generalized gauge transformation (\ref{fourpointggt}).
The amplitude (\ref{fourgluonamplitude}) of course remains unchanged.

\subsection{Kinematic numerators of the four-gluon amplitude}

Using the fact that the three-gluon vertex (\ref{threegluonvertex}) is linear in momenta,
we rewrite \eqn{attachleg} using 
\be
V^{\nu \mu_2 \mu_3} (k_1 + k_4, k_2, k_3)
~=~ \left( 1 +  k_{4\gamma} {\partial \over \partial  k_{1\gamma} } \right) 
V^{\nu \mu_2 \mu_3} (k_1 , k_2, k_3)
\label{taylor}
\ee
and then combine \eqns{attachleg}{attachvertex} to obtain
\begin{align}
n_{(1)}
&~=~ 
n_{(1)}\Big|_{\rm leg} + n_{(1)}\Big|_{\rm vertex}
\nn\\
&~=~
g^2 \ve_{1\mu_1} 
\left[  
   \ve_4 \cdot k_1 \delta^{\mu_1}_{~\nu} 
-i \ve_{4\alpha} k_{4\beta} ( S_1^{\alpha\beta} )^{\mu_1}_{~~ \nu} 
-i \ve_{4\alpha} k_{4\beta}  L_1^{\alpha\beta} \delta^{\mu_1}_{~\nu} 
 \right]  
V^{\nu \mu_2 \mu_3} (k_1 , k_2, k_3)
\ve_{2\mu_2} \ve_{3\mu_3} 
\nn\\
&\quad\quad +g^2
\ve_{1\mu_1} 
\left[ -i \ve_{4\alpha} k_{4\beta} ( S_1^{\alpha\beta} )^{\mu_1}_{~~ \nu} 
 \right]  
\left(  k_{4\gamma} {\partial \over \partial  k_{1\gamma} } \right) 
V^{\nu \mu_2 \mu_3} (k_1 , k_2, k_3)
\ve_{2\mu_2} \ve_{3\mu_3} 
\end{align}
where the orbital angular momentum operators are defined  as
\be
 L_r^{\alpha\beta}  ~=~ i \left( 
k_r^\alpha {\partial \over \partial  k_{r\beta} } - k_r^\beta {\partial \over \partial  k_{r\alpha} } \right) \,.
\label{L}
\ee
These satisfy the Lorentz algebra commutation relations 
\be
[ L_r^{\alpha\beta}, L_r^{\gamma\delta}  ]
~=~ 
-i \left[ 
 \eta^{\alpha\gamma} L_r^{\beta\delta}  
-\eta^{\alpha\delta} L_r^{\beta\gamma}  
-\eta^{\beta\gamma} L_r^{\alpha\delta}  
+ \eta^{\beta\delta} L_r^{\alpha\gamma} 
\right] \,.
\ee
Similar expressions are obtained for $n_{(2)}$ and $n_{(3)}$.
Finally, defining  the total angular momentum operator
\be 
( J_r^{\alpha\beta} )^{\mu_r}_{~~ \nu} ~=~ 
 L_r^{\alpha\beta} \delta^{\mu_r}_{~ \nu}  + 
( S_r^{\alpha\beta} )^{\mu_r}_{~~ \nu} 
\label{J}
\ee
we can write all the kinematic numerators as 
\begin{align}
\nr
&~=~
g^2 \left[  
   \ve_4 \cdot k_r 
-i \ve_{4\alpha} k_{4\beta}  J_r^{\alpha\beta} 
-i \ve_{4\alpha} k_{4\beta}  k_{4\gamma} 
S_r^{\alpha\beta} 
{\partial \over \partial  k_{r\gamma} }
 \right]  
V(k_1 , k_2, k_3)
\label{nr}
\end{align}
where we have suppressed the polarization vectors $\ve_1$, $\ve_2$, and $\ve_3$.
The subscripts on $J_r$ and $S_r$ indicate on which 
polarization indices these operators act.
\para

Note that under a gauge transformation 
$\ve_4 \to \ve_4 + \lambda k_4$ of gluon 4, 
the last two terms in \eqn{nr} 
vanish by virtue of the antisymmetry of $J_r^{\alpha\beta}$ and $S_r^{\alpha\beta}$, and the first term gives
$\nr \to \nr + \lambda g^2 k_4 \cdot k_r$ in accord with \eqn{fourpointggt}.
\para

We observe that the operators $L^{\alpha\beta}_r$ and $S^{\alpha\beta}_r$ 
are the same as those appearing in the Burnett-Kroll 
form \cite{Burnett:1967km} of the subleading terms 
of the Low soft-photon theorem \cite{Low:1958sn}
as applied to gluons \cite{DelDuca:1990gz,Laenen:2008gt,Laenen:2010uz,Casali:2014xpa,White:2014qia}. 
(See refs.~\cite{Bern:2014vva,Broedel:2014fsa,Luo:2014wea}
for recent derivations of the soft-gluon theorem from gauge invariance.)
The first two terms in \eqn{nr} correspond precisely 
to the leading and subleading terms in the $k_4 \to 0$ expansion of 
the four-gluon amplitude;
the third term is higher order in the soft momentum.
We emphasize, however, that \eqn{nr}, 
and the vertex expansion that we will derive in the next section, 
are exact, and not dependent on taking a soft limit.

\subsection{Kinematic Jacobi relation}

We will now show that the sum of the four-gluon kinematic numerators 
(\ref{nr}) vanishes.
This result, which has been known at least since 1980 \cite{Zhu:1980sz,Goebel:1980es}
inspired the conjecture of color-kinematic duality \cite{Bern:2008qj}.
We demonstrate it in a way that will facilitate 
the proof of color-factor symmetry of the $n$-gluon amplitude.
\para

Using  \eqn{nr}, we see that 
\begin{align}
\sumr \nr
~=~&
 g^2 \left( \sumr \ve_4 \cdot k_r \right) V(k_1 , k_2, k_3)
\nn\\
&-
ig^2  \ve_{4\alpha} k_{4\beta} \left( \sumr J_r^{\alpha\beta} \right) V(k_1 , k_2, k_3)
\label{sumnr}
\\
&-
ig^2  \ve_{4\alpha} k_{4\beta} k_{4\gamma} 
\sumr S_r^{\alpha\beta} {\partial \over \partial  k_{r\gamma}   }
V(k_1 , k_2, k_3) \,.
\nn
\end{align}
The first sum on the right-hand side of \eqn{sumnr} 
vanishes by momentum conservation $\sum_{r=1}^4 k_r =  0$ 
together with $\ve_4 \cdot k_4=0$.
The second sum on the right hand side of \eqn{sumnr},
which may be written more explicitly as
\be
(J_1^{\alpha\beta} )^{\mu_1}_{~~\nu} 
V^{\nu \mu_2 \mu_3} (k_1, k_2, k_3)
+ 
(J_2^{\alpha\beta} )^{\mu_2}_{~~ \nu} 
V^{\mu_1 \nu \mu_3} (k_1, k_2, k_3)
+ 
(J_3^{\alpha\beta} )^{\mu_3}_{~~ \nu} 
V^{\mu_1 \mu_2 \nu} (k_1, k_2, k_3)
\label{secondsum}
\ee
is the first-order Lorentz transformation of the three-gluon vertex.
This vanishes,
as may be verified by explicit computation,
because $V^{\mu_1 \mu_2 \mu_3} (k_1, k_2, k_3)$ is a Lorentz tensor.
Alternatively, we can define the spin-one angular momentum operator  to act on 
polarization indices 
\cite{Bern:2014vva,Broedel:2014fsa}
\be
 S_r^{\alpha\beta} ~=~ i \left( 
\ve_r^\alpha {\partial \over \partial  \ve_{r\beta} } - \ve_r^\beta {\partial \over \partial  \ve_{r\alpha} } \right) 
\label{altS}
\ee
in which case
\be
\left( \sumr J_r^{\alpha\beta} \right)
V(k_1, k_2, k_3)
~=~ 
i 
\sumr 
\left( k_r^\alpha {\partial \over \partial  k_{r\beta} } - k_r^\beta {\partial \over \partial  k_{r\alpha} } 
+ \ve_r^\alpha {\partial \over \partial  \ve_{r\beta} } - \ve_r^\beta {\partial \over \partial  \ve_{r\alpha} } \right) 
V(k_1, k_2, k_3)
\label{lorentzscalar}
\ee
where 
\be
V(k_1, k_2, k_3) ~=~ 
 \ve_1 \cdot \ve_2 ~\ve_3 \cdot (k_2 - k_1)
+\ve_2 \cdot \ve_3 ~\ve_1 \cdot (k_3 - k_2)
+\ve_3 \cdot \ve_1 ~\ve_2 \cdot (k_1 - k_3)\,.
\ee 
\Eqn{lorentzscalar} vanishes because $V(k_1, k_2, k_3)$ is a 
Lorentz-invariant function of $k_r$ and $\ve_r$.
The third sum in \eqn{sumnr} is proportional to 
\begin{align}
&
(S_1^{\alpha\beta} )^{\mu_1}_{~~ \nu} 
{ \partial \over \partial k_{1\gamma} } V^{\nu \mu_2 \mu_3} (k_1, k_2, k_3)
+ 
(S_2^{\alpha\beta} )^{\mu_2 }_{~~\nu} 
{\partial \over \partial k_{2\gamma} } V^{\mu_1 \nu\mu_3} (k_1, k_2, k_3)
+ 
(S_3^{\alpha\beta} )^{\mu_3 }_{~~\nu} 
{\partial \over \partial k_{3\gamma} } V^{\mu_1 \mu_2\nu} (k_1, k_2, k_3)
\nn\\
&~=~ 2 i 
\left(
-\eta^{\alpha\mu_1} \eta^{\beta\mu_2} \eta^{\gamma\mu_3}
+\eta^{\alpha\mu_1} \eta^{\gamma\mu_2} \eta^{\beta\mu_3} \right) 
 +\hbox{(cyclic permutations of $123$)} \,.
\label{thirdsum}
\end{align}
This expression is antisymmetric in $\beta$ and $\gamma$.
Since it multiplies $k_{4\beta} k_{4\gamma} $ in \eqn{sumnr}, the whole expression vanishes.
The cancellations that we have just exhibited 
were originally used in ref.~\cite{Brown:1982xx} to prove the 
radiation zero theorem.
\para

We have shown that the sum of kinematic numerators vanishes
(the kinematic Jacobi relation),
and thus have demonstrated the invariance of 
the four-gluon amplitude $\delta_4 \cA_4=0$
under the color-factor shift symmetry. 
We would like to emphasize that in proving the vanishing of 
\eqns{secondsum}{thirdsum}  we did not use that $k_r$ were on-shell,
nor did we use $\ve_r \cdot k_r=0$ for $r=1, 2, 3$.  
Thus we will be able to use these results in sec.~\ref{sec:rvegluon}
for an off-shell three-gluon vertex $V^{\mu_1 \mu_2 \mu_3} (k_1, k_2, k_3)$.

\section{Proof of color-factor symmetry for $n$-gluon amplitudes} 
\setcounter{equation}{0}
\label{sec:rvegluon}

We now turn to the proof that the tree-level 
$n$-gluon amplitude is invariant under a
color-factor shift associated with any of the gluons.
To do so, we employ a decomposition of the 
amplitude known as the radiation vertex expansion \cite{Brown:1982xx}.
This is a recursive approach
which constructs an $n$-point amplitude by attaching a massless vector boson 
to all possible $(n-1)$-point diagrams.
\para

Consider the set of all tree-level $(n-1)$-gluon diagrams with 
external legs $\{ 1, \cdots, n\} \setminus \{ a\}$
constructed using Feynman rules.
Label these diagrams by $\I$.
Please note that this set differs 
from the set of $(n-1)$-point diagrams 
appearing in the cubic vertex expansion (cf. sec.~\ref{sec:cfs})
because it includes not only cubic diagrams 
but also diagrams with four-gluon vertices.
(In the cubic decomposition (\ref{cubicdecomp}),
all diagrams containing quartic vertices 
are redistributed among the cubic diagrams.)
\para

We will construct all possible $n$-gluon diagrams by attaching
gluon $a$ to each $(n-1)$-gluon diagram $\I$ in all possible ways.
This includes:
(1) attaching gluon $a$ to an external leg, 
(2) attaching gluon $a$ to an internal line, or 
(3) attaching gluon $a$ to one of the 
three-gluon vertices of $\I$ to make a four-gluon vertex.
By rearranging terms and discarding pieces that vanish by Ward
identities, we obtain simple expressions for the contribution 
to the $n$-gluon amplitude from each vertex of diagram $\I$.
We then show that each such contribution is invariant under
the color-factor shift associated with gluon $a$.

\subsection{Attaching a gluon to an external leg}

First we single out one of the external legs, $b$, of $\I$, 
denoting the resulting expression as
\be
\ve_{b\mu_b} B^{\mu_b}_{\ta_b}  (k_b, \cdots )
\ee
where $k_b$ and $\ve_b$ are the momentum and polarization vector of gluon $b$,
and $\cdots$ denotes momenta belonging to gluons 
$\{ 1, \cdots, n\} \setminus \{ a, b \}$.
Attaching gluon $a$ to external leg $b$ and using \eqn{threegluonvertex}
we obtain 
\begin{align}
 - { g \over \sqrt2}  
& {f_{\ta_{b} \ta_a \tb}  \over (k_a + k_b)^2  }
V^{\mu_b \mu_a \nu} (k_b, k_a, -k_a-k_b) 
B_{\tb\nu}  (k_a + k_b, \cdots  )
\nn\\
~=~ 
& - { g \over \sqrt2}  
{f_{\ta_{b} \ta_a \tb}  \over (k_a + k_b)^2  }
\Big[   
\eta^{\mu_b \nu  } k_a^{\mu_a }
- \eta^{\mu_a \nu } k_b^{\mu_b }
- \eta^{\mu_b \mu_a } (k_a + k_b)^{\nu}
\\
& 
\quad\qquad\qquad\qquad
+ 2 \eta^{\mu_b \nu} k_b^{\mu_a} 
+ 2 \left( \eta^{\mu_b \mu_a} k_a^\nu - \eta^{\mu_a \nu} k_a^{\mu_b}  \right)
 \Big]  
B_{\tb \nu}  (k_a + k_b, \cdots  ) \,.
\nn
\end{align}
Contracting this with 
$\ve_{b\mu_b} \ve_{a\mu_a} $ eliminates the first two terms in the square brackets.
The third term is proportional to 
\be
(k_a + k_b)^{\nu}
B_{\tb \nu}  (k_a + k_b, \cdots  ) \,.
\label{Wardtwo}
\ee
This term does not vanish by itself as it did in the case of the four-gluon amplitude, 
but when we obtain the total $n$-gluon amplitude
by including all $(n-1)$-point diagrams $\I$, 
the sum of such terms vanishes due to gauge invariance (Ward identity).
Therefore we are left with the last two terms in square brackets,
which yield the two terms in the following expression
\be
 - {  \sqrt2 g}  
{f_{\ta_{b} \ta_a \tb}  \over (k_a + k_b)^2  }
\ve_{b\mu_b} 
\Big[  
   \ve_a \cdot k_b \eta^{\mu_b \nu} 
+ \left( \ve_a^{\mu_b} k_a^\nu - \ve_a^{\nu} k_a^{\mu_b}  \right)
 \Big]  
B_{\tb \nu}  (k_a + k_b, \cdots  ) \,.
\label{attachexternalgluon}
\ee
We set this expression aside for now.

\subsection{Attaching a gluon to an internal line}

Next we single out one of the internal lines of $\I$, 
which divides the diagram into two subdiagrams $B$ and $C$,
and splits the external legs $\{ 1, \cdots, n\} \setminus \{ a \}$
into two complementary sets $\SaB$ and $\SaC$.
The contribution of the diagram can thus be written as 
\be
B_{\tb}^\mu ( -K, \cdots ) 
{(-i \eta_{\mu\nu}) \delta_{\tb\tc} \over K^2 }
C_{\tc}^\nu( K,  \cdots  ) 
\ee
where $K = \sum_{d \in \SaB} k_d$ 
is the momentum running through the line,
and the $\cdots$ in $B$ and $C$ denote momenta belonging to 
$\SaB$ and $\SaC$ respectively.
Attaching gluon $a$ to the line connecting the two subgraphs yields
\be
{i g\over \sqrt2}  
B_{\tb\mu} (-K, \cdots ) 
{ f_{\tb \tc \ta_a}  V^{\mu \nu \mu_a} (K,-K-k_a,k_a) 
\over  K^2 (K+k_a)^2 } 
C_{\tc\nu}(K+k_a,   \cdots  ) \,.
\ee
Writing the three-gluon vertex (\ref{threegluonvertex}) as
\be
V^{\mu \nu \mu_a} (K, -K-k_a,  k_a) 
~=~ 
- \eta^{\mu \nu  } k_a^{\mu_a }
+ \eta^{\mu_a \nu } K^{\mu }
+ \eta^{\mu \mu_a } (K + k_a)^{\nu}
- 2 \eta^{\mu \nu} K^{\mu_a} 
- 2 \left( \eta^{\mu \mu_a} k_a^\nu - \eta^{\mu_a \nu} k_a^{\mu}  \right)
\ee
we see that the first term vanishes 
when contracted with  $\ve_{a \mu_a}$.
The second and third terms give terms proportional to 
\be
K^\mu B_{\tb\mu} (-K, \cdots ) , 
\qquad
(K+ k_a)^\nu C_{\tc\nu} (K + k_a,   \cdots  ) \,.
\label{Wardthree}
\ee
Again these terms do not vanish by themselves,
but when we include all $(n-1)$-gluon diagrams $\I$, 
$B_{\tb\mu} $
will be replaced by the sum over all diagrams
containing external legs $\SaB$ plus one additional off-shell line,
and similarly for $ C_{\tc\nu}$,
and these expressions will vanish by gauge invariance (Ward identity).
We are thus left with 
\be
- \sqrt2 i g
B_{\tb \mu} (-K, \cdots ) 
{ f_{\tb \tc \ta_a}  
\left[  \eta^{\mu \nu} \ve_a \cdot K
+  \left( \ve_a^\mu k_a^\nu - \ve_a^\nu k_a^{\mu}  \right)
\right]
\over K^2 (K+k_a)^2 } 
C_{\tc\nu} ( K+k_a,  \cdots  )  \,.
\ee
Now we use the identity (\ref{splitprop})
to rewrite this as
\begin{align}
&
B_{\tb \mu} ( -K, \cdots ) {-i \over K^2} 
\left\{  \sqrt2 g
{f_{\tb \tc \ta_a}  \over 2 k_a \cdot K}
\Big[  \eta^{\mu \nu} \ve_a \cdot K
+  \left( \ve_a^\mu k_a^\nu - \ve_a^\nu k_a^{\mu}  \right)
\Big]
C_{\tc\nu} ( K+k_a,  \cdots  ) 
\right\}
\label{attachinternalgluon}
\\
&
+
\left\{ 
- \sqrt2 g
{ f_{\tb \tc \ta_a}  \over 2 k_a \cdot K}
B_{\tb \mu} (\cdots ,  -K)  
\Big[  \eta^{\mu \nu} \ve_a \cdot K
+  \left( \ve_a^\mu k_a^\nu - \ve_a^\nu k_a^{\mu}  \right)
\Big]
\right \}{-i \over (K+k_a)^2} 
 C_{\tc\nu} ( K+k_a,  \cdots  )  \,.
\nn
\end{align}
We associate each of the terms in this equation with
one of the two vertices to which the line is attached.
\para

\subsection{Radiation vertex expansion}

In the previous subsections,
we showed that attaching gluon $a$ to an $(n-1)$-gluon diagram
yields one term (\ref{attachexternalgluon}) for each external leg
and two terms (\ref{attachinternalgluon}) for each internal line,
or in other words,
one term for each leg of each vertex of $\I$.
We can therefore reorganize these terms into 
a sum over the legs of the vertices of each of the
$(n-1)$-gluon diagrams $\I$. 
This is the radiation vertex expansion \cite{Brown:1982xx}.
\para

First we choose one of the three-gluon vertices $v$ of $\I$ (if it has any).
Such a vertex divides the external legs into three non-overlapping subsets
$S_{ \aIvr }$, $r=1,2,3$ such that
$\bigcup_{r=1}^3 S_{ \aIvr } = \{ 1, \cdots, n\} \setminus \{ a\}$.
The contribution of diagram $\I$ to the $(n-1)$-gluon amplitude can be expressed 
\be
-{i g \over \sqrt2}
f_{\tc_1\tc_2\tc_3}
V^{\mu_1 \mu_2 \mu_3} (K_1, K_2, K_3)
\prod_{r=1}^3   
A^\aIvrr_{\tc_r \mu_r} (- K_r,  \cdots )  
\label{threegluonfactors}
\ee
where $K_r = \sum_{d \in S_{ \aIvr }} k_d$
is the momentum flowing out of each leg of the vertex,
and $\cdots$ in $A^\aIvrr$ denotes momenta belonging to $S_\aIvr$.
If any of the legs is external, then 
$A^\aIvrr_{\tc_r \mu_r}$ is just 
$ \delta_{\tb \tc_r} \ve_{b  \mu_b} $.
\para

We now attach gluon $a$ to each of the legs of this three-gluon vertex,
either to an external leg or to an internal line.
{}From \eqns{attachexternalgluon}{attachinternalgluon} this yields
\begin{align}
 i g^2 
 \prod_{r=1}^3   
A^\aIvrr_{\tc_r \mu_r} (- K_r,  \cdots ) 
\bigg( 
& 
{ 
f_{ \tb \tc_{1} \ta_a } f_{\tb\tc_{2} \tc_{3} } 
\over 2 k_a \cdot K_1}
\Big[  \eta^{\mu_1 \nu} \ve_a \cdot K_1
+  \left( \ve_a^{\mu_1} k_a^\nu - \ve_a^\nu k_a^{\mu_1}  \right)
\Big]
V_{\nu}^{~ \mu_2 \mu_3} (K_1 + k_a, K_2, K_3)
\nn\\
+&
{ 
f_{ \tb \tc_{2} \ta_a } f_{\tb\tc_{3} \tc_{1} } 
\over 2 k_a \cdot K_2}
\Big[  \eta^{\mu_2 \nu} \ve_a \cdot K_2
+  \left( \ve_a^{\mu_2} k_a^\nu - \ve_a^\nu k_a^{\mu_2}  \right)
\Big]
V_{~~~\nu}^{\mu_1 ~~\mu_3} (K_1, K_2 + k_a, K_3)
\nn\\
+&
{ 
f_{ \tb \tc_{3} \ta_a } f_{\tb\tc_{1} \tc_{2} } 
\over 2 k_a \cdot K_3}
\Big[  \eta^{\mu_3 \nu} \ve_a \cdot K_3
+  \left( \ve_a^{\mu_3} k_a^\nu - \ve_a^\nu k_a^{\mu_3}  \right)
\Big]
V_{~~~~~~\nu}^{\mu_1 \mu_2} (K_1, K_2, K_3 + k_a)
\bigg)  \,.
\label{attachlegthree}
\end{align}
We can also attach gluon $a$ directly to the three-gluon vertex itself.
Using \eqn{fourgluonvertexderiv}, this yields 
\begin{align}
- {i g^2 \over 2}  
\prod_{r=1}^3   
A^\aIvrr_{\tc_r \mu_r} (- K_r,  \cdots ) 
\bigg( 
&
f_{ \tb \tc_{1} \ta_a } f_{\tb\tc_{2} \tc_{3}}
\ve_{a\mu_a}   {\partial \over \partial  K_{1\mu_a}    }
V^{\mu_1 \mu_2 \mu_3} (K_1, K_2, K_3) 
\nn\\
+&
f_{ \tb \tc_{2} \ta_a } f_{\tb\tc_{3} \tc_{1} } 
\ve_{a\mu_a}   {\partial \over \partial  K_{2\mu_a}   }
V^{\mu_1 \mu_2 \mu_3} (K_1, K_2, K_3) 
\label{attachvertexthree}
\\
+
&
f_{ \tb \tc_{3} \ta_a } f_{\tb\tc_{1} \tc_{2} } 
\ve_{a\mu_a}   {\partial \over \partial  K_{3\mu_a}   }
V^{\mu_1 \mu_2 \mu_3} (K_1, K_2, K_3) 
\bigg) \,.
\nn
\end{align}
We now use \eqn{taylor} in \eqn{attachlegthree},
and combine \eqns{attachlegthree}{attachvertexthree}
as we did in sec.~\ref{sec:fourgluon}.
Leaving the indices on
$ V^{\mu_1 \mu_2 \mu_3} (K_1, K_2, K_3) $
implicit, we obtain the contribution of the 
three-gluon vertex to the radiation vertex expansion
\begin{align}
  i g^2  
\prod_{r=1}^3   A^\aIvrr_{\tc_r} (- K_r,  \cdots ) 
\bigg( 
&
{ f_{ \tb \tc_{1} \ta_a } f_{\tb\tc_{2} \tc_{3} } \over 2 k_a \cdot K_1}
\left[  \ve_a \cdot K_1 
-i \ve_{a\alpha} k_{a\beta}  J_1^{\alpha\beta} 
-i \ve_{a\alpha} k_{a\beta}  k_{a\gamma} S_1^{\alpha\beta} 
{\partial \over \partial  K_{1\gamma} }
\right]
V(K_1, K_2, K_3)
\nn\\
+
&
{ f_{ \tb \tc_{2} \ta_a } f_{\tb\tc_{3} \tc_{1} } \over 2 k_a \cdot K_2}
\left[  \ve_a \cdot K_2 
-i \ve_{a\alpha} k_{a\beta}  J_2^{\alpha\beta} 
-i \ve_{a\alpha} k_{a\beta}  k_{a\gamma} S_2^{\alpha\beta} 
{\partial \over \partial  K_{2\gamma} }
\right]
V(K_1, K_2, K_3)
\nn\\
+
&
{ f_{ \tb \tc_{3} \ta_a } f_{\tb\tc_{1} \tc_{2} } \over 2 k_a \cdot K_3}
\left[  \ve_a \cdot K_3 
-i \ve_{a\alpha} k_{a\beta}  J_3^{\alpha\beta} 
-i \ve_{a\alpha} k_{a\beta}  k_{a\gamma} S_3^{\alpha\beta} 
{\partial \over \partial  K_{3\gamma} }
\right]
V(K_1, K_2, K_3)
\bigg)
\label{threegluoncontrib}
\end{align}
where the subscripts on $J_r$ and $S_r$ indicate on which indices of 
$ V^{\mu_1 \mu_2 \mu_3} (K_1, K_2, K_3) $ 
these operators act.
\para

Next we choose one of the four-gluon vertices $v$ of $\I$ (if it has any).
Such a vertex divides the external legs into four non-overlapping subsets
$S_{ \aIvr}$, $r=1, \cdots, 4 $ such that
$\bigcup_{r=1}^4 S_{ (a,\I,v,r)} = \{ 1, \cdots, n\} \setminus \{ a\}$.
The contribution of the diagram $\I$ can be expressed as
\be
V_{\tc_1 \tc_2 \tc_3 \tc_4}^{\mu_1 \mu_2 \mu_3 \mu_4} 
\prod_{r=1}^4   
A^\aIvrr_{\tc_r \mu_r} (- K_r,  \cdots )  \,.
\ee
We now attach gluon $a$ to each of the legs of this four-gluon vertex.
{}From \eqns{attachexternalgluon}{attachinternalgluon} we obtain
\begin{align}
 -\sqrt2 g  
\prod_{r=1}^4   
A^\aIvrr_{\tc_r \mu_r} (- K_r,  \cdots ) 
\bigg( 
&
{ f_{ \tb \tc_{1} \ta_a } \over 2 k_a \cdot K_1}
\Big[  \eta^{\mu_1 \nu} \ve_a \cdot K_1
+  \left( \ve_a^{\mu_1} k_a^\nu - \ve_a^\nu k_a^{\mu_1}  \right)
\Big]
\eta_{\nu\lambda} V_{\tb  \tc_2 \tc_3 \tc_4}^{\lambda \mu_2 \mu_3 \mu_4} 
\nn\\
+&
{ f_{ \tb \tc_{2} \ta_a } \over 2 k_a \cdot K_2}
\Big[  \eta^{\mu_2 \nu} \ve_a \cdot K_2
+  \left( \ve_a^{\mu_2} k_a^\nu - \ve_a^\nu k_a^{\mu_2}  \right)
\Big]
\eta_{\nu\lambda} V_{\tc_1 \tb \tc_3 \tc_4}^{\mu_1 \lambda \mu_3 \mu_4} 
\nn\\
+&
{ f_{ \tb \tc_{3} \ta_a } \over 2 k_a \cdot K_3}
\Big[  \eta^{\mu_3 \nu} \ve_a \cdot K_3
+  \left( \ve_a^{\mu_3} k_a^\nu - \ve_a^\nu k_a^{\mu_3}  \right)
\Big]
\eta_{\nu\lambda} V_{\tc_1 \tc_2 \tb \tc_4}^{\mu_1 \mu_2 \lambda \mu_4} 
\nn\\
+&
{ f_{ \tb \tc_{4} \ta_a } \over 2 k_a \cdot K_4}
\Big[  \eta^{\mu_4 \nu} \ve_a \cdot K_4
+  \left( \ve_a^{\mu_4} k_a^\nu - \ve_a^\nu k_a^{\mu_4}  \right)
\Big]
\eta_{\nu\lambda} V_{\tc_1 \tc_2 \tc_3 \tb}^{\mu_1 \mu_2 \mu_3 \lambda} 
\bigg) \,.
\label{attachlegfour}
\end{align}
One cannot attach gluon $a$ to the four-gluon vertex itself
since there are no five-gluon vertices.
Thus the contribution to the radiation vertex expansion 
from the four-gluon vertex is
\begin{align}
 -\sqrt2 g 
 \prod_{r=1}^4   
A^\aIvrr_{\tc_r \mu_r} (- K_r,  \cdots ) 
\bigg( 
&
{ f_{ \tb \tc_{1} \ta_a } \over 2 k_a \cdot K_1}
\Big[
\delta^{\mu_1 }_{~\nu} \ve_a \cdot K_1
-i \ve_{a\alpha} k_{a\beta} ( S_1^{\alpha\beta} )^{\mu_1 }_{~~\nu} 
\Big]
V_{\tb  \tc_2 \tc_3 \tc_4}^{\nu \mu_2 \mu_3 \mu_4} 
\nn\\
+&
{ f_{ \tb \tc_{2} \ta_a } \over 2 k_a \cdot K_2}
\Big[
\delta^{\mu_2 }_{~\nu} \ve_a \cdot K_2
-i \ve_{a\alpha} k_{a\beta} ( S_2^{\alpha\beta} )^{\mu_2 }_{~~\nu} 
\Big]
V_{\tc_1\tb  \tc_3 \tc_4}^{\mu_1 \nu \mu_3 \mu_4} 
\nn\\
+&
{ f_{ \tb \tc_{3} \ta_a } \over 2 k_a \cdot K_3}
\Big[
\delta^{\mu_3 }_{~\nu} \ve_a \cdot K_3
-i \ve_{a\alpha} k_{a\beta} ( S_3^{\alpha\beta} )^{\mu_3 }_{~~\nu} 
\Big]
V_{\tc_1  \tc_2 \tb \tc_4}^{\mu_1 \mu_2 \nu \mu_4} 
\nn\\
+&
{ f_{ \tb \tc_{4} \ta_a } \over 2 k_a \cdot K_4}
\Big[
\delta^{\mu_4 }_{~\nu} \ve_a \cdot K_4
-i \ve_{a\alpha} k_{a\beta} ( S_4^{\alpha\beta} )^{\mu_4 }_{~~\nu} 
\Big]
V_{\tc_1  \tc_2 \tc_3 \tb}^{\mu_1 \mu_2 \mu_3 \nu} 
\bigg) \,.
\label{fourgluoncontrib}
\end{align}

To summarize this section, we have expressed an $n$-gluon amplitude as a sum 
over the vertices of all of the $(n-1)$-gluon diagrams $\I$,
comprising a term (\ref{threegluoncontrib}) for each three-gluon vertex of $\I$ 
and a term (\ref{fourgluoncontrib}) for each four-gluon vertex of $\I$.

\subsection{Invariance of the radiation vertex expansion under color-factor symmetry}

Computing the variation of the radiation vertex expansion of the amplitude under 
a color-factor shift is somewhat more delicate than calculating 
the variation of the cubic vertex expansion of the amplitude  (as we did in sec.~\ref{sec:cfs})
because each factor $A^\aIvrr_{\tc_r \mu_r}$ in \eqns{threegluoncontrib}{fourgluoncontrib}
can contain more than one color factor $c_i$
due to the possible presence of four-gluon vertices.
\para

First let us consider the contribution (\ref{threegluoncontrib})
of a three-gluon vertex to the radiation vertex expansion,
and for the moment let us assume that the
subdiagrams corresponding to $A^\aIvrr_{\tc_r \mu_r}$
contain only three-gluon vertices.
Designate by $c_\aIvr$ with $r=1, 2, 3$ the color factor 
associated with each line of \eqn{threegluoncontrib}.
Thus, for example, $c_{(a,\I,v,1)}$ is the product of
$f_{ \tb \tc_{1} \ta_a } f_{\tb\tc_{2} \tc_{3} } $
and the structure constants from all the three-gluon vertices in 
$\prod_{r=1}^3 A^\aIvrr_{\tc_r \mu_r}$.
These color factors manifestly satisfy  $\sumr c_\aIvr=0$.
The variation of $c_\aIvr$ under the color-factor shift associated with gluon $a$ is
\be
\delta_a ~ c_\aIvr   ~=~ \alpha_{(a,\I,v)} ~   k_a \cdot K_r
\ee
which preserves $\sumr c_\aIvr=0$.
The variation of \eqn{threegluoncontrib} under the color-factor shift 
is therefore proportional to 
\be
\left[ \left(  \sumr  \ve_a \cdot K_r  \right)
-i \ve_{a\alpha} k_{a\beta}  
\left( \sumr J_r^{\alpha\beta}  \right)
-i \ve_{a\alpha} 
k_{a\beta}  
k_{a\gamma}
\left( \sumr S_r^{\alpha\beta} {\partial \over \partial  K_{r\gamma} } \right)
\right]
V(K_1, K_2, K_3) \,.
\label{threesums}
\ee
In sec.~\ref{sec:fourgluon}, we demonstrated that each of the three terms 
in \eqn{threesums} vanishes.
If the diagrams corresponding to 
$A^\aIvrr_{\tc_r \mu_r}$ contain four-gluon vertices, 
we can use \eqn{fourgluonvertex} to expand these expressions
and then use the argument above to show that each separate contribution will
vanish under the color-factor shift.
Therefore the contribution of the three-gluon vertices (\ref{threegluoncontrib})
to the radiation vertex expansion is invariant under the color-factor shift
associated with gluon $a$.
\para

Second let us consider the contribution (\ref{fourgluoncontrib})
of a four-gluon vertex to the radiation vertex expansion.
Again we begin by assuming that the diagrams corresponding 
to $A^\aIvrr_{\tc_r \mu_r}$ contain only three-gluon vertices.
We now expand the four-gluon vertices
$V_{\tc_1  \tc_2 \tc_3 \tc_4}^{\mu_1 \mu_2 \mu_3 \mu_4} $
in \eqn{fourgluoncontrib}
into several terms, one of which is 
\begin{align}
 -\sqrt2 g 
& \prod_{r=1}^4   
A^\aIvrr_{\tc_r \mu_r} (- K_r,  \cdots ) 
\nn\\
\times \bigg( 
&
{ 
f_{ \tb \tc_{1} \ta_a } f_{b \tc_{2} \td} f_{\td \tc_3 \tc_4}
\over 2 k_a \cdot K_1}
\Big[
\delta^{\mu_1}_{~ \nu} \ve_a \cdot K_1
-i \ve_{a\alpha} k_{a\beta} ( S_1^{\alpha\beta} )^{\mu_1}_{~~ \nu} 
\Big]
\left( \eta^{\nu \mu_3} \eta^{\mu_2 \mu_4} -\eta^{\nu \mu_4} \eta^{\mu_2 \mu_3} \right)
\nn\\
+&
{ f_{ \tb \tc_{2} \ta_a } f_{\tc_{1} \tb \td} f_{\td \tc_3 \tc_4}
\over 2 k_a \cdot K_2}
\Big[
\delta^{\mu_2}_{~ \nu} \ve_a \cdot K_2
-i \ve_{a\alpha} k_{a\beta} ( S_2^{\alpha\beta} )^{\mu_2}_{~~ \nu} 
\Big]
\left( \eta^{\mu_1 \mu_3} \eta^{\nu \mu_4} -\eta^{\mu_1 \mu_4} \eta^{\nu \mu_3} \right)
\nn\\
+&
{ f_{ \tb \tc_{3} \ta_a } f_{b \tc_{4} \td} f_{\td \tc_1 \tc_2}
\over 2 k_a \cdot K_3}
\Big[
\delta^{\mu_3}_{~ \nu} \ve_a \cdot K_3
-i \ve_{a\alpha} k_{a\beta} ( S_3^{\alpha\beta} )^{\mu_3}_{~~ \nu} 
\Big]
\left( \eta^{\mu_1 \nu} \eta^{\mu_2 \mu_4} -\eta^{\mu_1 \mu_4} \eta^{\mu_2 \nu} \right)
\nn\\
+&
{ f_{ \tb \tc_{4} \ta_a } f_{\tc_{3} \tb \td} f_{\td \tc_1 \tc_2}
\over 2 k_a \cdot K_4}
\Big[
\delta^{\mu_4}_{~ \nu} \ve_a \cdot K_4
-i \ve_{a\alpha} k_{a\beta} ( S_4^{\alpha\beta} )^{\mu_4}_{~~ \nu} 
\Big]
\left( \eta^{\mu_1 \mu_3} \eta^{\mu_2 \nu} -\eta^{\mu_1 \nu} \eta^{\mu_2 \mu_3} \right)
\bigg) \,.
\label{expandedfourgluoncontrib}
\end{align}
Designate by $c_\aIvr$ with $r=1, \cdots 4$  the color factor 
associated with each line of \eqn{expandedfourgluoncontrib},
including the structure constants from all the three-gluon vertices in 
$\prod_{r=1}^4 A^\aIvrr_{\tc_r \mu_r}$.
These color factors satisfy
$\sumrr c_\aIvr=0$ by virtue of 
\be
 f_{ \tb \tc_{1} \ta_a } f_{b \tc_{2} \td} f_{\td \tc_3 \tc_4}
+ f_{ \tb \tc_{2} \ta_a } f_{\tc_{1} \tb \td} f_{\td \tc_3 \tc_4}
+ f_{ \tb \tc_{3} \ta_a } f_{b \tc_{4} \td} f_{\td \tc_1 \tc_2}
+ f_{ \tb \tc_{4} \ta_a } f_{\tc_{3} \tb \td} f_{\td \tc_1 \tc_2}
~=~0 \,.
\ee
The variation of $c_\aIvr$ under the color-factor shift associated with gluon $a$ is
\be
\delta_a ~ c_\aIvr   ~=~ \alpha_{(a,\I,v)} ~   k_a \cdot K_r \,.
\ee
The variation of \eqn{expandedfourgluoncontrib} under the color-factor shift 
therefore contains two sums.
The first
\be
\left(  \sumrr  \ve_a \cdot K_r  \right)
\left( \eta^{\mu_1 \mu_3} \eta^{\mu_2 \mu_4} -\eta^{\mu_1 \mu_4} \eta^{\mu_2 \mu_3} \right)
\label{variationfourglounleading}
\ee
vanishes by momentum conservation, $k_a + \sumrr K_r = 0 $,  
together with $\ve_a \cdot k_a = 0$.
The second 
\begin{align}
( S_1^{\alpha\beta} )^{\mu_1}_{~~ \nu} 
\left( \eta^{\nu \mu_3} \eta^{\mu_2 \mu_4} -\eta^{\nu \mu_4} \eta^{\mu_2 \mu_3} \right)
&+( S_2^{\alpha\beta} )^{\mu_2}_{~~ \nu} 
\left( \eta^{\mu_1 \mu_3} \eta^{\nu \mu_4} -\eta^{\mu_1 \mu_4} \eta^{\nu \mu_3} \right)
\nn\\
+( S_3^{\alpha\beta} )^{\mu_3}_{~~ \nu} 
\left( \eta^{\mu_1 \nu} \eta^{\mu_2 \mu_4} -\eta^{\mu_1 \mu_4} \eta^{\mu_2 \nu} \right)
&+( S_4^{\alpha\beta} )^{\mu_4}_{~~ \nu} 
\left( \eta^{\mu_1 \mu_3} \eta^{\mu_2 \nu} -\eta^{\mu_1 \nu} \eta^{\mu_2 \mu_3} \right)
\label{variationfourglounsubleading}
\end{align} 
is the first-order Lorentz transformation of the tensor
$\eta^{\mu_1 \mu_3} \eta^{\mu_2 \mu_4} -\eta^{\mu_1 \mu_4} \eta^{\mu_2 \mu_3}$,
which vanishes.  
The variation under the color-factor shift
of the other two terms from the expansion of
$V_{\tc_1  \tc_2 \tc_3 \tc_4}^{\mu_1 \mu_2 \mu_3 \mu_4} $
similarly vanishes.
Furthermore, the same argument applies when the diagrams corresponding to 
$A^\aIvrr_{\tc_r \mu_r}$ contain four-gluon vertices, 
by expanding these expressions using \eqn{fourgluonvertex}.
Therefore the contribution of the four-gluon vertices (\ref{fourgluoncontrib})
to the radiation vertex expansion is invariant under the color-factor shift
associated with gluon $a$.
\para

In fine, we have shown that each contribution to the radiation vertex expansion
is invariant under the color-factor shift associated with gluon $a$,
and therefore the entire $n$-gluon amplitude is invariant under this
shift. QED

\section{Color-factor symmetry for more general amplitudes}
\setcounter{equation}{0}
\label{sec:fund}

In secs.~\ref{sec:fourgluon} and \ref{sec:rvegluon}, 
we proved the color-factor symmetry of $n$-gluon amplitudes, 
from which follow the BCJ relations for color-ordered amplitudes.
Color-factor symmetry is a property of a much larger class 
of tree-level gauge-theory amplitudes, 
namely those containing at least one gluon together 
with massless or massive particles in 
arbitrary representations of the gauge group
with arbitrary spin $\le 1$,
with the usual gauge-theory couplings.
We will establish the invariance of this larger class of
gauge-theory amplitudes under a color-factor shift
in secs.~\ref{sec:fourfund} and \ref{sec:rvefund}.
\para

Consider a tree-level $n$-point gauge-theory amplitude $\cAn$
with gluons as well as particles $\psi$ and $\barpsi$,
either massless or massive, 
with spin zero, one-half, or one,
in an arbitrary representation of the gauge group.
For convenience, throughout the next three sections 
we refer to $\psi$ (and $\barpsi$) 
as fundamentals (and antifundamentals), 
but they can be in any representation.
This amplitude has the cubic decomposition
\be
\cAn
~=~ \sum_i 
{c'_i ~ n'_i 
\over d'_i } 
\label{cubicdecompfund}
\ee
where we decorate the color factors, kinematic numerators, and 
denominators with primes to distinguish them from the 
analogous quantities for $n$-gluon amplitudes.
The denominator $d'_i$ now consists of the product 
of inverse propagators for both massless and massive particles. 
The color factor $c'_i$ associated with each cubic diagram is 
obtained by sewing together $ggg$ vertices $f_{\textsf{abc}}$
and $\barpsi g \psi $ vertices $(T^{\textsf{a}})^{\textsf{i}}_{~\textsf{j}} $
where $T^\textsf{a}$ denote the generators in 
the appropriate representation.
Contributions from Feynman diagrams with quartic vertices 
(either $gggg$ or $\barpsi g g \psi $ in the case of a scalar or 
vector $\psi$)
are parceled out among the cubic diagrams.
The number of cubic diagrams in the sum (\ref{cubicdecompfund})
will generally be fewer than for the $n$-gluon amplitude, 
as some will be excluded for violating fermion number or flavor symmetry.
\para

Just as in the case of the $n$-gluon amplitude,
the amplitude $\cAn$ can be written in a cubic vertex expansion
with respect to gluon $a$:
\be
\cAn ~=~ 
\sum_I  
\sum_v  {1 \over \prod_{s=1}^3 d'_{(a,I,v,s)}  }
\sumr 
{ c'_{(a,I,v,r)} n'_{(a,I,v,r)} \over 2 k_a \cdot K_{(a,I,v,r)}   } \,.
\ee
The only difference between the derivation of this expression
and that for the $n$-gluon amplitude given 
in sec.~\ref{sec:cfs}  is that we must use the modified identity
\be
{ 1\over [K^2 -m^2] [ (K+k_a)^2 -m^2] }
~=~ 
{1 \over [K^2 -m^2]  (2 k_a \cdot K) } 
~+~
{1 \over  (-2 k_a \cdot K) [(K+k_a)^2 - m^2]} 
\label{massivesplitprop}
\ee
when gluon $a$ is attached to a propagator of a massive field.
\para

As usual, the color-factor shift associated with gluon $a$
is defined by two requirements: 
(I) that it satisfy all the algebraic symmetries (\eg Jacobi relations)
obeyed by the color factors $c'_i$, and 
(II) that it satisfy
\be
\delta_a c'_i  ~\propto~  \sum_{c\in \Sai } k_a \cdot k_c  
\ee
where $\Sai$ denotes the subset of the external particles on one side of the point
at which $a$ is attached to $c'_i$.
These together imply that 
\be
\delta_a ~ c'_{(a,I,v,r)} ~=~   \alpha_{(a,I,v)} ~k_a \cdot K_{(a,I,v,r)} \,.
\ee
Therefore the invariance of the amplitude under the color-factor shift implies
the constraint
\be
\sum_I  \sum_v
{\alpha_{(a,I,v)}  \over \prod_{s=1}^3 d'_{(a,I,v,s)}  }
\sumr 
n'_{(a,I,v,r)}  ~ = ~ 0 
\ee
on the sums of kinematic numerators appearing in the cubic 
decomposition.
\para

As in the case of $n$-gluon amplitudes,
color-factor symmetry can be used to derive BCJ relations 
among the color-ordered amplitudes associated with $\cAn$.
BCJ relations for $n$-point amplitudes with gluons 
and a single pair of massive fundamentals were conjectured 
in ref.~\cite{Naculich:2014naa}
and more generally for amplitudes containing an arbitrary number of 
pairs of fundamentals in ref.~\cite{Johansson:2015oia},
based on the assumption of color-kinematic duality.
A proof of these BCJ relations using BCFW on-shell recursion 
was given in ref.~\cite{delaCruz:2015dpa}.
In order to derive these relations, however,
it is necessary to write the amplitude in a {\it proper} 
decomposition \cite{Melia:2015ika}, \ie
in terms of an independent set of color factors
and generalized-gauge-invariant primitive amplitudes. 
For general amplitudes, this is a subtle problem, 
which was recently solved for the case of multiple pairs 
of distinct-flavor fundamentals by 
Melia \cite{Melia:2013bta,Melia:2013epa,Melia:2015ika}
and Johansson and Ochirov \cite{Johansson:2015oia}.
In a sequel to this paper \cite{Brown:2016hck}, we review their solution,
and then derive the BCJ relations using the color-factor symmetry.
\para

There is one class of amplitudes, however, 
for which the story is practically identical to the $n$-gluon case,
namely, $n$-point amplitudes with $n-2$ gluons 
and a single pair of fundamentals.
For that case, 
an independent set of $(n-2)!$ color factors is given by the 
half-ladders
\be
\bc'_{1 \gamma n}
~\equiv~
\left( {T}^{\textsf{a}_{\gamma(2)}}{T}^{\textsf{a}_{\gamma(3)}}
\cdots {T}^{\textsf{a}_{\gamma(n-1)}} \right)^{\textsf{i}_1}_{~~ \textsf{i}_n} 
\, .
\ee
All other color factors $c_i'$ can be reduced to half ladders 
\be
c'_i ~=~ \sum_{\gamma \in S_{n-2}}  M_{i, 1\gamma n} \bc'_{1 \gamma n } 
\ee
by repeatedly applying
$
f_{\textsf{abc}} \left( T^{\textsf{c}} \right)^{\textsf{i}}_{~\textsf{j}} 
= \left[ T^{\textsf{a}}, T^{\textsf{b}} \right]^{\textsf{i}}_{~\textsf{j}}   \,,
$
similar to the case of $n$-gluon amplitudes \cite{DelDuca:1999rs}.
The coefficients $M_{i, 1\gamma n}$ are precisely the same as 
in the $n$-gluon case.
The $n$-point amplitude can then be written 
in a proper decomposition \cite{Kosower:1987ic,Mangano:1988kk}
\ba
\cAn ( \barpsi_1, g_2, g_3, \cdots, g_{n-1}, \psi_n)
&=& 
\sum_{\gamma \in S_{n-2} }
\bc'_{1 \gamma n} ~A' (1, \gamma(2),  \cdots, \gamma(n-1), n)  
\label{ddmfund}
\ea
where the primitive amplitudes are given by
\ba
A' (1, \gamma(2),  \cdots, \gamma(n-1), n)  
&=&
\sum_i 
{M_{i, 1\gamma n}  ~ n'_i 
\over d'_i }  \,.
\ea
We define an $(n-3)!$-parameter family of shifts 
associated with each gluon $a \in \{2, \cdots, n-1\}$ via 
\begin{align}
\delta_a~ \bc'_{1 \sigma(2) \cdots \sigma(b-1) a \sigma(b) \cdots  \sigma(n-1) n } 
~&=~ 
 \alpha_{a,\sigma} \left( 
k_a \cdot k_1  +  \sum_{c=2}^{b-1} k_a \cdot k_{\sigma(c)} \right), 
\qquad a, b \in \{2, \cdots, n-1\}, \qquad b \neq a 
\nn\\
\delta_a\, c'_i ~&=~ \sum_{\gamma \in S_{n-2}}  M_{i, 1\gamma n} 
~\delta_a\, \bc'_{1 \gamma  n }  
\label{halfladdershiftfund}
\end{align}
where $\gamma$ is a permutation of 
$ \{2, \cdots, n-1\}$,
$\sigma$ is a permutation of 
$ \{2, \cdots, n-1\} \setminus  \{ a\}$,
and $\alpha_{a,\sigma}$ is a set of $(n-3)!$ arbitrary constants for each $a$.
As in the case of the $n$-gluon amplitude,
the dimension of the (abelian) group of color-factor shifts is $(n-3) (n-3)!$.
We show in secs.~\ref{sec:fourfund} and \ref{sec:rvefund} that the amplitude 
$\cAn ( \barpsi_1, g_2, g_3, \cdots, g_{n-1}, \psi_n) $
is invariant under the color-factors shifts (\ref{halfladdershiftfund}).
As a consequence, the color-ordered amplitudes defined in \eqn{ddmfund}
obey BCJ relations that
have the same form 
(when expressed in terms of invariants $k_a \cdot k_b$, 
where $k_a$ is the momentum of a gluon)
as those for the $n$-gluon amplitude,
namely
\be
\sum_{b=3}^n \left( k_1 \cdot k_2 + \sum_{c=3}^{b-1} k_2 \cdot k_{\sigma(c)}  \right) 
A' (1, \sigma(3), \cdots, \sigma(b-1), 2, \sigma(b), \cdots, \sigma(n-1), n) 
~=~ 0
\ee
together with all permutations of this equation with $2$ replaced by $a$,
as conjectured in refs.~\cite{Naculich:2014naa,Johansson:2015oia}.

\section{Proof of color-factor symmetry for the 
$ \cA_4 (\barpsi_1, \psi_2, g_3, g_4)  $ 
amplitude}
\setcounter{equation}{0}
\label{sec:fourfund}

In this section, we prove that the tree-level four-point amplitude 
with two gluons and two massive particles 
in an arbitrary representation of the gauge group
(which for convenience we refer to as fundamentals)
with spin zero, one-half, or one
is invariant under the color-factor symmetry.
We will use these results for the proof of the invariance of the 
more general $n$-point amplitude in sec.~\ref{sec:rvefund}. 
\para

The four-point amplitude
$ \cA_4 (\barpsi_1, \psi_2, g_3, g_4)  $ 
can be constructed from the $\barpsi \psi g$ vertex by attaching a 
gluon to a propagator emanating from each of the external legs of the vertex,
or (in the case of a spin-zero or spin-one fundamental) to the vertex itself.
This yields
\be
\cA_4 (\barpsi_1, \psi_2, g_3, g_4)  
~=~ \sumr  {\crp \nrp \over 2 k_4 \cdot k_r}
\label{fourfundamplitude}
\ee
where the color factors
\be
c'_{(1)}  ~=~  - \left( {T}^{\ta_4} {T}^{\ta_3} \right)^{\ti_1}_{~~ \ti_2} ,\qquad
c'_{(2)}  ~=~  \left( {T}^{\ta_3} {T}^{\ta_4} \right)^{\ti_1}_{~~ \ti_2} , \qquad 
c'_{(3)}  ~=~  f_{\ta_4 \ta_3 \tb} ( {T}^{\tb} )^{\ti_1}_{~~ \ti_2} 
\label{fourfundcolorfactors}
\ee
obey $\sum_{r=1}^3 \crp = 0$
using $[T^\ta, T^\tb] = f_{\ta\tb\tc} T^\tc $.
The color-factor symmetry (associated with gluon 4) acts on the 
color factors (\ref{fourfundcolorfactors}) and the four-point amplitude 
(\ref{fourfundamplitude}) as 
\be
\delta_4 \crp  ~=~ \alpha_4  \, k_4 \cdot k_r 
\qquad \implies \qquad 
\delta_4 \cA_4 ~=~ 
{1\over 2} \alpha_4
\sumr \nrp  \,.
\ee
We will establish that $\delta_4 \cA_4 =0$ by showing that $\sumr \nrp =0$,
a result that has long been known \cite{Zhu:1980sz,Goebel:1980es}.
We will do this separately for 
spin zero, spin one-half, and spin one fundamentals.

\subsection{Kinematic numerators for spin-one-half fundamentals}

We begin with the case of a spin-one-half fundamental,
which is simpler  due to the absence of a $\barpsi \psi g g $ vertex. 
The $\barpsi \psi g$ 
vertex\footnote{ Recall that $\Tr( T^\ta T^\tb ) = \delta^{\ta\tb}$.}
and Dirac propagator are
\be
{i  g \over \sqrt2} ( {T}^{\ta_3} )^{\ti_1}_{~~ \ti_2} \V^{\mu_3} , 
\qquad \qquad \qquad 
{i \delta^\ti_{~\tj} \over  \kslash - m}
\,.
\ee
Attaching gluon 4 to (fermion) leg 1 yields the expression
\begin{align}
-  { ig^2 \over 2} 
& 
{  ({T}^{\ta_4} T^{\ta_3} )^{\ti_1}_{~~ \ti_2} 
 \over (k_1 + k_4)^2- m^2  }
\gamma^{\mu_4} (\kslash_1 + \kslash_4 + m) \V^{\mu_3} 
~=~ 
{i g^2 \over 2} {  c'_{(1)} \over 2 k_4 \cdot k_1 }
\left[ (-\kslash_1 + m) \gamma^{\mu_4} + 2 k_1^{\mu_4} + {1\over 2} \left[ \gamma^{\mu_4} , \kslash_4 \right] \right]
\V^{\mu_3}  \,.
\label{fermionlegone}
\end{align}
The contribution of this diagram to \eqn{fourfundamplitude} 
is obtained by sandwiching \eqn{fermionlegone} 
between $\bar{u} (k_1)$ and $u(-k_2)$,
contracting with $\ve_{3\mu_3} \ve_{4\mu_4} $,
and dividing by $i$.
The first term in the square brackets vanishes using $\bar{u} (k_1) (-\kslash_1 + m) = 0$, 
leaving
\be
n'_{(1)}
~=~  g^2  \bar{u} (k_1)
\left[ \ve_4 \cdot k_1 -i \ve_{4\alpha} k_{4\beta}  \Sigma^{\alpha\beta} 
\right] \epsslash u(-k_2) \,,
\qquad\qquad
\Sigma^{\alpha\beta} \equiv {i \over 4} \left[ \gamma^\alpha, \gamma^\beta \right]
\ee 
where the spin-one-half angular momentum  matrices 
$\Sigma^{\alpha\beta}$ 
satisfy  the Lorentz algebra commutation relations
\be
[ \Sigma^{\alpha\beta}, \Sigma^{\gamma\delta}  ]
~=~ 
-i \left[ 
 \eta^{\alpha\gamma} \Sigma^{\beta\delta}  
-\eta^{\alpha\delta} \Sigma^{\beta\gamma}  
-\eta^{\beta\gamma} \Sigma^{\alpha\delta}  
+\eta^{\beta\delta} \Sigma^{\alpha\gamma} 
\right] \,.
\ee
Similarly, attaching gluon 4 to (fermion) leg 2 yields
\begin{align}
-  { ig^2 \over 2} 
& 
{  ({T}^{\ta_3} T^{\ta_4} )^{\ti_1}_{~~ \ti_2} 
 \over (k_2 + k_4)^2- m^2  }
\V^{\mu_3} (-\kslash_2 - \kslash_4 + m) \gamma^{\mu_4} 
~=~ 
- {i g^2 \over 2} {  c'_{(2)} \over 2 k_4 \cdot k_2 }
\V^{\mu_3} 
\left[ \gamma^{\mu_4}  (\kslash_2 + m) - 2 k_2^{\mu_4} 
+ {1\over 2} \left[  \gamma^{\mu_4} ,\kslash_4   \right] \right]\,.
\end{align}
Using  $(\kslash_2 + m) u(-k_2)=0$, we obtain
\be
n'_{(2)}
~=~  g^2  \bar{u} (k_1)
\epsslash
   \left[ 
\ve_4 \cdot k_2 +i \ve_{4\alpha} k_{4\beta}  \Sigma^{\alpha\beta} 
\right]
u(-k_2) \,.
\ee 
Finally, attaching gluon 4 to (gluon) leg 3 yields
\be
n'_{(3)}
~=~  g^2  \ve_{3\mu_3} 
\bar{u} (k_1)
   \left[ 
   \ve_4 \cdot k_3 \delta^{\mu_3}_{~\nu }
-i \ve_{4\alpha} k_{4\beta} ( S_3^{\alpha\beta} )^{\mu_3}_{~~ \nu} 
\right]
\V^{\nu} 
u(-k_2) \,.
\ee 
The sum of the kinematic numerators is thus
\begin{align}
\sumr \nrp
~=~&
g^2  \bar{u} (k_1) \epsslash u(-k_2)
\left( \sumr    \ve_4 \cdot k_r \right) 
\nn\\
& 
-i g^2  \ve_{4\alpha} k_{4\beta}  \ve_{3\mu_3} \bar{u} (k_1) 
\left(
\Sigma^{\alpha\beta} \V^{\mu_3} 
- \V^{\mu_3} \Sigma^{\alpha\beta} 
+ ( S_3^{\alpha\beta} )^{\mu_3}_{~~ \nu} 
\V^{\nu} 
\right)
u(-k_2) \,.
\end{align}
The first sum on the right-hand side of this equation 
vanishes by momentum conservation $\sum_{r=1}^4 k_r =  0$ 
together with  $\ve_4 \cdot k_4=0$.   
The second sum vanishes because
\be
\Sigma^{\alpha\beta} \V^{\mu_3} 
- \V^{\mu_3} \Sigma^{\alpha\beta} 
+ ( S_3^{\alpha\beta} )^{\mu_3}_{~~ \nu} \V^{\nu} 
\ee
is the first-order Lorentz transformation of $\gamma^\mu$
(acting on both spinor indices as well as the vector index)
and hence vanishes.
Thus the sum of kinematic numerators for the amplitude 
$ \cA_4 (\barpsi_1, \psi_2, g_3, g_4)  $ is zero
for spin-one-half fundamentals.

\subsection{Kinematic numerators for spin-zero fundamentals}

Next we turn to the case of spin-zero fundamentals.
The $\barpsi \psi g$ vertex is
\be
    {i  g \over \sqrt2} ( {T}^{\ta_3} )^{\ti_1}_{~~ \ti_2} V^{\mu_3} (k_1, k_2, k_3) , 
\qquad
V^{\mu_3} (k_1, k_2, k_3) ~=~ (k_1 - k_2)^{\mu_3}
\label{scalarscalargluonvertex}
\ee
where $k_a$ are outgoing momenta.
The scalar propagator is $i \delta^\ti_{~\tj} /(k^2-m^2)$.
Thus attaching gluon 4 to (scalar) leg 1 yields the expression
\begin{align}
-  { ig^2 \over 2} 
& 
{  ({T}^{\ta_4} T^{\ta_3} )^{\ti_1}_{~~ \ti_2} 
 \over (k_1 + k_4)^2- m^2  }
V^{\mu_4} (k_1, -k_1-k_4, k_4) 
V^{\mu_3} (k_1 + k_4, k_2, k_3)
\nn\\
~=~ &
   {i g^2 \over 2} {  c'_{(1)} \over 2 k_4 \cdot k_1 }
\left[   
k_4^{\mu_4 } + 2 k_1^{\mu_4} 
 \right]  
{V^{\mu_3} (k_1 + k_4, k_2, k_3) } \,.
\end{align}
The contribution of this diagram to \eqn{fourfundamplitude} 
is obtained by contracting with $\ve_{3\mu_3} \ve_{4\mu_4} $
and dividing by $i$.
The first term in the square brackets vanishes  
using $\ve_4 \cdot k_4=0$, 
leaving
\be
n'_{(1)}\Big|_{\rm leg} 
~=~  g^2 
   \ve_4 \cdot k_1 
\ve_{3\mu_3} 
V^{\mu_3} (k_1 + k_4, k_2, k_3) \,.
\ee 
Similarly, attaching gluon 4 to (scalar) leg 2 yields
\be
n'_{(2)}\Big|_{\rm leg} 
~=~  g^2 
   \ve_4 \cdot k_2 
\ve_{3\mu_3} 
V^{\mu_3} (k_1, k_2+k_4, k_3)\,.
\ee 
Attaching gluon 4 to (gluon) leg 3 yields
\be
n'_{(3)}\Big|_{\rm leg} 
~=~  g^2 \ve_{3\mu_3} 
\left[  
   \ve_4 \cdot k_3 \delta^{\mu_3}_{~\nu }
-i \ve_{4\alpha} k_{4\beta} ( S_3^{\alpha\beta} )^{\mu_3}_{~~ \nu} 
 \right]  
V^{\nu} (k_1, k_2, k_3 + k_4) \,.
\ee 
Attaching gluon 4 directly to the $\barpsi \psi g$ vertex, 
we obtain the $\barpsi \psi g g $ vertex
\be
{i g^2 \over 2} \eta^{\mu_3 \mu_4} 
\left( \{ {T}^{\ta_3},  {T}^{\ta_4} \} \right)^{\ti_1}_{~~ \ti_2} 
\label{ssgg}
\ee
which can be written as
\be
-  {i g^2 \over 2} \left( 
 c'_{(1)} {\partial \over \partial  k_{1\mu_4}   }
+c'_{(2)} {\partial \over \partial  k_{2\mu_4}   }
+c'_{(3)} {\partial \over \partial  k_{3\mu_4}   }
\right) 
V^{\mu_3} (k_1, k_2, k_3)  \,.
\label{scalarscalargluongluonvertex}
\ee
The rest of the story proceeds exactly as in sec.~\ref{sec:fourgluon},
allowing us to write the kinematic numerators as 
\begin{align}
\nrp
&~=~
g^2 \left[  
   \ve_4 \cdot k_r 
-i \ve_{4\alpha} k_{4\beta}  J_r^{\alpha\beta} 
-i \ve_{4\alpha} k_{4\beta}  k_{4\gamma} 
S_r^{\alpha\beta} 
{\partial \over \partial  k_{r\gamma} }
 \right]  
V(k_1 , k_2, k_3)
\end{align}
where we have suppressed the polarization vector $\ve_3$.
Because two of the legs are scalars, we have $S^{\alpha\beta}_1 = S^{\alpha\beta}_2 = 0$,
whereas $S^{\alpha\beta}_3$ is given by \eqn{S}.
\para

Now consider the sum of kinematic numerators 
\begin{align}
\sumr \nrp
~=~&
 g^2 \left( \sumr \ve_4 \cdot k_r \right) V(k_1 , k_2, k_3)
\nn\\
&-
ig^2  \ve_{4\alpha} k_{4\beta} \left( \sumr J_r^{\alpha\beta} \right) V(k_1 , k_2, k_3)
\label{sumnrscalar}
\\
&-
ig^2  \ve_{4\alpha} k_{4\beta} k_{4\gamma} 
S_3^{\alpha\beta} {\partial \over \partial  k_{3\gamma}   }
V(k_1 , k_2, k_3)
\nn
\end{align}
where the scalar-scalar-gluon vertex is 
$V^{\mu_3} (k_1, k_2, k_3) ~=~ (k_1 - k_2 + \lambda[k_1 + k_2 + k_3])^{\mu_3}$
with $\lambda$ arbitrary due to momentum conservation.
The first sum on the right-hand side of this equation 
vanishes as usual by momentum conservation.
The second sum,
which may be written more explicitly as
\be
L_1^{\alpha\beta} 
V^{\mu_3} (k_1, k_2, k_3)
+ 
L_2^{\alpha\beta} V^{\mu_3} (k_1, k_2, k_3)
+ 
(J_3^{\alpha\beta} )^{\mu_3}_{~~ \nu} 
V^{\nu} (k_1, k_2, k_3)
\label{secondsumscalar}
\ee
is the first-order Lorentz transformation of $\barpsi \psi g$ vertex. 
This vanishes,
as may be seen by explicit computation,
because $V^{\mu_3} (k_1, k_2, k_3)$ is a Lorentz tensor.
The third term on the right-hand side of \eqn{sumnrscalar} is 
\be
\ve_{4\alpha} k_{4\beta} k_{4\gamma} 
(S_3^{\alpha\beta})^{\mu_3}_{~~\nu}  {\partial \over \partial  k_{3\gamma}   }
V^\nu (k_1 , k_2, k_3)
= 
\lambda \left(
k_4^2 \, \ve_{4}^{\mu_3} - \ve_4 \cdot k_4 \, k_{4}^{\mu_3}  \right)
\label{thirdtermscalar}
\ee
which automatically vanishes due to $k_4^2=0$ and $\ve_4 \cdot k_4 =0$.
Again, we emphasize that in proving the vanishing of 
\eqns{secondsumscalar}{thirdtermscalar} we did not use that $k_r$ were on-shell,
nor did we use $\ve_r \cdot k_r=0$, for $r=1, 2, 3$.  
Thus these results remain valid for an off-shell vertex $V^{\mu_3} (k_1, k_2, k_3)$.

\subsection{Kinematic numerators for spin-one fundamentals}

Finally we consider massive spin-one fundamentals
with $\barpsi g  \psi$ vertex of the form 
\be
{ i g \over \sqrt2} ( {T}^{\ta_3} )^{\ti_1}_{~~ \ti_2} 
V^{\mu_1 \mu_2 \mu_3} (k_1, k_2, k_3) 
\label{spinonevertex}
\ee
where $ V^{\mu_1 \mu_2 \mu_3} (k_1, k_2, k_3) $
is given by \eqn{threegluonvertex}.
We emphasize that although we refer to the vector particles
as fundamentals, they could be in any representation, 
including the adjoint, in which case
$( {T}^{\ta_3} )^{\ta_1}_{~~ \ta_2}  = f_{\ta_1 \ta_3 \ta_2}$
and \eqn{spinonevertex} is equal to \eqn{fV},
except that now the vector boson is massive.
The propagator for a massive spin-one particle is 
\be
{-i \delta^\ti_{~\tj} P_{\mu\nu} (k)
\over k^2-m^2 }, 
\qquad\qquad
P_{\mu\nu}(k)  = \eta_{\mu\nu} - { k_\mu k_\nu \over m^2} \,.
\label{spinonepropagator}
\ee
In the case in which the vector boson
gets its mass from a spontaneously-broken symmetry,
\eqn{spinonepropagator} is the propagator in unitary 
gauge;
this is most convenient for tree-level calculations
as we need not compute contributions involving Goldstone bosons.
\para

Attaching gluon 4 to leg 1 yields the expression
\be
 { ig^2 \over 2} 
{  c'_{(1)}  
\over (k_1 + k_4)^2 - m_1^2 }
V^{\mu_1 \mu_4 \nu} (k_1, k_4, -k_1-k_4) 
P_{\nu\lambda} (k_1+k_4) V^{\lambda \mu_2 \mu_3} (k_1 + k_4, k_2, k_3) \,.
\ee
The contribution of this diagram to \eqn{fourfundamplitude} 
is obtained by contracting with $\prod_{a=1}^4 \ve_{a\mu_a} $
and dividing by $i$,
giving 
\be
n'_{(1)}\Big|_{\rm leg} 
~=~  g^2 \ve_{1\mu_1} 
\left[  
   \ve_4 \cdot k_1 \delta^{\mu_1}_{~\nu }
-i \ve_{4\alpha} k_{4\beta} ( S_1^{\alpha\beta} )^{\mu_1}_{~~ \nu} 
 \right]  
V^{\nu \mu_2 \mu_3} (k_1 + k_4, k_2, k_3)
\ve_{2\mu_2} \ve_{3\mu_3} 
\ee
where we have used 
$0= k_4^2 =  \ve_4 \cdot k_4= \ve_1 \cdot k_1$ and $k_1^2 = m_1^2$
but we did not use \eqn{Wardone},
which is not valid in this case because  $k_2^2 \neq k_3^2$.
Analogous expressions are obtained for
$n'_{(2)}$ and $n'_{(3)}$.
Attaching gluon 4 directly to the $\barpsi \psi g$ vertex, 
we obtain the $\barpsi \psi g g $ vertex
\be
-  {i g^2 \over 2} \left( 
 c'_{(1)} {\partial \over \partial  k_{1\mu_4}   }
+c'_{(2)} {\partial \over \partial  k_{2\mu_4}   }
+c'_{(3)} {\partial \over \partial  k_{3\mu_4}   }
\right) 
V^{\mu_1\mu_2\mu_3} (k_1, k_2, k_3)  \,.
\ee
Again, the rest of the story proceeds exactly 
as in sec.~\ref{sec:fourgluon},
allowing us to write the kinematic numerators as 
\begin{align}
\nrp
&~=~
g^2 \left[  
   \ve_4 \cdot k_r 
-i \ve_{4\alpha} k_{4\beta}  J_r^{\alpha\beta} 
-i \ve_{4\alpha} k_{4\beta}  k_{4\gamma} 
S_r^{\alpha\beta} 
{\partial \over \partial  k_{r\gamma} }
 \right]  
V(k_1 , k_2, k_3)
\end{align}
where we have suppressed the polarization vectors 
$\ve_r$ for $r=1, 2, 3$. 
Note that the kinematic numerators $\nrp$ have
exactly the same form as the kinematic 
numerators for the four-gluon amplitude,
even though the masses for particles 1 through 3 can be nonzero.
The proof of the vanishing of the sum of numerators proceeds 
exactly as in sec.~\ref{sec:fourgluon}.
\para

In this section, we have explicitly shown that the 
sum of kinematic numerators for four-point amplitudes 
$ \cA_4 (\barpsi_1, \psi_2, g_3, g_4) $ vanishes
(where $\psi$ can have spin zero, one-half, or one)
and thus have demonstrated the invariance of 
the four-point amplitude under the color-factor symmetry
associated with gluon 4. 
As the results of the last subsection have shown,
this result remains valid even when particles 
1 through 3 are massive;
only the gluon associated with the color-factor symmetry
need be massless.
The results we have derived will be used in the next 
section to prove a more general result.

\section{Proof of color-factor symmetry for more general amplitudes}
\setcounter{equation}{0}
\label{sec:rvefund}

In this section, we use the radiation vertex expansion 
to demonstrate the invariance under the color-factor symmetry
of tree-level gauge-theory amplitudes 
containing at least one gluon together 
with massless or massive particles 
in arbitrary representations of the gauge group
(but referred to as fundamentals for convenience)
and with arbitrary spin $\le 1$.
For concreteness, we focus on the $n$-point amplitude
$\cAn ( \barpsi_1, g_2, g_3, \cdots, g_{n-1}, \psi_n)$ 
with $n-2$ gluons and a pair of fundamentals $\psi$,
but it will be clear that the proof applies to more
general amplitudes.
The proof is very similar to that given in 
sec.~\ref{sec:rvegluon} for $n$-gluon amplitudes, 
and so we only highlight the differences.
\para

The radiation vertex expansion constructs the $n$-point amplitude
$\cAn ( \barpsi_1, g_2, g_3, \cdots, g_{n-1}, \psi_n)$ 
by attaching gluon $a \in \{ 2, \cdots, n-1 \}$ to all possible $(n-1)$-point diagrams $\I$,
with two fundamentals and $n-3$ gluons,
in all possible ways, 
and reorganizing this as a sum over all the vertices of the
$(n-1)$-point  diagram.
We have already shown in sec.~\ref{sec:rvegluon} that the contributions of 
the three- and four-gluon vertices are invariant under the color-factor symmetry,
so we only need to demonstrate the same for 
vertices involving two fundamentals.
We do this separately for fundamentals with spin zero, one-half, and one.

\subsection{Vertices involving spin-one-half fundamentals}

To derive the contribution of the $\barpsi \psi g$ vertices 
to the radiation vertex expansion for spin-one-half fundamentals $\psi$,
we examine the effect of attaching a gluon to a fermion leg,
either external or internal.
\para

First we single out (fermion) leg 1, 
denoting the contribution of an $(n-1)$-point diagram  $\I$ to the amplitude as
$\bar{u} (k_1) C^{\ti_1} (k_1, \cdots)$
where $\cdots$ denotes momenta belonging to $ \{ 2, \cdots, n \} \setminus \{ a \}$.
Attaching gluon $a$ to external fermion leg $1$,
using $\bar{u} (k_1) (-\kslash_1 + m) = 0$, 
and contracting with $\ve_{a\mu_a} $, we obtain
\be
 - {  \sqrt2 g}  
{ (T^{\ta_a})^{\ti_1}_{ ~~\tj}  \over 2 k_a \cdot k_1 }
\bar{u} (k_1) 
\left[   \ve_a \cdot k_1 -i \ve_{a\alpha} k_{a\beta}  \Sigma^{\alpha\beta} \right]
C^{\tj} (k_1+k_a, \cdots )  \,.
\label{attachexternalpsibar}
\ee
Next, we single out (fermion) leg $n$, denoting the contribution of the diagram $\I$ 
to the amplitude as 
$ B_{\ti_n} (k_n, \cdots) u(-k_n)$,
where $\cdots$ denotes momenta belonging to $ \{ 1, \cdots, n-1 \} \setminus \{ a \}$.
Attach gluon $a$ to external fermion leg $n$, 
use  $ (\kslash_n + m) u(-k_n)=0$,
and contract with $\ve_{a\mu_a} $
to obtain
\be
 + {  \sqrt2 g}  
B_{\tj} (k_n + k_a, \cdots) 
{ (T^{\ta_a})^\tj_{~~\ti_n}  \over 2 k_a \cdot k_n  }
\left[ \ve_4 \cdot k_n +i \ve_{4\alpha} k_{4\beta}  \Sigma^{\alpha\beta}  \right] u(-k_n) \,.
\label{attachexternalpsi}
\ee
Now  we single out one of the internal fermion lines of $\I$, 
which divides the diagram into two subdiagrams
$B_{\tj}$ and $C^{\tj} $,
and splits the external legs $\{1, \cdots, n\} \setminus \{ a \}$ into two 
complementary sets $\SaB$ and $\SaC$.
The contribution of the diagram $\I$ can thus be written as 
\be
B_{\tj} ( -K, \cdots) 
{i \delta^\tj_{~\tk} \over \Kslash - m} 
C^{\tk} ( K,  \cdots  ) 
\ee
where $K = \sum_{d \in \SaB} k_d$ 
is the momentum running through the line,
and the $\cdots$ in $B$ and $C$ denote momenta belonging to 
$\SaB$ and $\SaC$ respectively.
Attaching gluon $a$ to the line connecting the two subgraphs
and contracting with  $\ve_{a \mu_a}$, we have
\be
- {i g\over \sqrt2}  
B_{\tj} ( -K, \cdots) 
{1 \over \Kslash - m} 
 (T^{\ta_a})^\tj_{~\tk} \epsaslash
{1 \over \Kslash + \kaslash - m} 
C^{\tk} (K+k_a ,  \cdots  )  \,.
\ee
Now we use the identity\cite{Brown:1982xx} 
\begin{align}
{1 \over \Kslash - m} 
\epsaslash
{1 \over \Kslash + \kaslash - m}  
&~=~
{1 \over \Kslash - m}  
{
\left( \ve_a \cdot K  + {1\over 4} \left[ \epsaslash, \kaslash \right]\right)
 \over   k_a \cdot K} 
~-~ 
{
\left( \ve_a \cdot K  + {1\over 4} \left[ \epsaslash, \kaslash \right] \right)
\over  k_a\cdot K } 
{1 \over \Kslash + \kaslash - m}
\label{fermionpropidentity}
\end{align}
to rewrite this as
\begin{align}
&
B_{\tj} (  -K, \cdots) 
{i \over \Kslash - m}  
\left\{
-\sqrt2 g
{ (T^{\ta_a})^\tj_{~\tk} \over  2 k_a \cdot K} 
\left[ 
\ve_a \cdot K  -i \ve_{a\alpha} k_{a\beta}  \Sigma^{\alpha\beta} 
\right]
C^{\tk} ( K + k_a,  \cdots  ) 
\right\}
\nn\\
&
+ 
\left\{ 
\sqrt2 g
{ (T^{\ta_a})^\tj_{~\tk} \over 2 k_a\cdot K } 
~B_{\tj} (  -K, \cdots) 
\left[ 
\ve_a \cdot K  -i \ve_{a\alpha} k_{a\beta}  \Sigma^{\alpha\beta}
  \right]
\right\}
{i \over \Kslash +\kaslash - m}
C^{\tk} (K+k_a ,  \cdots  )  \,.
\label{attachinternalfermion}
\end{align}
Each term can be associated with one of the two vertices to which the line is attached.
\para

We now choose one of the $\barpsi \psi g$  vertices $v$ of $\I$. 
Such a vertex divides the external legs into three non-overlapping subsets
$S_{ \aIvr }$, $r=1,2,3$ such that
$\bigcup_{r=1}^3 S_{ (a,\I,v,r)} = \{ 1, \cdots, n\} \setminus \{ a\}$.
The contribution of the diagram $\I$ to the $(n-1)$-point amplitude can be expressed as
\begin{align}
{i  g \over \sqrt2} 
A^\aIvrt_{\tc_3 \mu_3} (- K_3, \cdots) 
B_{\tj_1} ( -K_1, \cdots)  
\V^{\mu_3} ( {T}^{\tc_3} )^{\tj_1}_{~~ \tj_2} 
C^{\tj_2} ( -K_2, \cdots ) 
\end{align}
where $K_r = \sum_{d \in S_{ \aIvr }} k_d$
is the momentum flowing out of each leg of the vertex,
and the $\cdots$ in $B$, $C$, and $A$ denote momenta belonging to 
$S_{ (a,\I,v,1)}$, $S_{ (a,\I,v,2)}$, and $S_{ (a,\I,v,3)}$ respectively.
If either fermion leg is external, 
then
$ B_{\tj_1}  = \bar{u}(k_1) \delta^{\ti_1}_{~~ \tj_1}$
or
$C^{\tj_2} = u(-k_n) \delta^{\tj_2}_{~~ \ti_n} $.
If the gluon leg is external, then 
$A^\aIvrt_{\tc_3 \mu_3} $ is  $\delta_{\tb \tc_3} \ve_{b \mu_b} $.
\para

We now attach gluon $a$ to each of the legs of this $\barpsi \psi g $ vertex,
either to an external leg or to an internal line.
Using the expressions above as well as those in sec.~\ref{sec:rvegluon}, 
we obtain 
\begin{align}
i g^2 
&
A^\aIvrt_{\tc_3 \mu_3} (- K_3, \cdots) 
B_{\tj_1} (-K_1,  \cdots ) 
\nn\\
\times 
\bigg( 
& 
- { (  T^{\ta_a} T^{\tc_3} )^{\tj_1}_{~~ \tj_2} \over 2 k_a \cdot K_1 }
\Big[  \ve_a \cdot K_1 - i \ve_{a\alpha} k_{a\beta}  \Sigma^{\alpha\beta} \Big]
\V^{\mu_3} 
+
{ (T^{\tc_3} T^{\ta_a} )^{\tj_1}_{~~ \tj_2} \over 2 k_a \cdot K_2 }
\V^{\mu_3} 
\Big[ \ve_a \cdot K_2 + i \ve_{a\alpha} k_{a\beta}  \Sigma^{\alpha\beta} \Big]
\nn\\
&
+
{ f_{ \ta_a \tc_{3} \tb } (T^{\tb} )^{\tj_1}_{~~ \tj_2} \over 2 k_a \cdot K_3 }
\Big[  \eta^{\mu_3 \nu} \ve_a \cdot K_3
-i \ve_{a\alpha} k_{a\beta}  (S_3^{\alpha\beta} )^{\mu_3\nu} \Big]
\V_{\nu} 
\bigg) 
C^{\tj_2} ( -K_2,  \cdots)   
\label{fermioncontrib}
\end{align}
which is the contribution of the $\barpsi \psi g $ vertex to the radiation vertex expansion.
\para

We now wish to show that \eqn{fermioncontrib}  is invariant under the color-factor symmetry.
As in sec.~\ref{sec:rvegluon}, 
we first assume that the  subdiagrams corresponding to $A$, $B$, and $C$ 
contain no four-gluon vertices.
Designate by $c_\aIvr$ with $r=1, 2, 3$ the color factors
associated with each of the three terms in \eqn{fermioncontrib},
including factors of $f_{\ta \tb \tc}$ and $(T^\ta)^\tj_{~\tk}$ 
in the subdiagrams.
These color factors manifestly satisfy  $\sumr c_\aIvr=0$,
and the variation of $c_\aIvr$ under the color-factor shift associated with gluon $a$ is
\be
\delta_a ~ c_\aIvr   ~=~ \alpha_{(a,\I,v)} ~   k_a \cdot K_r
\label{psicolorfactorshift}
\ee
which preserves $\sumr c_\aIvr=0$.
The variation of \eqn{fermioncontrib} under \eqn{psicolorfactorshift}
is then proportional to 
\be
\left( \sumr \ve_a \cdot K_r \right)  \V^{\mu_3} 
- i \ve_{a\alpha} k_{a\beta}  
\left[ \Sigma^{\alpha\beta} \V^{\mu_3} 
- \V^{\mu_3} \Sigma^{\alpha\beta} 
+ (S_3^{\alpha\beta} )^{\mu_3\nu} \V_{\nu} 
\right] \,.
\label{variationspinhalf}
\ee
The first term vanishes by momentum conservation, 
and the second by the transformation properties of $\gamma_\nu$, 
as we saw in sec.~\ref{sec:fourfund}.
If the subdiagrams $A$, $B$, and $C$ do contain four-gluon vertices, we can expand \eqn{fermioncontrib} into
individual pieces, each of which is invariant under the color-factor symmetry.
Together with the result from sec.~\ref{sec:rvegluon} that the contributions to the
radiation vertex expansion from the three- and four-gluon vertices are also invariant,
we have thus shown that the amplitude 
$\cAn ( \barpsi_1, g_2, g_3, \cdots, g_{n-1}, \psi_n)$ 
with spin-one-half fundamentals is invariant under the color-factor symmetry.
In fact, this proof applies to an amplitude with an 
arbitrary number of pairs of fundamentals, 
and will be used in the sequel \cite{Brown:2016hck}
to prove the BCJ relations \cite{Johansson:2015oia} 
for that class of amplitudes.

\subsection{Vertices involving spin-zero fundamentals}

To derive the contribution of the $\barpsi \psi g$ 
and $\barpsi \psi g g $ vertices to the radiation vertex expansion
for spin-zero fundamentals $\psi$,
we examine the effect of attaching a gluon to a scalar leg,
either external or internal.
\para

First we single out (scalar) leg 1, 
denoting the contribution of an $(n-1)$-point diagram  $\I$ to the amplitude as
$C^{\ti_1}  (k_1, \cdots)$
where $\cdots$ denotes momenta belonging to 
$ \{ 2, \cdots, n \} \setminus \{ a \}$.
Attaching gluon $a$ to external scalar leg $1$
we obtain 
\begin{align}
-  { g \over \sqrt2}  
& 
{ (T^{\ta_a})^{\ti_1}_{ ~~\tj}  \over (k_a + k_1)^2 - m^2 }
V^{\mu_a } (k_1,  -k_1 - k_a, k_a) 
C^{\tj} (k_1+k_a, \cdots )
\nn\\
~=~ 
& - { g \over \sqrt2}  
{ (T^{\ta_a})^{\ti_1}_{ ~~\tj}  \over 2 k_a \cdot k_1 }
\left[   
k_a^{\mu_a} + 2 k_1^{\mu_a} 
 \right]  
C^{\tj} (k_1+k_a, \cdots ) \,.
\end{align}
Contracting this with $\ve_{a\mu_a} $ 
eliminates the first term in the square brackets, leaving 
\be
 - {  \sqrt2 g}  
{ (T^{\ta_a})^{\ti_1}_{ ~~\tj}  \over 2 k_a \cdot k_1 }
   \ve_a \cdot k_1 
C^{\tj} (k_1+k_a, \cdots ) \,.
\ee
Similarly, attaching gluon $a$ to external scalar leg $n$, we obtain
\be
 + {  \sqrt2 g}  
   \ve_a \cdot k_n 
B_{\tj} (k_n + k_a, \cdots) 
{ (T^{\ta_a})^\tj_{~~\ti_n}  \over 2 k_a \cdot k_n  } \,.
\ee
Next we single out one of the internal scalar lines of $\I$, 
which divides the diagram into two subdiagrams
$B_{\tj}$ and $C^{\tj} $,
and splits the external legs into two 
complementary sets $\SaB$ and $\SaC$.
The contribution of the diagram $\I$ can thus be written as 
\be
B_{\tj} ( -K, \cdots) 
{i \delta^\tj_\tk \over K^2-m^2} 
C^{\tk} (  K,  \cdots  ) 
\ee
where $K = \sum_{d \in \SaB} k_d$ 
is the momentum running through the line,
and the $\cdots$ in $B$ and $C$ denote momenta belonging to 
$\SaB$ and $\SaC$ respectively.
Attaching gluon $a$ to the line connecting the two subgraphs yields
\be
- {i g\over \sqrt2}  
B_{\tj} ( -K, \cdots )
{ (T^{\ta_a})^\tj_{~\tk} V^{\mu_a} (K,-K-k_a, k_a) 
\over  [K^2-m^2]  [ (K+k_a)^2-m^2]} 
C^{\tk} ( K+k_a,  \cdots  )  \,.
\ee
Contracting with  $\ve_{a \mu_a}$ and using $\ve_a \cdot k_a=0$, we have
\be
- \sqrt2 i  g
B_{\tj} ( -K, \cdots )
{ (T^{\ta_a})^\tj_{~\tk} \ve_a \cdot K
\over   [K^2-m^2] [(K+k_a)^2-m^2]     }
C^{\tk} ( K+k_a,  \cdots  )  \,.
\ee
Now we use the identity (\ref{massivesplitprop}) to rewrite this as
\begin{align}
&
B_{\tj} ( -K, \cdots )
 {i \over K^2-m^2}
\left\{
-\sqrt2 g
{ (T^{\ta_a})^\tj_{~\tk} \over  2 k_a \cdot K} 
\ve_a \cdot K ~
C^{\tk} (K+k_a ,  \cdots  ) 
\right\}
\nn\\
&+ 
\left\{ 
\sqrt2 g
 B_{\tj} ( -K, \cdots)
{(T^{\ta_a})^\tj_{~\tk} \over 2 k_a\cdot K } 
\ve_a \cdot K 
\right\}
{ i \over (K+k_a)^2 -m^2}
C^{\tk} (K+k_a ,  \cdots  )  \,.
\end{align}
We associate each of the terms in this equation with
one of the two vertices to which the line is attached.
\para

Next we choose one of the scalar-scalar-gluon vertices $v$ of $\I$ (if it has any).
Such a vertex divides the external legs into three non-overlapping subsets
$S_{ \aIvr }$, $r=1,2,3$ such that
$\bigcup_{r=1}^3 S_{ (a,\I,v,r)} = \{ 1, \cdots, n\} \setminus \{ a\}$.
The contribution of the diagram $\I$ can be expressed as
\begin{align}
{i  g \over \sqrt2} 
V^{\mu_3} (K_1, K_2, K_3)
A^\aIvrt_{\tc_3 \mu_3} (- K_3, \cdots) 
B_{\tj_1} ( -K_1, \cdots)  
( {T}^{\tc_3} )^{\tj_1}_{~~ \tj_2} 
C^{\tj_2} ( -K_2, \cdots ) 
\end{align}
where $K_r = \sum_{d \in S_{ \aIvr }} k_d$
is the momentum flowing out of each leg of the vertex,
and the $\cdots$ in $B$, $C$, and $A$ denote momenta belonging to 
$S_{ (a,\I,v,1)}$, $S_{ (a,\I,v,2)}$, and $S_{ (a,\I,v,3)}$ respectively.
If either scalar leg is external, then
$ B_{\tj_1}  = \delta^{\ti_1}_{~~ \tj_1}$
or
$C^{\tj_2} = \delta^{\tj_2}_{~~ \ti_n} $.
If the gluon leg is external, then $A^\aIvrt_{\tc_3 \mu_3} $ is  
$\delta_{\tb \tc_3} \ve_{b \mu_b} $.
\para

We now attach gluon $a$ to each of the legs of this scalar-scalar-gluon vertex,
either to an external leg or to an internal line.
This yields
\begin{align}
i g^2 
&
A^\aIvrt_{\tc_3 \mu_3} (- K_3, \cdots) 
B_{\tj_1} (-K_1,  \cdots ) 
\nn\\
\times 
\bigg( 
& 
- { 
  (  T^{\ta_a} T^{\tc_3} )^{\tj_1}_{~~ \tj_2} 
\over 2 k_a \cdot K_1
}
\Big[  \ve_a \cdot K_1 \Big]
V^{\mu_3} (K_1 + k_a, K_2, K_3)
+
{ 
(T^{\tc_3} T^{\ta_a} )^{\tj_1}_{~~ \tj_2} 
\over 2 k_a \cdot K_2
}
\Big[  \ve_a \cdot K_2
\Big]
V^{\mu_3} (K_1, K_2 + k_a, K_3)
\nn\\
&
+
{ 
f_{ \ta_a \tc_{3} \tb } (T^{\tb} )^{\tj_1}_{~~ \tj_2} 
\over 2 k_a \cdot K_3
}
\Big[  \eta^{\mu_3 \nu} \ve_a \cdot K_3
+  \left( \ve_a^{\mu_3} k_a^\nu - \ve_a^\nu k_a^{\mu_3}  \right)
\Big]
V_{\nu} (K_1, K_2, K_3 + k_a)
\bigg) 
C^{\tj_2} ( -K_2,  \cdots)   \,.
\label{attachscalarleg}
\end{align}
We can also attach gluon $a$ directly to the scalar-scalar-gluon vertex itself.
Using \eqn{scalarscalargluongluonvertex}, this yields 
\begin{align}
-{i g^2  \over 2}
& 
A^\aIvrt_{\tc_3 \mu_3} (- K_3,  \cdots) 
B_{\tj_1} ( -K_1, \cdots ) 
\nn\\
\times 
\bigg( 
&
- (T^{\ta_a} T^{\tc_3} )^{\tj_1}_{~~ \tj_2} 
\ve_{a\mu_a}   {\partial \over \partial  K_{1\mu_a}    }
V^{\mu_3} (K_1, K_2, K_3) 
+
  (T^{\tc_3} T^{\ta_a} )^{\tj_1}_{~~ \tj_2} 
\ve_{a\mu_a}   {\partial \over \partial  K_{2\mu_a}   }
V^{\mu_3} (K_1, K_2, K_3) 
\nn\\
&
+
f_{ \ta_a \tc_{3} \tb } (T^{\tb} )^{\tj_1}_{~~ \tj_2} 
\ve_{a\mu_a}   {\partial \over \partial  K_{3\mu_a}   }
V^{\mu_3} (K_1, K_2, K_3) 
\bigg)
C^{\tj_2} ( -K_2,  \cdots)   \,.
\label{attachscalarvertex}
\end{align}
We now use \eqn{taylor} in \eqn{attachscalarleg},
and combine \eqns{attachscalarleg}{attachscalarvertex}
as we did in sec.~\ref{sec:fourfund}. 
Leaving the $\mu_3$ index implicit, 
we obtain the contribution of the 
scalar-scalar-gluon vertex to the radiation vertex expansion
\begin{align}
i g^2 
& 
A^\aIvrt_{\tc_3} (- K_3,   \cdots) 
B_{\tj_1} ( -K_1, \cdots ) 
\nn\\
\times 
\bigg( 
& 
- { (  T^{\ta_a} T^{\tc_3} )^{\tj_1}_{~~ \tj_2} \over 2 k_a \cdot K_1 }
\left[  \ve_a \cdot K_1 
-i \ve_{a\alpha} k_{a\beta}  L_1^{\alpha\beta} 
\right]
V(K_1, K_2, K_3)
\label{threescalarcontrib}
\\
&
+
{ (T^{\tc_3} T^{\ta_a} )^{\tj_1}_{~~ \tj_2} \over 2 k_a \cdot K_2 }
\left[  \ve_a \cdot K_2 
-i \ve_{a\alpha} k_{a\beta}  L_2^{\alpha\beta} 
\right]
V(K_1, K_2, K_3)
\nn\\
&
+
{ f_{ \ta_a \tc_{3} \tb } (T^{\tb} )^{\tj_1}_{~~ \tj_2} \over 2 k_a \cdot K_3 }
\left[  \ve_a \cdot K_3 
-i \ve_{a\alpha} k_{a\beta}  J_3^{\alpha\beta} 
-i \ve_{a\alpha} k_{a\beta}  k_{a\gamma} S_3^{\alpha\beta} 
{\partial \over \partial  K_{3\gamma} }
\right]
V(K_1, K_2, K_3)
\bigg) 
C^{\tj_2} ( -K_2,  \cdots)  
\nn
\end{align}
where $J_3$ and $S_3$ act on the $\mu_3$ index of $ V (K_1, K_2, K_3) $ .
\para

At the end of the last subsection, we discussed 
the color-factor symmetry acting on the $\barpsi \psi g$ vertex contribution
to the radiation vertex expansion.
Under \eqn{psicolorfactorshift}, the variation of \eqn{threescalarcontrib} is 
proportional to 
\be
\left[ \left(  \sumr  \ve_a \cdot K_r  \right)
-i \ve_{a\alpha} k_{a\beta}  
\left( \sumr J_r^{\alpha\beta}  \right)
-i \ve_{a\alpha} 
k_{a\beta}  
k_{a\gamma}
S_3^{\alpha\beta} {\partial \over \partial  K_{3\gamma} } 
\right]
V(K_1, K_2, K_3) \,.
\label{threesumsfund}
\ee
In sec.~\ref{sec:fourfund}, 
we demonstrated that each of the three terms 
in \eqn{threesumsfund} vanishes.
Therefore the contributions of the $\barpsi \psi g$ vertices 
(\ref{threescalarcontrib})
to the radiation vertex expansion are invariant under the color-factor shift
associated with gluon $a$.
\para

Finally we choose one of the scalar-scalar-gluon-gluon vertices $v$ of $\I$ (if it has any).
Such a vertex divides the external legs into four non-overlapping subsets
$S_{ \aIvr}$, $r=1, \cdots, 4 $ such that
$\bigcup_{r=1}^4 S_{ (a,\I,v,r)} = \{ 1, \cdots, n\} \setminus \{ a\}$.
Using \eqn{ssgg}, the contribution of diagram $\I$ can be expressed as
\begin{align}
&
{i g^2 \over 2} \eta^{\mu_3 \mu_4} 
A^\aIvrt_{\tc_3 \mu_3} (- K_3,  \cdots) 
A^\aIvrf_{\tc_4 \mu_4} (- K_4, \cdots) 
B_{\tj_1} ( -K_1, \cdots ) 
\left( \{ {T}^{\tc_3},  {T}^{\tc_4} \} \right)^{\tj_1}_{~~ \tj_2} 
C^{\tj_2} ( -K_2,  \cdots)   \,.
\end{align} 
We now attach gluon $a$ to each of the legs of this scalar-scalar-gluon-gluon vertex,
either to an external leg or to an internal line.
This yields
\begin{align}
 {i g^3 \over \sqrt2} 
&
A^\aIvrt_{\tc_3 \mu_3} (- K_3,  \cdots) 
A^\aIvrf_{\tc_4 \mu_4} (- K_4, \cdots) 
B_{\tj_1} ( -K_1, \cdots ) 
\nn\\
\times \bigg( 
&
- { (T^{\ta_a} \{ T^{\tc_3} , T^{ \tc_4} \} )^{\tj_1}_{~~ \tj_2} \over 2 k_a \cdot K_1 }
\Big[
\ve_a \cdot K_1
\Big]
\eta^{\mu_3 \mu_4} 
+
{ ( \{ T^{\tc_3} , T^{ \tc_4} \} T^{\ta_a} )^{\tj_1}_{~~ \tj_2} \over 2 k_a \cdot K_2 }
\Big[
\ve_a \cdot K_2
\Big]
\eta^{\mu_3 \mu_4} 
\nn\\
&+
{ f_{ \ta_a  \tc_{3} \tb } (\{ T^{\tb} , T^{ \tc_4} \} )^{\tj_1}_{~~ \tj_2} 
\over 2 k_a \cdot K_3 }
\Big[
\delta^{\mu_3 }_{~\nu} \ve_a \cdot K_3
-i \ve_{a\alpha} k_{a\beta} ( S_3^{\alpha\beta} )^{\mu_3 }_{~~\nu} 
\Big]
\eta^{\nu \mu_4} 
\nn\\
&+
{ f_{ \ta_a  \tc_{4} \tb } (\{ T^{\tc_3} , T^{ \tb} \} )^{\tj_1}_{~~ \tj_2} 
\over 2 k_a \cdot K_4 }
\Big[
\delta^{\mu_4 }_{~\nu} \ve_a \cdot K_4
-i \ve_{a\alpha} k_{a\beta} ( S_4^{\alpha\beta} )^{\mu_4 }_{~~\nu} 
\Big]
\eta^{\mu_3 \nu } 
\bigg)
C^{\tj_2} ( -K_2,  \cdots)  \,.
\label{fourscalarcontrib}
\end{align}
We now need to consider the variation under the color-factor symmetry of this
contribution to the radiation vertex expansion.
Designate by $c_\aIvr$ with $r=1, \cdots 4$  the color factor 
associated with each of the terms in \eqn{fourscalarcontrib},
including the factors of $f_{\ta \tb \tc}$ and $(T^\ta)^\tj_{~\tk}$ 
in the subdiagrams.
These color factors satisfy
$\sumrr c_\aIvr=0$ by virtue of 
\begin{align} 
- (T^{\ta_a} \{ T^{\tc_3} , T^{ \tc_4} \} )^{\tj_1}_{~~ \tj_2} 
+ ( \{ T^{\tc_3} , T^{ \tc_4} \} T^{\ta_a} )^{\tj_1}_{~~ \tj_2} 
+ f_{ \ta_a  \tc_{3} \tb } (\{ T^{\tb} , T^{ \tc_4} \} )^{\tj_1}_{~~ \tj_2} 
+ f_{ \ta_a  \tc_{4} \tb } (\{ T^{\tc_3} , T^{ \tb} \} )^{\tj_1}_{~~ \tj_2} 
=0 \,.
\end{align}
The variation of $c_\aIvr$ under the color-factor shift associated with gluon $a$ is
\be
\delta_a ~ c_\aIvr   ~=~ \alpha_{(a,\I,v)} ~   k_a \cdot K_r \,.
\label{scalarshift}
\ee
The variation of \eqn{fourscalarcontrib} under \eqn{scalarshift}
is therefore proportional to 
\be
\left(  \sumrr  \ve_a \cdot K_r  \right) \eta^{\mu_3 \mu_4} 
-i \ve_{a\alpha} k_{a\beta} \left[ 
( S_3^{\alpha\beta} )^{\mu_3}_{~~ \nu} \eta^{\nu \mu_4} 
+( S_4^{\alpha\beta} )^{\mu_4}_{~~ \nu} \eta^{\mu_3 \nu} 
\right] \,.
\label{variationquarticspinzero}
\ee
The first term vanishes by momentum conservation, $k_a + \sumrr K_r = 0 $,  
and $\ve_a \cdot k_a = 0$.
The second term is the first-order Lorentz transformation of the tensor
$\eta^{\mu_3 \mu_4} $, which vanishes.  
\para

Thus each vertex involving scalars that contributes to 
the radiation vertex expansion 
is invariant under the color-factor shift associated with gluon $a$.
Together with the result from sec.~\ref{sec:rvegluon} that the contributions 
from the three- and four-gluon vertices 
to the radiation vertex expansion are also invariant,
we have shown that the full amplitude 
$\cAn ( \barpsi_1, g_2, g_3, \cdots, g_{n-1}, \psi_n)$ 
with spin-zero fundamentals is invariant under the color-factor shift.
In fact, any amplitude built with 
$\barpsi \psi g$ and $\barpsi \psi g g$ vertices
for scalar $\psi$ will have the color-factor symmetry.

\subsection{Vertices involving spin-one fundamentals}

To derive the contribution of the $\barpsi \psi g$ 
and $\barpsi \psi g g $ vertices to the radiation vertex expansion
for spin-one fundamentals $\psi$,
we examine the effect of attaching a gluon 
to a massive vector leg,
either external or internal.
\para

First we single out (vector) leg 1, 
denoting the contribution of an $(n-1)$-point diagram  $\I$ to the amplitude as
$\ve_{1 \mu_1} C^{\ti_1\mu_1}  (k_1, \cdots)$
where $ \cdots$ denotes momenta belonging to 
$ \{ 2, \cdots, n \} \setminus \{ a \}$.
Attaching gluon $a$ to external vector leg $1$
we obtain 
\be
-  { g \over \sqrt2}  
{ (T^{\ta_a})^{\ti_1}_{ ~~\tj}  \over (k_a + k_1)^2 - m_1^2 }
V^{\mu_1 \nu \mu_a} (k_1,  -k_1 - k_a, k_a) 
P_{\nu\lambda} (k_1+k_a)  C^{\tj\lambda} (k_1+k_a, \cdots ) \,.
\ee
Contracting with 
$\ve_{1\mu_1} \ve_{a\mu_a} $,
we obtain 
\be
 - {  \sqrt2 g}  
{ (T^{\ta_a})^{\ti_1}_{ ~~\tj}  \over 2 k_a \cdot k_1 }
\ve_{1\mu_1} 
\Big[  
   \ve_a \cdot k_1 \eta^{\mu_1 \nu} 
+ \left( \ve_a^{\mu_1} k_a^\nu - \ve_a^{\nu} k_a^{\mu_1}  \right)
 \Big]  
C^{\tj}_\nu  (k_1 + k_a, \cdots  ) 
\label{attachexternalmassivevector}
\ee
where we used $k_4^2 =  \ve_4 \cdot k_4= \ve_1 \cdot k_1=0$ 
and $k_1^2 = m_1^2$.
We did {\it not} use the vanishing of \eqn{Wardtwo}.
Similar expressions result from attaching gluon $a$ 
to the other legs. 
By comparing \eqns{attachexternalgluon}{attachexternalmassivevector},
we observe that the expression is the same for a massless
or a massive vector particle.
\para

Next we single out one of the internal lines of $\I$, 
which divides the diagram into two subdiagrams $B$ and $C$,
and splits the external legs $\{ 1, \cdots, n\} \setminus \{ a \}$
into two complementary sets $\SaB$ and $\SaC$.
The contribution of the diagram can thus be written as 
\be
B_{\tj}^\mu ( -K, \cdots ) 
{(-i \delta^{\tj}_{\tk} )  P_{\mu\nu}(K)  \over K^2 - m^2 }
C^{\tk\nu} ( K,  \cdots  ) 
\ee
where $K = \sum_{d \in \SaB} k_d$ 
is the momentum running through the line,
and the $\cdots$ in $B$ and $C$ denote momenta belonging to 
$\SaB$ and $\SaC$ respectively.
Attaching gluon $a$ to the line connecting the two subgraphs
and contracting with $\ve_{a\mu_a} $, we obtain
\be
- {i g\over \sqrt2}  
B_{\tj}^\mu ( -K, \cdots )
{ (T^{\ta_a})^\tj_{~\tk} 
P_{\mu\lambda} (K) 
\ve_{a\mu_a} V^{\lambda\kappa\mu_a} (K,-K-k_a, k_a) 
P_{\kappa\nu} (K+k_a) \over  [K^2-m^2]  [ (K+k_a)^2-m^2]} 
C^{\tk\nu} ( K+k_a,  \cdots  )  \,.
\ee

Now we use the identity  \cite{Brown:1982xx}
\begin{align}
&
{ P_{\mu\lambda} (K) 
\ve_{a\mu_a} V^{\lambda\kappa\mu_a} (K,-K-k_a, k_a)  
P_{\kappa\nu} (K+k_a) 
\over  
[K^2-m^2]  [ (K+k_a)^2-m^2] }
\label{vectorpropidentity}
\\
&=
{1 \over k_a \cdot K }
\left\{ 
{- 
P_{\mu\lambda} (K) 
\left[  \delta^\lambda_{~\nu} \ve_a \cdot K
+  \ve_a^\lambda k_{a\nu} - k_a^{\lambda} \ve_{a\nu} \right]
\over   K^2-m^2 } 
+
{ \left[  \delta_\mu^{~\kappa} \ve_a \cdot K
+  \ve_{a\mu} k_a^\kappa - k_{a\mu} \ve_a^\kappa \right] 
P_{\kappa\nu} (K+k_a) 
\over (K+k_a)^2 -m^2}
\right\}
\nn
\end{align}
to rewrite this as 
\begin{align}
&
B_\tj^\mu ( -K, \cdots ) 
{-i P_{\mu\lambda} (K) \over K^2-m^2} 
\left\{  -\sqrt2 g
{ (T^{\ta_a})^\tj_{~\tk} \over  2 k_a \cdot K} 
\left[ 
 \eta^{\lambda \nu} \ve_a \cdot K
+  \left( \ve_a^\lambda k_a^\nu - \ve_a^\nu k_a^{\lambda}   \right)
\right]
C^{\tk}_{\nu} ( K+k_a,  \cdots  ) 
\right\}
\\
&
+
\left\{ 
  \sqrt2 g
B_{\tj \mu} (\cdots ,  -K)  
{ (T^{\ta_a})^\tj_{~\tk} \over  2 k_a \cdot K} 
\Big[  \eta^{\mu \kappa} \ve_a \cdot K
+  \left( \ve_a^\mu k_a^\kappa - \ve_a^\kappa k_a^{\mu}  \right)
\Big]
\right \}{-i P_{\kappa\nu} (K+k_a) \over (K+k_a)^2 -m^2} 
 C^{\tk\nu} ( K+k_a,  \cdots  )  \,.
\nn
\end{align}
We did {\it not} use the vanishing of \eqn{Wardthree}.
We associate each of the terms in curly brackets 
with one of the two vertices to which the line is attached.
\para

For the rest of the discussion, we can be brief.
The expressions for the contributions 
to the radiation vertex expansion 
from $\barpsi \psi g$ and $\barpsi \psi g g$ vertices 
are similar to those for spin-zero fundamentals,
except that we must include $J_r$ and $S_r$ terms for $r=1$ and 2.
The proof that these vertex contributions are invariant under
the color-factor symmetry relies on some of the results from
sec.~\ref{sec:rvegluon}.
\para

Thus we have shown that the full amplitude 
$\cAn ( \barpsi_1, g_2, g_3, \cdots, g_{n-1}, \psi_n)$ 
with massive spin-one fundamentals 
is invariant under the color-factor shift.
In fact, any amplitude built with 
$\barpsi \psi g$ and $\barpsi \psi g g$ vertices
with a massive vector particle $\psi$ will have the color-factor symmetry. 
\para

We can go even further and state that any amplitude 
built from $ggg$ and $\barpsi g \psi$ vertices
(with $\psi$ having arbitrary spin $\le 1$)
where not only $\psi$ but also some of the gluons are massive
(\ie through spontaneous symmetry breaking) will be invariant
under the color-factor shift.
The only particle in the amplitude
that {\it must} be massless 
is the gluon associated with the color-factor symmetry.

\section{Null eigenvectors of the propagator matrix}
\setcounter{equation}{0}
\label{sec:prop}

The symmetry that we have introduced in this paper 
is possessed not only by gauge-theory amplitudes 
but also by the amplitudes of the much simpler theory \cite{Cachazo:2013iea}
of massless scalars $ \phi^{\ta \ta'} $
transforming in the adjoint of 
the color group $U(N) \times U(\tilde{N})$.
These bi-adjoint scalars have only cubic interactions of the form
\be
f_{\ta\tb\tc} \tilde f_{\ta' \tb' \tc'} \phi^{\ta \ta'} \phi^{\tb \tb'} \phi^{\tc \tc'} 
\ee
where 
$f_{\ta\tb\tc} $
and 
$\tilde f_{\ta' \tb' \tc'}$
are the structure constants of $U(N)$ and $U(\tilde{N})$.
The tree-level $n$-point amplitude
is given by the sum over cubic diagrams
\be
\cAbi_n
~=~ \sum_i 
{c_i  {\tilde c}_i \over d_i } \,.
\label{biadjointamp}
\ee 
Using \eqn{dependent},
the bi-adjoint scalar tree amplitude (\ref{biadjointamp}) 
can be written as
\be
\cAbi_n 
~=~  \sum_{\gamma \in S_{n-2}}  \sum_{\delta \in S_{n-2}}  
{\bc}_{1 \gamma n } ~m( 1 \gamma n| 1 \delta n) ~\tilde{\bc}_{1 \delta n }  
\label{biadjointprop}
\ee
where
\be 
m( 1 \gamma n| 1 \delta n) 
~=~ \sum_i 
{M_{i,1\gamma n} M_{i, 1\delta n} \over d_i} 
\label{propagator}
\ee
are double-partial amplitudes of the bi-adjoint scalar theory \cite{Cachazo:2013iea}.
The $m( 1 \gamma n| 1 \delta n)$
are also the entries of $(n-2)! \times (n-2)!$ propagator matrix
defined in ref.~\cite{Vaman:2010ez}. 
Vaman and Yao argued that the propagator matrix has rank $(n-3)!$ by virtue
of momentum conservation, using explicit low $n$ examples.
Cachazo, He, and Yuan
confirmed this for general $n$ by expressing 
the double-partial amplitudes
as a sum over the  $(n-3)!$ solutions of the scattering equations \cite{Cachazo:2013iea}.
\para

The cubic vertex expansion of the $n$-point bi-adjoint scalar amplitude 
with respect to external scalar $a$ (see sec.~\ref{sec:cfs}) is given by
\be
\cAbi_n ~=~ 
\sum_I  
\sum_{v \in V_{(a,I)}}  {1 \over \prod_{s=1}^3 d_{(a,I,v,s)}  }
\sumr 
{ c_{(a,I,v,r)} {\tilde c}_{(a,I,v,r)} \over 2 k_a \cdot K_{(a,I,v,r)}   }
\label{biadjointvertexexpansion}
\ee
where
\be 
\sumr c_{(a,I,v,r)} ~=~ 0  \,, 
\qquad\qquad
\sumr {\tilde c}_{(a,I,v,r)} ~=~ 0  \,.
\label{biadjointjacobi}
\ee 
The color-factor shift with respect to massless scalar $a$ acts on the color factors 
appearing in \eqn{biadjointvertexexpansion}  as
\be
\delta_a ~ c_{(a,I,v,r)} ~=~   \alpha_{(a,I,v)} ~k_a \cdot K_{(a,I,v,r)} \,.
\ee
The variation of \eqn{biadjointvertexexpansion}
under this shift 
\be
\delta_a ~ \cAbi_{n}  ~ = ~
{1\over 2} \sum_I  
\sum_{v \in V_{(a,I)}}  
{\alpha_{(a,I,v)}  \over \prod_{s=1}^3 d_{(a,I,v,s)}  }
\sumr 
{\tilde c}_{(a,I,v,r)} 
\ee
vanishes by virtue of \eqn{biadjointjacobi},
thus establishing that the 
amplitudes of the bi-adjoint scalar theory 
possess the color-factor symmetry.
\para

Now we consider the variation of the amplitude (\ref{biadjointprop})
under the shift (\ref{halfladdershift})
associated with $a=2$,
\begin{align}
&\deltatwo  \cAbi_n \\
&= \sum_{\sigma  \in S_{n-3} } \alpha_{2,\sigma}
\sum_{b=3}^n \left( k_1 \cdot k_2 + \sum_{c=3}^{b-1} k_2 \cdot k_{\sigma(c)}  \right) 
m(1, \sigma(3), \cdots, \sigma(b-1), 2, \sigma(b), \cdots, \sigma(n-1), n | 1 \delta n)
) ~\tilde{\bc}_{1 \delta n }  \,. \nn
\end{align}
Using the invariance of $\cAbi_n$, 
together with independence of $\alpha_{2,\sigma}$ and $\tilde{\bc}_{1 \delta n }$,
we obtain 
\be
\sum_{b=3}^n \left( k_1 \cdot k_2 + \sum_{c=3}^{b-1} k_2 \cdot k_{\sigma(c)}  \right) 
m(1, \sigma(3), \cdots, \sigma(b-1), 2, \sigma(b), \cdots, \sigma(n-1), n | 1 \delta n)
~=~ 0
\label{propnullvectors}
\ee
i.e., we have derived a set of $(n-3)!$ 
null eigenvectors of the propagator matrix.
Other sets of null eigenvectors are obtained from the color-factor shifts 
associated the other massless scalars in the amplitude.
\para

Since the $(n-2)! \times (n-2)!$ propagator matrix
is known \cite{Cachazo:2013iea}  to have rank $(n-3)!$, 
at most $(n-3) (n-3)!$ of these null eigenvectors can be independent.
Thus, the color-factor symmetry associated with $n-3$ 
of the massless scalars suffices to guarantee the reduced
rank of the propagator matrix.
This nicely explains a result found by one of the present authors
in ref.~\cite{Naculich:2015zha}, \viz 
that the propagator matrix persists in having rank $(n-3)!$ 
even when up to three of the external particles in the amplitude are massive,
but if more than three particles are massive, 
the rank of the propagator matrix is greater.
We now see that for $m \ge 3$ massive and $n-m$ massless particles,
the number of null eigenvectors generated by the color-factor
symmetry will be $(n-m) (n-3)!$ , and therefore the rank of the
propagator matrix will be $(m-2) (n-3)!$  for $m=3$ through $m=n$.
\para

Returning now to Yang-Mills theory, 
if we assume that the numerators of the $n$-gluon amplitude obey
color-kinematic duality, 
i.e. they obey the same Jacobi relations as $c_i$, 
then they can similarly be expressed in terms of 
$(n-2)!$ half-ladder numerators ${\bf n}_{1 \gamma n }$ via
 \be
n_i ~=~ \sum_{\gamma \in S_{n-2}}  M_{i, 1\gamma n} 
~{\bf n}_{1 \gamma n } \,.
\label{nMn}
\ee
Using \eqns{colorordered}{nMn}, 
the color-ordered amplitudes in the Kleiss-Kuijf basis
can be written in terms of the propagator matrix as 
\be 
A(1, \gamma(2), \cdots, \gamma(n-1), n)  
~=~  \sum_{\delta  \,\in\, S_{n-2}} 
m( 1 \gamma n| 1 \delta n) ~{\bf n}_{1 \delta n }    \,.
\label{Amn}
\ee
Thus the null eigenvectors of the propagator matrix (\ref{propnullvectors})
imply that the color-ordered $n$-gluon amplitudes obey the BCJ relations (\ref{fundbcj}).
\para

As we saw in sec.~\ref{sec:cfs}, however,
it is not necessary to require color-kinematic duality in order
to prove the BCJ relations.
The BCJ relations follow from the weaker constraint (\ref{sumoverdelta}),
and both \eqn{sumoverdelta} and the BCJ relations are a consequence of the color-factor
symmetry of the amplitude, which is established through the radiation vertex expansion.

\section{Loop-level amplitudes}
\setcounter{equation}{0}
\label{sec:loop}

Given the connection established in this paper between the
color-factor symmetry of tree-level gauge-theory amplitudes 
and color-kinematic duality/BCJ relations, 
it is naturally of great interest to see whether these ideas can 
be extended to loop level.
In this section, we generalize the 
cubic vertex expansion introduced in sec.~\ref{sec:cfs} to loop-level amplitudes.
We then define a set of shifts of one-loop color factors that depend 
on the loop momentum as well as the momenta of external particles,
and ask whether the cubic vertex expansion of the 
one-loop amplitude is invariant under these shifts.
\para

It is obvious from the variation of the cubic vertex expansion (\ref{variation})
that the tree-level amplitude will be invariant under a color-factor shift
if the numerators satisfy color-kinematic duality.
The reader may have noticed, however, 
that up until now we have taken great pains 
not to invoke color-kinematic duality to prove color-factor symmetry.
We have chosen rather to show that it follows directly
from a Lagrangian approach.
At loop level, we no longer have that luxury, at least at the current stage
of development.
Instead we will explicitly invoke loop-level color-kinematic duality 
(for the theories in which it has been shown to hold
\cite{Bern:2008qj,Bern:2010ue,Carrasco:2011mn,Bern:2011rj,Bern:2012uf,Boels:2012ew,Carrasco:2012ca,Yuan:2012rg,Bjerrum-Bohr:2013iza,Bern:2013uka,Bern:2014sna,Bern:2013yya,Nohle:2013bfa})
in order to demonstrate the color-factor symmetry of the one-loop amplitude.
We then show that the invariance of the amplitude
under color-factor shifts
implies a set of relations among the integrands of 
its color-ordered amplitudes.

\subsection{Cubic vertex expansion for loop-level amplitudes}

\begin{figure}
\begin{center}
\includegraphics[width=5.0cm]{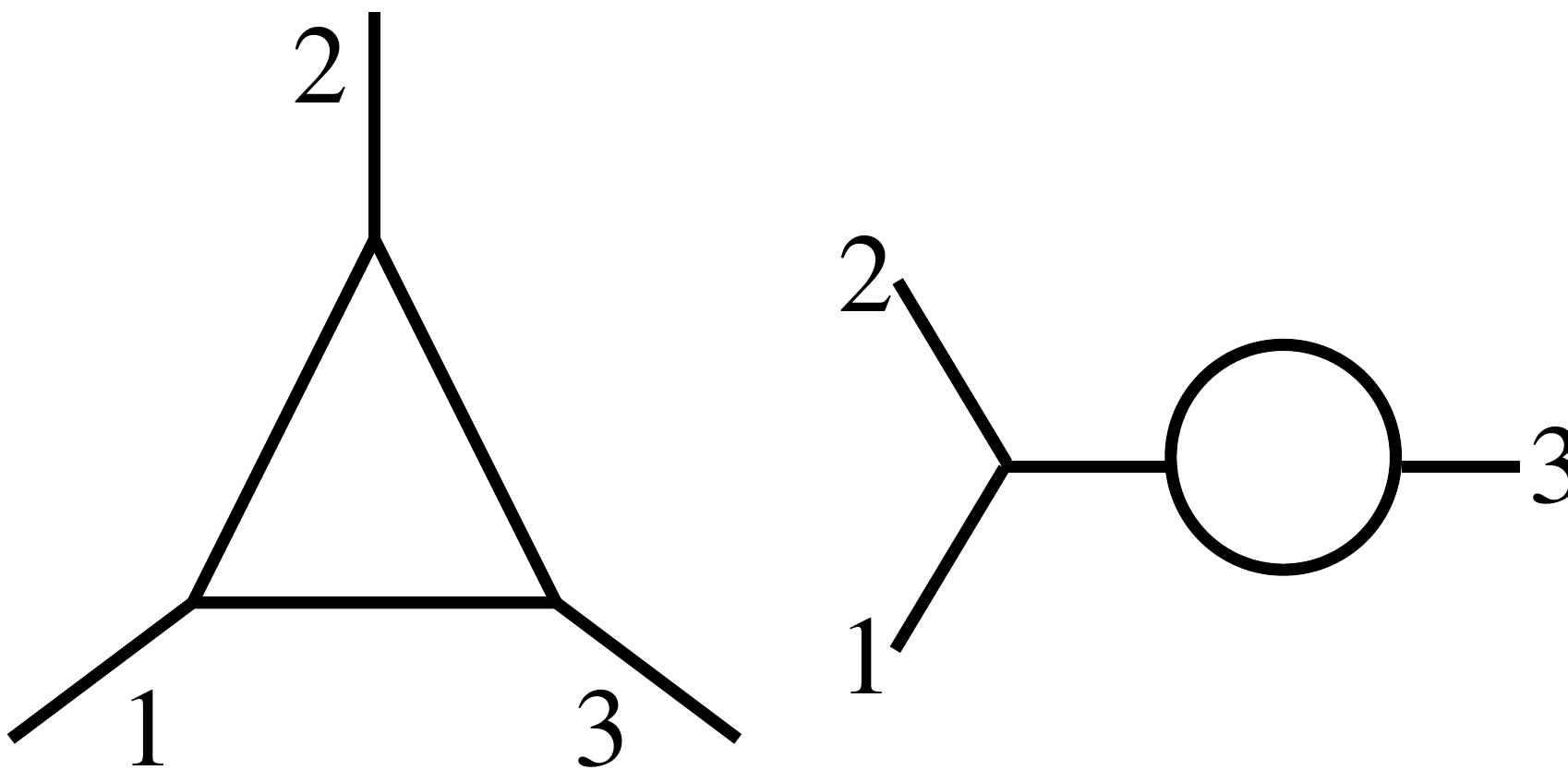}
\caption{Some of the diagrams to which a gluon is attached
to obtain the one-loop four-point cubic decomposition.}
\label{fig:three}
\end{center}
\end{figure}

\begin{figure}[b]
\begin{center}
\includegraphics[width=16.0cm]{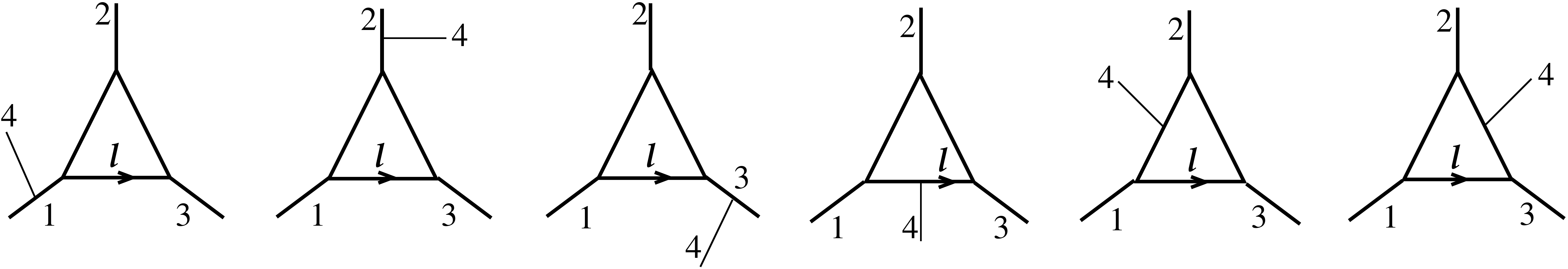}
\caption{Diagrams ${[14]23}$, ${1[24]3}$, ${12[34]}$, ${1234}$, ${1423}$, and ${1243}$.}
\label{fig:four}
\end{center}
\end{figure}

To construct the cubic vertex expansion of an $L$-loop $n$-gluon amplitude
with respect to gluon $a$,
we begin with the set of $L$-loop $(n-1)$-point cubic diagrams $I$
with external legs $\{ 1, \cdots, n \}  \setminus \{ a \}$.
For example, for the one-loop four-gluon amplitude, 
two of the three-point diagrams are shown in fig.~\ref{fig:three};
the rest are obtained from relabelings of the external legs.
Then we attach gluon $a$ in all possible ways, 
either to the external legs or to the internal lines of $I$.
For example, attaching gluon 4 to the triangle diagram in fig.~\ref{fig:three}
in all possible ways, we obtain the diagrams in fig.~\ref{fig:four}.
The diagrams in fig.~\ref{fig:four} 
correspond to the following terms in the cubic decomposition 
of the one-loop four-point amplitude \cite{Bern:2008qj,Bern:2010ue}
\begin{align}
\cA_{4,{\rm tri}}^{(1)}
=
\int {d^D \ell \over (2\pi)^D} 
&
\bigg[  
   { c_{[14]23} \,\, n_{[14]23} \over d_{[14]23} }
 ~+~ { c_{1[24]3} \,\, n_{1[24]3} \over d_{1[24]3} }
 ~+~ { c_{12[34]} \,\, n_{12[34]} \over d_{12[34]} }
\nn\\
&
 ~+~ { c_{1234} \,\, n_{1234} \over d_{1234} }
 ~+~ { c_{1423} \,\, n_{1423} \over d_{1423} }
 ~+~ { c_{1243} \,\, n_{1243} \over d_{1243} }
\bigg]
\label{looptrianglecontrib}
\end{align}
where the denominators in \eqn{looptrianglecontrib} are the products
of inverse propagators associated with the diagrams.
(There could be different sets of denominators 
depending on the mass of the particle circulating in the loop.)
Explicit definitions of the color factors are given below in \eqn{loopcolorfactors}.
The terms obtained by attaching the gluon to an internal line
 --- in this case, the last three terms of \eqn{looptrianglecontrib} ---
are split into two by applying the identity (\ref{splitprop}) 
or,  in the case of massive internal lines, \eqn{massivesplitprop}.
The terms are then reorganized into a sum over the vertices of $I$.
For example, the terms in \eqn{looptrianglecontrib} are reorganized into
\begin{align}
\cA_{4,{\rm tri}}^{(1)}
&=
\int {d^D \ell \over (2\pi)^D}
 \Bigg\{ {1 \over \ell^2 (\ell-k_2 -k_3)^2 (\ell-k_3)^2}
\left[  
  { c_{[14]23} \,\, n_{[14]23} \over 2 k_4 \cdot k_1}
~+~ { c_{1234} \,\, n_{1234} \over 2 k_4 \cdot \ell}
~-~ { c_{1423} \,\, n_{1423} \over 2 k_4 \cdot (\ell+k_1)}
\right]
\nn\\
& +{1 \over \ell^2 (\ell+k_1)^2 (\ell-k_3)^2}
\left[  
  { c_{1[24]3} \,\, n_{1[24]3} \over 2 k_4 \cdot k_2}
~+~ { c_{1423} \,\, n_{1423} \over 2 k_4 \cdot (\ell+k_1)}
~-~ { c_{1243} \,\, n_{1243} \over 2 k_4 \cdot (\ell+k_1+k_2)}
\right]
\label{loopfourpoint}
\\
& +{1 \over \ell^2 (\ell+k_1)^2 (\ell+k_1 +k_2)^2 }
\left[  
  { c_{12[34]} \,\, n_{12[34]} \over 2 k_4 \cdot k_3}
~+~ { c_{1243} \,\, n_{1243} \over 2 k_4 \cdot (\ell+k_1+k_2)}
~-~ { c_{1234} \,\, n_{1234} \over 2 k_4 \cdot (\ell+k_1+k_2+k_3)}
\right]
\Bigg\}
\nn
\end{align}
where $\ell$ in the last term of the last line of \eqn{loopfourpoint}
differs from the label in fig.~\ref{fig:four}  by a shift.\footnote{This
points up an inherent ambiguity in defining a common loop momentum 
when adding together different loop-level diagrams, as in 
\eqns{looptrianglecontrib}{loopfourpoint}.  This requires further study.
For now, we simply note that there exists a choice of loop momentum
for each diagram such that (part of) the one-loop amplitude has the form 
(\ref{loopfourpoint}).
With this choice, the numerator shifts defined in \eqn{loopkinematicshift}
correspond to a generalized gauge transformation.
\label{loopfootnote}}
If the particle circulating in the loop has mass $m$,
the expressions $(\ell + \cdots)^2$ in \eqn{loopfourpoint}
(\ie those outside the square brackets)
are all replaced by $(\ell + \cdots)^2-m^2$. 
We hasten to remind the reader that 
$ \cA_{4,{\rm tri}}^{(1)}$
is only one part of the one-loop four-gluon amplitude;
similar expressions are obtained by attaching gluon 4 
to the other one-loop three-point diagrams $I$. 
\para

Observe that if we apply a shift to the numerators
\begin{align}
\delta_4 n_{[14]23} &~=~ \beta k_4 \cdot k_1, &
\delta_4 n_{1[24]3} &~=~ \beta k_4 \cdot k_2, &
\delta_4 n_{12[34]} &~=~ \beta k_4 \cdot k_3,
\nn
\\
\delta_4 n_{1234}   &~=~ \beta k_4 \cdot \ell , &
\qquad 
\delta_4 n_{1423}    &~=~ \beta k_4 \cdot (\ell + k_1),&
\qquad
\delta_4 n_{1243}    &~=~ \beta k_4 \cdot (\ell + k_1+k_2),
\label{loopkinematicshift}
\end{align}
the expression (\ref{loopfourpoint}) remains unchanged 
as a result of the Jacobi identities
\be
0 ~=~   c_{[14]23} + c_{1234} - c_{1423} 
  ~=~ c_{1[24]3} + c_{1423} - c_{1243} 
  ~=~ c_{12[34]} + c_{1243} - c_{1234}
\label{loopcolorjacobi}
\ee
which means that \eqn{loopkinematicshift}
corresponds to a generalized gauge transformation.
\para

\begin{figure}
\begin{center}
\includegraphics[width=7.0cm]{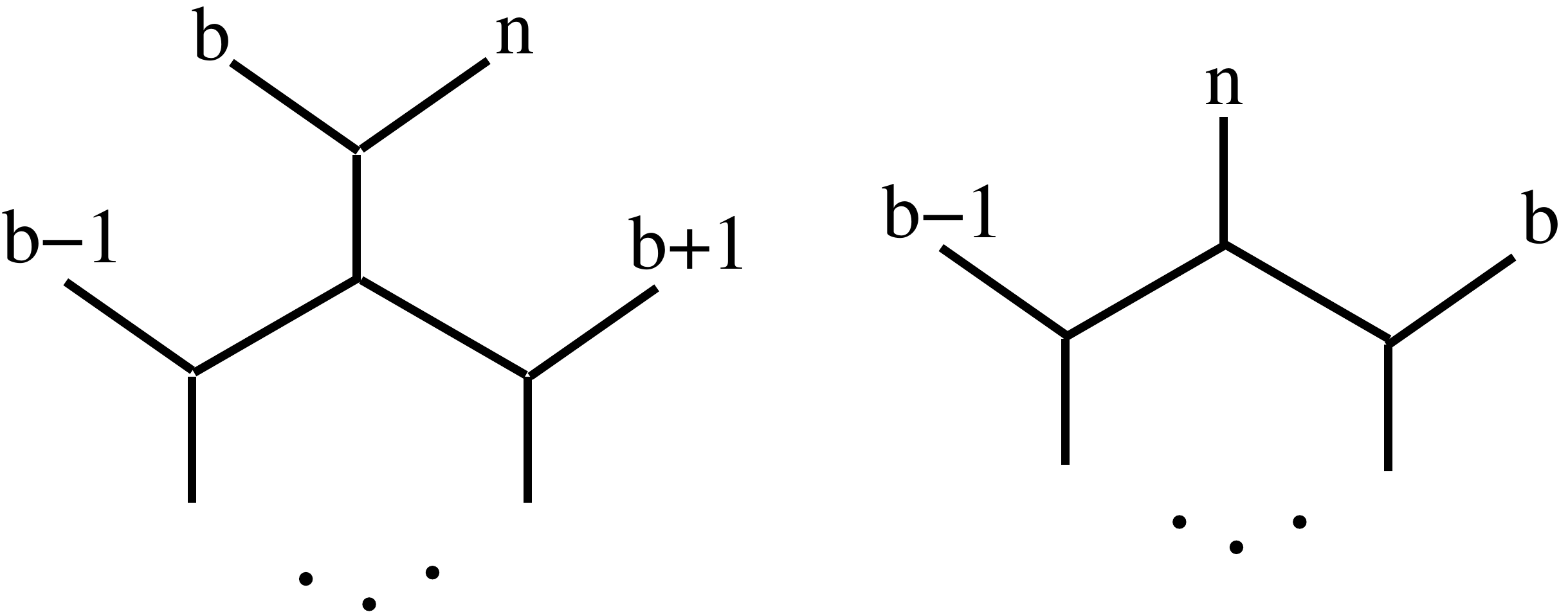}
\caption{Diagrams with color factors 
 $c_{\cdots [bn] \cdots}$ and  $ c_{\cdots b-1,n,b \cdots}$ } 
\label{fig:five}
\end{center}
\end{figure}

The $n$-point generalization of \eqn{loopfourpoint},
obtained by attaching gluon $n$ to the $(n-1)$-gon 
diagram, is given by 
\begin{align}
\cA_{n, (n-1){\rm gon}}^{(1)}
 ~=~& \int {d^D \ell \over (2\pi)^D} \sum_{b=1}^{n-1}
 \Bigg\{  { 1 \over 
\prod_{c=1}^{b} 
\left( \ell + \sum_{a=1}^{c-1} k_a\right)^2
\prod_{d=b+1}^{n-1} 
\left( \ell - \sum_{a=d}^{n-1} k_a\right)^2
}
\nn\\
&
\times \left[  
   { 
c_{\cdots [bn] \cdots} n_{\cdots [bn] \cdots}
\over 2 k_n \cdot k_b
}
~+~ { 
c_{\cdots b-1,n,b \cdots} n_{\cdots b-1,n,b \cdots}
\over 2 k_n \cdot \left( \ell + \sum_{a=1}^{b-1}k_a \right) 
}
~-~ { 
c_{\cdots b,n,b+1 \cdots} n_{\cdots b,n,b+1 \cdots}
\over 2 k_n \cdot \left( \ell + \sum_{a=1}^{b} k_a \right) 
}
\right]
\Bigg\}
\label{loopnpoint}
\end{align}
where 
\begin{align}
c_{12 \cdots n} 
& ~\equiv~  \sum_{\tb_1,\ldots,\tb_n} 
f_{\tb_1 \ta_1 \tb_2}
f_{\tb_2 \ta_2 \tb_3}
\cdots 
f_{\tb_n \ta_n \tb_1} \,,
\nn\\
c_{[12]3 \cdots n} 
& ~\equiv~  \sum_{\tb_1,\ldots,\tb_n} 
f_{\ta_1 \ta_2 \tb_2}
f_{\tb_1 \tb_2 \tb_3}
f_{\tb_3 \ta_3 \tb_3}
\cdots 
f_{\tb_n \ta_n \tb_1} \,.
\label{loopcolorfactors}
\end{align}
The ``ring'' diagrams $c_{12 \cdots n}$ are cyclically symmetric
and reflection symmetric 
\be 
c_{12\dots n} = (-1)^n c_{n\cdots 21} \,.
\label{reflection}
\ee
The color factors appearing in \eqn{loopnpoint} are shown in fig.~\ref{fig:five}
and satisfy the Jacobi relations
\be
0 ~=~ 
c_{\cdots [bn] \cdots} 
~+~ 
c_{\cdots b-1,n,b \cdots} 
~-~ 
c_{\cdots b,n,b+1 \cdots}   \,.
\label{loopnpointjacobi}
\ee
Again, if a particle of mass $m$ is circulating in the loop, 
we replace terms of the form $(\ell + \sum k)^2$ 
with $(\ell + \sum k)^2-m^2$.
Similar expressions are obtained by attaching gluon $n$ to the other
one-loop $(n-1)$-point diagrams $I$,
including $(n-1)$-gon diagrams with other permutations of the external legs
$\{ 1, \cdots, n-1 \}$,
\para

Recently, partial fraction identities similar to \eqn{splitprop}
have been employed to recast one-loop amplitudes into a 
new form whose denominators contain factors linear in the loop momentum
\cite{Geyer:2015bja,Baadsgaard:2015hia,He:2015yua,Baadsgaard:2015twa,Geyer:2015jch,Cachazo:2015aol},
somewhat analogous to \eqn{loopnpoint}.
These new expressions are those that 
naturally emerge from a scattering-equation approach 
to loop-level amplitudes.

\subsection{Color-factor symmetry at one-loop level}

Next we consider the behavior of one-loop amplitudes
under momentum-dependent shifts of its color factors.
First we must define a set of shifts 
consistent with the requirements elucidated  earlier in the paper.
For the one-loop four-gluon amplitude,
the color-factor shift associated with gluon 4
must satisfy
\be
\delta_4 c_{[14]23}  ~=~ \alpha\,  k_4 \cdot k_1, \qquad
\delta_4 c_{1[24]3}  ~=~ \alpha\,  k_4 \cdot k_2, \qquad
\delta_4 c_{12[34]} ~=~ \alpha\,  k_4 \cdot k_3
\label{loopfourcommshift}
\ee
since these diagrams have gluon 4 attached to an external leg.
Requiring the color-factor shift to respect the Jacobi relations
(\ref{loopcolorjacobi}) implies that
\be
\delta_4 c_{1423} ~=~ \delta_4 c_{1234} + \alpha\,  k_4 \cdot k_1
\qquad
\qquad
\delta_4 c_{1243} ~=~ \delta_4 c_{1234} +  \alpha\,  k_4 \cdot (k_1+k_2)  \,.
\label{loopfourjacobishift}
\ee
Unlike the tree-level case, however, these requirements alone are not sufficient
to fix the values of all the shifts;
one of them, $\delta_4 c_{1234}$, remains arbitrary.
In analogy with \eqn{loopkinematicshift},
we define the remaining shift to be
\be
\delta_4 c_{1234}   ~=~ \alpha\,  k_4 \cdot \ell 
\label{loopfourshift}
\ee
where $\ell$ is the loop momentum.\footnote{See footnote \ref{loopfootnote}.}
The effect of this shift on \eqn{loopfourpoint} is
\begin{align}
\delta_4 \cA_{4,{\rm tri}}^{(1)}
~=~{\alpha \over 2}
\int {d^D \ell \over (2\pi)^D}
\bigg[
& { n_{[14]23} +n_{1234} -n_{1423} \over \ell^2 (\ell-k_2 -k_3)^2 (\ell-k_3)^2}
~+~{ n_{1[24]3} +n_{1423} -n_{1243} \over \ell^2 (\ell+k_1)^2 (\ell-k_3)^2}
\nn\\
&
~+~{ n_{12[34]} +n_{1243} -n_{1234} \over \ell^2 (\ell+k_1)^2 (\ell+k_1 +k_2)^2 }
\bigg] \,.
\label{loopsumoverdelta}
\end{align}
Similar expressions are obtained 
for the contributions to the cubic vertex expansion 
from the other three-point diagrams.
\para

A goal consistent with the development in this paper 
would be to prove by some alternative means 
(such as the radiation vertex expansion)
that $\delta_4 \cA^{(1)}_{4}$ vanishes under the one-loop color-factor shift.   
That would imply the vanishing of the r.~h.~s.~of \eqn{loopsumoverdelta}
plus the expressions obtained from the other three-point diagrams,
imposing a generalized-gauge-invariant constraint on the 
one-loop kinematic numerators (namely, that the terms in
the square brackets in \eqn{loopsumoverdelta}
add up to something that integrates to zero).
But at this point in the development of the subject, 
we have no such proof, and therefore we will turn the argument around,
and use the knowledge that there exist kinematic numerators 
for the one-loop four-gluon amplitude that obey color-kinematic duality 
\be
0 ~=~   n_{[14]23} + n_{1234} - n_{1423} 
  ~=~ n_{1[24]3} + n_{1423} - n_{1243} 
  ~=~ n_{12[34]} + n_{1243} - n_{1234}
\label{loopkinematicjacobi}
\ee
for pure Yang-Mills theory (with only gluons circulating
in the loop) \cite{Bern:2013yya}
as well as for theories with other particles 
circulating in the loop \cite{Bern:2008qj,Nohle:2013bfa}.
In these cases, the 
kinematic Jacobi identities (\ref{loopkinematicjacobi}) 
imply that \eqn{loopsumoverdelta} vanishes,
as do the other contributions to the cubic vertex expansion.
Thus, the one-loop four-gluon 
amplitude in these theories 
is invariant under the color-factor shift specified by 
eqs.~(\ref{loopfourcommshift}),
(\ref{loopfourjacobishift}),
and (\ref{loopfourshift}).
\para

The cubic vertex expansion of the one-loop four-point amplitude 
of the bi-adjoint scalar theory may be obtained by 
replacing the kinematic numerators $n_i$ with
a second copy of the color factors ${\tilde c}_i$. 
Since the latter obey the one-loop color Jacobi 
identities (\ref{loopcolorjacobi}), 
the bi-adjoint scalar one-loop four-point amplitude 
is also invariant under the color-factor shift. 
\para

It is known that an independent basis of one-loop color factors 
are those associated with ring diagrams 
modulo cyclic permutations and reflections \cite{DelDuca:1999rs}.
Therefore the one-loop $n$-gluon amplitude can be written 
\be
\cA_n^{(1)} ~=~ 
\int {d^D \ell \over (2\pi)^D}
\sum_{\gamma \in S_{n-1}/\Z_2}  c_{1\gamma}
~ I(1,  \gamma(2), \cdots, \gamma(n)) 
\label{ringbasis}
\ee
where $\gamma$ is a permutation of $\{2, \cdots, n \}$,
$\Z_2$ denotes the reflection symmetry
$c_{123\dots n} \to c_{1n \cdots 32}$,
and  $I(1,  \gamma(2), \cdots, \gamma(n))$
are the integrands of the one-loop color-ordered amplitudes.
\Eqn{ringbasis} may be regarded as the result 
of a generalized gauge transformation in which the kinematic
numerators associated with the non-ring color factors are
set to zero \cite{Bern:2011rj}.
Specializing to the one-loop four-gluon amplitude, \eqn{ringbasis} gives 
\be
\cA_4^{(1)} ~=~ 
\int {d^D \ell \over (2\pi)^D}
\left[ 
  c_{1234} I(1,2,3,4)
~+~ c_{1423} I(1,4,2,3)
~+~ c_{1243} I(1,2,4,3)
\right] \,.
\ee
For theories whose numerators respect one-loop color-kinematic duality,
the invariance of the one-loop amplitude under the color-factor shift implies
the following condition on the integrands
\be
0 ~=~ 
\int {d^D \ell \over (2\pi)^D}
\left[ 
k_4 \cdot \ell ~I(1,2,3,4)
~+~ k_4 \cdot (\ell + k_1) ~I(1,4,2,3)
~+~k_4 \cdot (\ell + k_1+k_2) ~I(1,2,4,3)
\right] \,.
\label{integrandfourpointconstraint}
\ee
Relations of this form were first uncovered in 
refs.~\cite{Boels:2011tp,Boels:2011mn,Du:2012mt} 
from the perspective of on-shell recursion relations,
and revisited recently using 
monodromy relations in string theory \cite{Tourkine:2016bak}.
(See also ref.~\cite{Chester:2016ojq} for 
BCJ-type relations among loop-level integrands.)
Conversely, if the integrands of the one-loop amplitude 
of a theory can be shown to satisfy \eqn{integrandfourpointconstraint},
then it follows that the one-loop amplitude is invariant 
under the color-factor shift, 
avoiding the need to invoke color-kinematic duality. 
\para

It is straightforward to generalize these considerations 
to one-loop $n$-gluon amplitudes.
We may define a color-factor shift associated with any external gluon $a$, 
but for simplicity of presentation we will focus on gluon $n$.
Let $\sigma$ denote a permutation of $\{ 2, \cdots, n-1 \}$.
The color factor $c_{\cdots [\sigma(b) n] \cdots}$ 
shown in fig.~\ref{fig:five} undergoes a shift 
\be
\delta_n c_{\cdots [\sigma(b) n] \cdots} ~\propto~ k_n \cdot k_{\sigma(b)} 
\ee
because gluon $n$ is attached to gluon leg $\sigma(b)$.
Requiring the shifts to respect the Jacobi relation (\ref{loopnpointjacobi})
implies
\be
\delta_n c_{ \cdots \sigma(b) n \sigma(b+1) \cdots }
~=~ \delta_n c_{ \cdots \sigma(b-1) n \sigma(b) \cdots }
~+~ \delta_n c_{\cdots [\sigma(b) n] \cdots} 
\ee
We must additionally define the shifts 
\be
\delta_n c_{ 1 \sigma(2) \cdots \sigma(n-1) n } = \alpha_{n,\sigma} \, k_n \cdot \ell
\label{additionalshift}
\ee
for a set of half\footnote{The shifts of the other half are then determined
by the reflection symmetry (\ref{reflection}).  
For example for the four-point case 
$\delta_4 c_{1234} = \alpha_{4,23}  k_4 \cdot \ell$
but
$\delta_4 c_{1324} = \delta_4 c_{1423} 
= \alpha_{4,23}  k_4 \cdot (\ell + k_1)$, 
which is not of the form (\ref{additionalshift}).
}
of the permutations $\sigma$,
where $\alpha_{n,\sigma}$ are a set of $(n-2)!/2$ 
independent arbitrary constants.
Together these conditions imply that the shifts of the ring color factors are given by
\be
\delta_n c_{1 \sigma(2)  \cdots \sigma(b-1) n \sigma(b) \cdots \sigma(n)}
~=~ 
\alpha_{n,\sigma} 
\left( k_n \cdot  \ell+  k_n \cdot k_1  +  \sum_{c=2}^{b-1} k_n \cdot  k_{\sigma(c)} \right), 
\qquad b \in \{2, \cdots, n-1\}
\label{loopnpointshift}
\ee
and the shifts of all other one-loop color factors are fixed 
by requiring that they respect the Jacobi relations.
Thus there is an $(n-2)!/2$-dimensional family of 
one-loop color-factor shifts associated with gluon $n$.
\para

Applying the shift (\ref{loopnpointshift}) to \eqn{loopnpoint}, we obtain
\be
\delta_n \cA_{n, (n-1){\rm gon}}^{(1)}
  ~\propto~
\int {d^D \ell \over (2\pi)^D} \sum_{b=1}^{n-1}
{ 
\left( n_{\cdots [bn] \cdots} ~+~  n_{\cdots b-1,n,b \cdots} ~-~ n_{\cdots b,n,b+1 \cdots}
\right)
 \over 
\prod_{c=1}^{b} 
\left( \ell + \sum_{a=1}^{c-1} k_a\right)^2
\prod_{d=b+1}^{n-1} 
\left( \ell - \sum_{a=d}^{n-1} k_a\right)^2
}  \,.
\label{loopsumoverdeltanpoint}
\ee
For $\cN=4$ supersymmetric Yang-Mills theory,
numerators for
the one-loop five-gluon \cite{Carrasco:2011mn}
and higher-point \cite{Yuan:2012rg,Bjerrum-Bohr:2013iza} amplitudes
have been constructed which satisfy color-kinematic duality
\be
0 ~=~ 
n_{\cdots [bn] \cdots} 
~+~ 
n_{\cdots b-1,n,b \cdots} 
~-~ 
n_{\cdots b,n,b+1 \cdots}  
\ee
and which therefore imply that \eqn{loopsumoverdeltanpoint} vanishes.
The shifts of the terms in the cubic vertex expansion 
obtained from other $(n-1)$-point diagrams $I$
similarly vanish.
Thus, we have established that these amplitudes 
possess one-loop color-factor symmetry.
The one-loop $n$-point amplitudes of the bi-adjoint scalar theory
also possess this symmetry because the second copy 
of the color factors ${\tilde c}_i$ obey Jacobi identities
(\ref{loopnpointjacobi}).
\para

As we did above for the four-gluon amplitude, 
we can use this invariance 
to derive constraints on the integrands of color-ordered amplitudes.
\Eqn{ringbasis}  can be rewritten  as 
\be
\cA_n^{(1)} ~=~ 
\sum_{\sigma \in S_{n-2}/\Z_2}  
\int {d^D \ell \over (2\pi)^D}
\sum_{b=2}^n 
c_{1 \sigma(2) \cdots \sigma(b-1)n \sigma(b) \cdots \sigma(n-1)}
~ I(1, \sigma(2), \cdots ,\sigma(b-1), n , \sigma(b) ,\cdots, \sigma(n-1))
\ee
Invariance of this expression under \eqn{loopnpointshift} 
together the independence of the parameters $\alpha_{n,\sigma}$
yields
\be
0=
\int {d^D \ell \over (2\pi)^D}
\sum_{b=2}^n 
k_n \cdot \left( \ell+  k_1  
+  \sum_{c=2}^{b-1} k_{\sigma(c)} \right)
~ I(1, \sigma(2), \cdots ,\sigma(b-1), n , \sigma(b) ,\cdots, \sigma(n-1))
\label{integrandnpointconstraint}
\ee
the relations uncovered in 
refs.~\cite{Boels:2011tp,Boels:2011mn,Du:2012mt,Tourkine:2016bak}.
Conversely, if we were to establish that the
integrands of the color-ordered amplitudes of a given theory
satisfy \eqn{integrandnpointconstraint},
we would have proven that the one-loop amplitude is invariant under
the color-factor shift, 
independently of the assumption of color-kinematic duality.  
Further study of this alternate path is merited.

\section{Discussion and conclusions}
\setcounter{equation}{0}
\label{sec:concl}

In this paper, we have introduced a new set of 
symmetries of gauge-theory amplitudes, 
which act as momentum-dependent shifts on the color factors
appearing in the cubic decomposition of the amplitude.
These symmetries are intimately linked to the presence 
of massless gauge bosons in the amplitude (or massless adjoint 
scalars in the case of the bi-adjoint scalar theory)
and can be considered generalizations of the radiation symmetry 
of ref.~\cite{Brown:1983pn}.
We demonstrated that a wide class of 
tree-level gauge-theory amplitudes 
are invariant under these shifts, 
using a representation of the amplitude known as 
the radiation vertex expansion \cite{Brown:1982xx}.  
We also introduced a related but 
distinct cubic vertex expansion of the amplitude,
and used this to derive a set of generalized-gauge-invariant
constraints on the kinematic numerators appearing 
in the cubic decomposition of the amplitude.
All known BCJ relations for tree-level gauge-theory amplitudes 
\cite{Bern:2008qj,BjerrumBohr:2009rd,Feng:2010my,Naculich:2014naa,Johansson:2015oia}
follow as a direct consequence of the color-factor symmetry
(this paper and ref.~\cite{Brown:2016hck}).
Finally, we generalized the cubic vertex expansion and color-factor
symmetry to loop level. 
We showed that one-loop amplitudes that 
satisfy color-kinematic duality are invariant 
under the one-loop color-factor symmetry,
and derived a set of relations among the integrands 
of one-loop color-ordered amplitudes.
\para

Let us take a look at the connection between the color-factor symmetry 
and more fundamental symmetries of the Lagrangian,
gauge and Poincar\'e invariance  \cite{Brown:1982xx}. 
The color-factor symmetry follows as a result of 
the vanishing of certain expressions,  namely 
eqs.~(\ref{threesums}), (\ref{variationspinhalf}),
and (\ref{threesumsfund}),
associated with the cubic vertices of a gauge-theory amplitude,
and eqs.~(\ref{variationfourglounleading}),
(\ref{variationfourglounsubleading}),
and (\ref{variationquarticspinzero}),
associated with the quartic vertices.
It is illustrative to examine the various contributions
in a soft expansion in the gluon momentum $k_a$,
even though the color-factor symmetry is exact in $k_a$.
\para

The leading term in the soft expansion corresponds to 
the $\cO(k_a^0)$ term in each of eqs.~(\ref{threesums}), 
(\ref{variationfourglounleading}),
(\ref{variationspinhalf}), (\ref{threesumsfund}),
and (\ref{variationquarticspinzero}). 
These are all proportional to 
$\sum_r \ve_a \cdot K_r$ where $K_r$ are the momenta
flowing out of each leg of the vertex.
This vanishes by $\ve_a \cdot k_a = 0$
together with momentum conservation 
$ k_a + \sum_r K_r=0 $ ---
a result of symmetry under spacetime translations.
\para
 
The subleading term in the soft expansion corresponds to
the $\cO(k_a^1)$ term in each of 
eqs.~(\ref{threesums}), 
(\ref{variationspinhalf}), 
(\ref{threesumsfund}), and
(\ref{variationquarticspinzero}),
and to 
\eqn{variationfourglounsubleading}.
These expressions are all given by a sum of
angular momentum generators $J_r^{\alpha\beta}$,
which act as a first-order Lorentz transformation 
on the relevant vertex factors.
They vanish by Lorentz invariance.
\para

Thus the first two terms in the soft expansion 
vanish as a result of Poincar\'e invariance.
It is a little more difficult to pin down the underlying symmetry
responsible for the vanishing of the sub-subleading terms in 
\eqns{threesums}{threesumsfund},
together with an analogous expression for spin-one particles.
The $\cO(k_a^2)$ term in \eqn{threesums}
is proportional to 
\begin{align}
&
(S_1^{\alpha\beta} )^{\mu_1}_{~~ \nu} 
{ \partial \over \partial K_{1\gamma} } V^{\nu \mu_2 \mu_3} (K_1, K_2, K_3)
+ 
(S_2^{\alpha\beta} )^{\mu_2 }_{~~\nu} 
{\partial \over \partial K_{2\gamma} } V^{\mu_1 \nu\mu_3} (K_1, K_2, K_3)
+ 
(S_3^{\alpha\beta} )^{\mu_3 }_{~~\nu} 
{\partial \over \partial K_{3\gamma} } V^{\mu_1 \mu_2\nu} (K_1, K_2, K_3)
\nn\\
&~=~ 2 i 
\left(
-\eta^{\alpha\mu_1} \eta^{\beta\mu_2} \eta^{\gamma\mu_3}
+\eta^{\alpha\mu_1} \eta^{\gamma\mu_2} \eta^{\beta\mu_3} \right) 
 +\hbox{(cyclic permutations of $123$)} 
\end{align}
and the $\cO(k_a^2)$ term in \eqn{threesumsfund}
is proportional to 
\be
(S_3^{\alpha\beta})^{\mu_3}_{~~\nu}  {\partial \over \partial  K_{3\gamma}   }
V^\nu (K_1 , K_2, K_3)
= 
\lambda \left(
\eta^{\alpha\mu_3} \eta^{\beta \gamma}
-\eta^{\beta\mu_3} \eta^{\alpha \gamma}
\right) \,.
\ee
Neither expression vanishes by itself
but both do when contracted with 
$\ve_{a\alpha} k_{a\beta} k_{a\gamma} $ for gluon $a$.
These identities,
which go beyond the first-order Poincar\'e cancellation
and are connected to Yang-Mills gauge symmetry,
are key ingredients contributing to the color-factor symmetry.
\para

Returning to our discussion in the introduction
of the connection of the color-factor symmetry
with the photon radiation symmetry and radiation zeros 
(for a collection of early references, 
see refs.~\cite{Zhu:1980sz,Goebel:1980es,Brown:1982xx,Brown:1983pn,Brodsky:1982sh,Brown:1983vk,Mikaelian:1977ux,Brown:1979ux,Mikaelian:1979nr,Samuel:1983eg,Naculich:1983pg}),
we have uncovered some additional analogs.  
For example, we can write a factorized form 
for the four-gluon amplitude (\ref{fourpointamplitude})
\be
\cA_4 = s 
\left( {c_s \over s} - {c_t \over t} \right)
\left( {n_s \over s} - {n_u \over u} \right)
\ee
which vanishes at $ c_s/s = c_t / t =c_u/u =  $ const. 
This is a non-abelian version of the original radiation zero 
studied almost forty years ago in
$q \bar{q} W \gamma$ and $e \nu W \gamma$ reactions,
with a zero at $Q_c /k_c \cdot k_a =$ constant.  
This original radiation factorization and its zero 
led to the prediction of a measurable experimental dip,
which has now been 
confirmed \cite{Abazov:2008ad,Chatrchyan:2011rr}.
The analogous zero in the four-gluon amplitude is washed out, 
however, by the color averaging that must be performed in 
observable quark-gluon processes.
\para
 
In the generalization to tree-level $n$-point amplitudes, 
the abelian radiation symmetry 
and existence of zeros for $Q_c /k_c \cdot k_a = $ constant
(for photon momentum $k_a$)
rest on having gauge-theory couplings, as noted earlier.  
The non-abelian color-factor symmetry uncovered in this paper
can also lead to zeros in $n$-point amplitudes, 
but with an important complication.
The invariance under
$ c_i  \to  c_i + \alpha_i \sum_{c \in \Sai} k_c \cdot k_a $ 
for the attachment of a gluon with momentum $k_a$ 
cannot be used to systematically cancel out the complete
$n$-gluon amplitude using an overall common value for $\alpha_i$.  
Because of the Jacobi relations, that overall common value must vanish.  
There are in principle zeros, however,  for separate islands of 
$\alpha_i$ values.  
Although they are again washed out by color-averaging, 
the generalized factorization coming from the color-factor
symmetry remains useful for theoretical analysis
of tree amplitudes.  
In a different direction, note that the BCJ form of the gluon 
amplitudes has been utilized in the planar zeros studied recently 
in refs.~\cite{Harland-Lang:2015faa,Jimenez:2016sgc}.
\para

The analogs described above are a bridge to a final overall remark.  
It has been helpful to think of gluon emission or absorption 
as effecting a (first-order) transformation in both 
color and kinematic space simultaneously 
on the graph to which it is attached.
In particular, the attachments lead to transformations 
of the various legs and vertices of the 
``parent'' diagram in either momentum or space-time representations.  
All the parent wave functions end up transformed,  
and identities derived from \eqn{splitprop} for the different spins, 
\eg \eqns{fermionpropidentity}{vectorpropidentity},
yield exactly the two terms expected from the propagator 
with its bilinearity in the wave functions.  
The cancellations highlighted throughout this paper 
arise precisely because we consider those theories whose 
amplitudes transform covariantly under color and 
kinematic transformations.  
Adding all possible massless gluon attachments 
to a complete set of parent graphs leads to a sum of
corresponding color shifts that vanishes because of invariance.  
Such a picture should help in finding directions 
in the diagrammatic analyses of a variety of extensions
of the gauge theories considered in this paper,
supersymmetric and otherwise. 

\section*{Acknowledgments}
This material is based upon work supported by the 
National Science Foundation 
under Grants Nos.~PHY14-16123 and PFI:BIC 1318206.
RWB is also supported by funds 
made available through a CWRU Institute Professorship Chair.
SGN gratefully acknowledges sabbatical support
from the Simons Foundation (Grant No.~342554 to Stephen Naculich).
He would also like to thank the Michigan Center for Theoretical Physics
and the Physics Department of the University of Michigan 
for generous hospitality 
and for providing a welcoming and stimulating sabbatical environment.

\begin{appendix}

\section{Five-gluon amplitudes}
\setcounter{equation}{0}
\label{sec:app}

In this appendix, we use the five-gluon amplitude to 
provide an explicit example of the cubic vertex expansion (\ref{cve})
introduced in sec.~\ref{sec:cfs},
and the relations among kinematic numerators (\ref{sumoverdelta})
resulting from the color-factor symmetry.
\para

\begin{figure}
\begin{center}
\includegraphics[width=3.0cm]{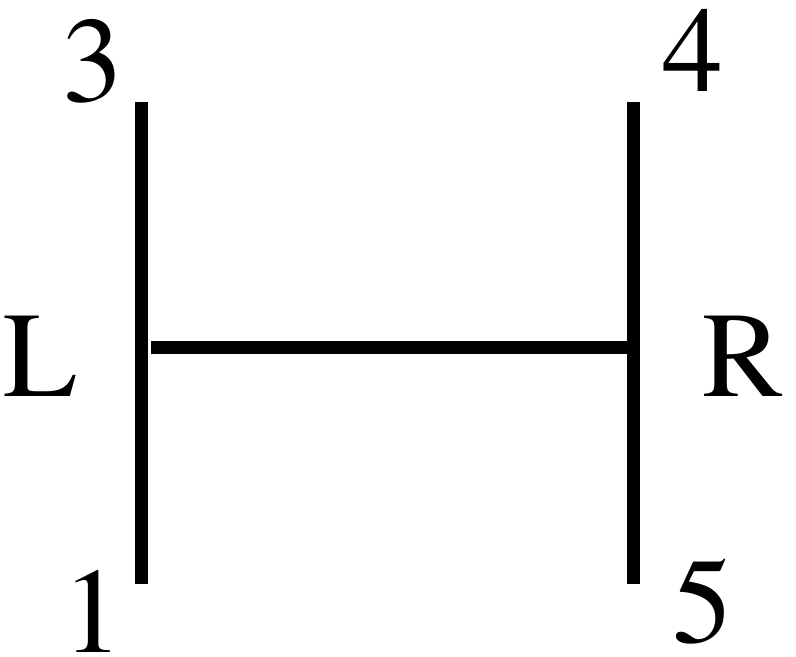}
\caption{One of the four-gluon diagrams to which gluon 2 
is added in all possible ways. }
\label{fig:seven}
\end{center}
\end{figure}

The cubic decomposition (\ref{cubicdecomp}) 
of the five-gluon amplitude is given by
\begin{eqnarray}
\cA_5
&=&
{\bc_{12345} ~n_{12345} \over s_{12} s_{45} } ~+~     
{\bc_{32145} ~n_{32145} \over s_{23} s_{45} } ~+~    
{\bc_{13245} ~n_{13245} \over s_{13} s_{45} } ~+~    
{\bc_{13425} ~n_{13425} \over s_{13} s_{25} } ~+~    
{\bc_{13524} ~n_{13524} \over s_{13} s_{24} }     
\nn\\
&+&
{\bc_{12435} ~n_{12435} \over s_{12} s_{35} } ~+~  
{\bc_{42135} ~n_{42135} \over s_{24} s_{35} } ~+~  
{\bc_{14235} ~n_{14235} \over s_{14} s_{35} } ~+~  
{\bc_{14325} ~n_{14325} \over s_{14} s_{25} } ~+~  
{\bc_{14523} ~n_{14523} \over s_{14} s_{23} }   
\nn\\
&+&
{\bc_{42315} ~n_{42315} \over s_{24} s_{15} } ~+~  
{\bc_{32415} ~n_{32415} \over s_{23} s_{15} } ~+~  
{\bc_{34215} ~n_{34215} \over s_{34} s_{15} } ~+~  
{\bc_{34125} ~n_{34125} \over s_{34} s_{25} } ~+~  
{\bc_{34512} ~n_{34512} \over s_{34} s_{12} }   
\nn\\
\label{fivegluoncubicdecomp}
\end{eqnarray}
where $\bc_{\alpha}$ are half-ladder color factors defined in \eqn{halfladder}.
Let us recast this amplitude in a 
cubic vertex expansion with respect to gluon 2.
We have already arranged the terms in \eqn{fivegluoncubicdecomp}
so that each line corresponds to 
one of the four-gluon diagrams $I$ obtained by omitting gluon 2.
We rewrite the denominator of the third term of the first line as 
\be
{ 1\over s_{13} s_{45} }
~=~ {1\over s_{45} (-s_{12} - s_{23}) }  ~+~ {1\over s_{13} (-s_{24} - s_{25}) } 
\ee
and similarly the denominators of the third terms of the other two lines to obtain
\begin{eqnarray}
\cA_5
&=&
{1 \over s_{45}} \left( 
{\bc_{12345} ~n_{12345} \over s_{12} } ~+~     
{\bc_{32145} ~n_{32145} \over s_{23} } ~-~    
{\bc_{13245} ~n_{13245} \over s_{12} + s_{23}  }      
\right)  \nn\\
&+& 
{1 \over s_{13}} \left( 
-~ {\bc_{13245} ~n_{13245} \over s_{24}+ s_{25} } ~+~    
{\bc_{13425} ~n_{13425} \over s_{25} } ~+~    
{\bc_{13524} ~n_{13524} \over s_{24} }     
\right)
\nn\\
&+&
{1 \over s_{35}} \left( 
{\bc_{12435} ~n_{12435} \over s_{12} } ~+~  
{\bc_{42135} ~n_{42135} \over s_{24} } ~-~  
{\bc_{14235} ~n_{14235} \over s_{12}~+~ s_{24} }	 
\right)
\nn\\
&+&
{1 \over s_{14}} \left( 
-~{\bc_{14235} ~n_{14235} \over s_{23}+ s_{25} } ~+~  
{\bc_{14325} ~n_{14325} \over s_{25} } ~+~  
{\bc_{14523} ~n_{14523} \over s_{23} }   
\right) 
\nn\\
&+&
{1 \over s_{15}} \left( 
{\bc_{42315} ~n_{42315} \over s_{24} } ~+~  
{\bc_{32415} ~n_{32415} \over s_{23} } ~-~  
{\bc_{34215} ~n_{34215} \over s_{23}+ s_{24} }   
\right)
\nn\\
&+&
{1 \over s_{34}} \left( 
-~ {\bc_{34215} ~n_{34215} \over s_{12}+s_{25}  } ~+~  
{\bc_{34125} ~n_{34125} \over s_{25} } ~+~  
{\bc_{34512} ~n_{34512} \over s_{12} }   
\right)
\label{fivegluonvertexexpansion}
\end{eqnarray}
which is precisely of the form of the cubic vertex expansion (\ref{cve}).
To make this connection more explicit,
note that the first two lines of 
\eqn{fivegluonvertexexpansion} correspond to adding gluon 2 to
the four-gluon diagram shown in fig.~\ref{fig:seven}
which we label as $I=1$.
The color factors $c_{(a,I,v,r)}$  (see fig.~\ref{fig:one})
associated with the left- and right-hand vertices of this diagram are
\begin{align}
c_{(2,1,L,1)} &= \bc_{12345} &  
c_{(2,1,L,2)} &= -~\bc_{32145} &  
c_{(2,1,L,3)} &= -~\bc_{13245}  \nn\\
c_{(2,1,R,1)} &= \bc_{13524} &  
c_{(2,1,R,2)} &= -~\bc_{13425} &  
c_{(2,1,R,3)} &= \bc_{13245}  
\end{align}
and obey $\sumr c_{(a,I,v,r)} ~=~ 0  $.
The relative signs result from flipping legs across lines.
Because $\bc_{13245}$ is associated with both left- and right-hand vertices,
we have $c_{(2,1,L,3)} = -c_{(2,1,R,3)}$.
Since $K_{(2,1,L,3)} = -K_{(2,1,R,3)}$, this implies
that $\alpha_{(2,I,L)}=  \alpha_{(2,I,R)}$ as discussed in subsection
\ref{subsec:kinematicnumeratorconstraint}.
\para

The six independent five-gluon half-ladder color factors 
(in the notation of ref.~\cite{Bern:2008qj}) are 
\begin{align}
c_{1\phantom{0}} &= \bc_{12345} &  c_{15}&= \bc_{13245 }& c_{9\phantom{1}} &= \bc_{13425}
\nn\\
c_{12}&= \bc_{12435} & c_{14}&= \bc_{14235} &c_{6\phantom{1}} &= \bc_{14325}. 
\end{align}
According to \eqn{halfladdershift},
the color-factor shifts associated with gluon 2 act as 
\begin{align}
\deltatwo  c_{1\phantom{0}} &= \alp s_{12}  &
\deltatwo  c_{15}&= \alp (s_{12}+s_{23}) & 
\deltatwo  c_{9\phantom{1}} &= - \alp s_{25} \nn\\
\deltatwo  c_{12}&= \bet s_{12} &
\deltatwo  c_{14}&= \bet (s_{12}+s_{24}) &
\deltatwo  c_{6\phantom{1}} &= - \bet s_{25}
\end{align}
where $\alp = \alpha_{2,34}$ and $\bet=\alpha_{2,43}$
are arbitrary constants.
The nine remaining five-gluon color factors, 
and the action of the color-factor shifts thereon,
are determined by the Jacobi relations to be
\begin{align}
c_2 &= \bc_{23451} = c_1 + c_6 - c_{14} - c_{15}   , & 
\deltatwo  c_2 &=(\bet-\alp)  s_{23} \nn\\
c_3 &= \bc_{34512}= c_1 - c_{12}    , & 
\deltatwo  c_3 &=(\alp-\bet)  s_{12} \nn\\
c_4 &= \bc_{45123}   = c_1 - c_{15} , & 
\deltatwo  c_4 &=- \alp  s_{23} \nn\\
c_5 &= \bc_{51234} = c_1 + c_6 - c_{9} - c_{12} , & 
\deltatwo  c_5 &=(\alp-\bet)  (s_{12}+s_{25})  \nn\\
c_7 &= \bc_{32514}  = c_6 - c_{14}  , & 
\deltatwo  c_7 &=\bet  s_{23} \nn\\
c_8 &= \bc_{25143}  = c_6 - c_9 , & 
\deltatwo  c_8 &=(\alp-\bet)  s_{25} \nn\\
c_{10} &=  \bc_{42513} = c_9 - c_{15} , & 
\deltatwo  c_{10} &=\alp   s_{24}  \nn\\
c_{11} &=  \bc_{51342} = c_9 + c_{12} - c_{14} - c_{15} , & 
\deltatwo  c_{11} &=(\alp-\bet)  s_{24} \nn\\
c_{13} &=  \bc_{35124} =c_{12} - c_{14} , &
\deltatwo  c_{13} &=-\bet  s_{24}  \,.
\end{align}
Applying this shift to \eqn{fivegluonvertexexpansion}, 
we obtain
\begin{eqnarray}
\deltatwo \cA_5
&=& 
\alpha_{2,34} \left( 
  { n_1 - n_4 - n_{15} \over s_{45}  }
~+~ {  n_{15} - n_9 + n_{10} \over s_{13} }
~+~ {  n_{11} - n_2 + n_5 \over s_{15} }
~+~ {   -n_5 + n_8 + n_3 \over s_{34} }
\right)
\\
&+&
\alpha_{2,43} \left( 
  { n_{12}-n_{13} - n_{14}\over s_{35}  }
~+~ {  n_{14} - n_6 + n_7 \over s_{14}  }
~+~{ - n_{11} + n_2 - n_5 \over s_{15} }
~+~{   n_5 - n_8 - n_3 \over s_{34} }
\right)
~=~ 0
\quad
\nn
\end{eqnarray}
which is precisely of the form of \eqn{sumoverdelta}.
The color-factor shift with respect to gluon 3 instead yields
\begin{eqnarray}
{ \delta_3 ~ } \cA_5
&=& 
\alpha_{3,24}  \left( 
    { n_1 - n_4 - n_{15} \over s_{45}  }
~+~ {  n_{3} - n_1 + n_{12} \over s_{12} }
~+~ {  n_{11} - n_2 + n_5 \over s_{15} }
~+~ {  n_{10} - n_{11} + n_{13} \over s_{24} }
\right)
\\
&+&
\alpha_{3,42}  \left( 
  { n_{6}-n_{8} - n_{9}\over s_{25}  }
~+~ {  n_{14} - n_6 + n_7 \over s_{14}  }
~+~{ - n_{11} + n_2 - n_5 \over s_{15} }
~+~ { - n_{10} + n_{11} - n_{13} \over s_{24} }
\right)
~=~ 0 \,.
\quad
\nn
\end{eqnarray}
Since 
$\alpha_{2,34}$,
$\alpha_{2,43}$,
$\alpha_{3,24}$, and
$\alpha_{3,42}$
are independent arbitrary constants, each expression in parentheses vanishes,
yielding four independent constraint equations on the kinematic numerators
of the five-gluon amplitude.
No additional independent constraints are obtained from the 
color-factor symmetries associated with the other three gluons.
These ``generalized Jacobi relations'' for five-gluon amplitudes
were previously derived in refs.~\cite{BjerrumBohr:2010zs,Tye:2010dd} 
by using the properties of string theory amplitudes.

\end{appendix}

\vfil\break


\end{document}